\documentclass[useamsfonts]{pasj01}

\Received{$\langle$reception date$\rangle$}
\Accepted{$\langle$acception date$\rangle$}
\Published{$\langle$publication date$\rangle$}

\begin{document}

\title{
Transonic galactic wind model including stellar feedbacks and application to outflows in high/low-$z$ galaxies
}

%%%%
\author{Asuka \textsc{Igarashi}\altaffilmark{1}}%
\altaffiltext{1}{National Institute of Advanced Industrial Science and Technology (AIST) Tsukuba Central 2, 1-1-1 Umezono, Tsukuba, Ibaraki, 305-8568, Japan}
\author{Masao \textsc{Mori}\altaffilmark{2}}%
\altaffiltext{2}{Center for Computational Sciences, University of Tsukuba, 1-1-1, Tennodai, Tsukuba, Ibaraki, 305-8577, Japan}
\email{mmori@ccs.tsukuba.ac.jp}
\author{Shin-ya \textsc{Nitta}\altaffilmark{3,4,5}}%
\altaffiltext{3}{Tsukuba University of Technology, 4-3-15, Amakubo, Tsukuba, Ibaraki, 305-8520, Japan}
\altaffiltext{4}{Solar Science Observatory, National Astronomical Observatory of Japan, 2-21-1 Osawa, Mitaka, Tokyo, 181-8588, Japan}
\altaffiltext{5}{Institute of Space and Astronautical Science, Japan Aerospace Exploration Agency, 3-1-1 Yoshinodai, Sagamihara, Kanagawa 229-8510, Japan}


\maketitle
%%%%

\begin{abstract}
Galactic winds play a crucial role in the ejection of the interstellar medium (ISM) into intergalactic space. This study presents a model that classifies possible transonic solutions of galactic winds in the gravitational potential of the dark matter halo and stellar component under spherically symmetric and steady assumptions. Our model includes injections of mass and energy resulting from supernovae feedback along a flow line. The mass flux in galactic winds is a critical factor in determining the acceleration process of the flow and revealing the impact of galactic winds on galaxy evolution. We apply the transonic galactic wind model to the observed outflow velocities of star-forming galaxies to estimate the mass flux. Dividing the mass flux by the star formation rate (SFR) yields the mass loading rate (and mass loading factor), which indicates the entrainment effect of the ISM by the hot gas flow. Our results demonstrate that the mass loading rate is inversely correlated with galaxy mass and SFR. In less massive galaxies (stellar mass $\sim 10^{7-8} M_\odot$), the mass loading rate exceeds unity, indicating effective ejection of the ISM into intergalactic space. However, in massive galaxies (stellar mass $\sim 10^{10-11} M_\odot$), the mass loading rate falls below unity, meaning that the mass flux cannot exceed the injected mass by supernovae, thus resulting in the ineffective ejection of the ISM.
\end{abstract}

%\pagewiselinenumbers

\section{Introduction}

% galactic winds 
In the evolution of galaxies, baryons captured in the gravitational potential well of dark matter have evolved through various physical processes such as star formation, gas accretion, gas stripping by the intergalactic medium (IGM), and galaxy mergers \citep{Gunn72, Dekel86, Dekel87, Efstathiou92}.
In particular, gas outflows have a significant impact on the evolution of galaxies.
Such gas outflows are called galactic winds and are observed ubiquitously in star-forming galaxies \citep{Veilleux05, Rupke18}.
Galactic winds eject a large amount of interstellar medium, which strongly suppresses star formation activity \citep{Larson74, Mori97, Mori99, Mori02}.
There have been many studies about the effect of galactic winds on the relation between stellar mass and dark matter halo mass \citep{Behroozi10, Behroozi13, Moster13, Brook14} and between stellar mass and metallicity \citep{Tremonti04, Andrews13, Zahid14}.
Galactic winds also affect metal enrichment in intergalactic space because they contain gas enriched with metals ejected from massive stars.
Observational studies have shown that galactic winds contribute to metal enrichment in the circum-galactic and intergalactic space \citep{Peeples14, Werk14, Songaila97, Ellison00}.
Theoretical studies have also suggested that galactic winds remove large amounts of metals from star-forming regions and determine the metallicity of local and high-redshift galaxies \citep{Dalcanton07, Erb08, Finlator08, Arrigoni10, Peeples11, Dave11, Dayal13, Creasey15, Christensen16}.
Although galactic winds are critical to galaxy evolution, numerical studies frequently parametrize the effects of galactic winds, still leaving them as an elusive problem.

% mass flux of galactic winds 
Mass flux is key in evaluating the impact of galactic winds.
The degree of star formation suppression depends on the mass flux ejected from star-forming galaxies.
Theoretical studies of galaxy formation predicted the amount of mass flux but required some unknown parameters \citep{Hopkins12, Barai15, Mitra15, Muratov15, Christensen16}.
To determine these parameters, they used observational results: the Tully-Fisher relation, the stellar-to-halo mass ratio, and the mass-metallicity relation.
Although they did not reach the same conclusion in detail, they suggested a negative correlation between the mass loading rate (ratio of mass flux to star formation rate SFR) and the dark matter halo mass.
This indicates that the gravity of the host galaxy powerfully brakes the galactic wind with large mass loading rates.

% observation for galactic winds 
From the observational viewpoint, we need both observed velocities and a simple theoretical model to evaluate the mass flux of the galactic wind.
The spectral shift of absorption lines can be used to deduce the velocity along the line of sight \citep{Rupke18}.
This estimation uses several atomic tracers with different ionization potentials.
Some tracers indicate the velocity of warm gas ($\sim$ 10$^{4-5}$ K).
For local star-forming galaxies, the velocity of the galactic wind, traced by the centroid or maximum extent of absorption lines, correlates positively with the SFR and the stellar mass \citep{Martin05, Rupke05a, Rupke05b, Chisholm15, Arribas14, Heckman15, Heckman16}.
Recent observations have also found these correlations in high-$z$ star-forming galaxies \citep{Sugahara17, Sugahara19}.

% shell model 
Observational studies often use the shell model to evaluate the mass flux of the galactic wind from these velocities \citep{Rupke05b, Heckman15}.
Hot gas expands from star-forming regions and forms a shock front.
The hot gas sweeps the ambient cold gas and compresses it into the shock front.
%The result is an expanding thin shell filled with hot gas.
%The shell model evaluates the mass of this expanding structure as the mass flux. The mass flux becomes 
Eventually, an expanding thin shell filled with hot gas is formed. The shell model evaluates the mass of this expanding structure as the mass flux, 
\begin{equation}
\dot{m} = \Omega \: C_{\rm f} \: \mu \: m_{\rm p} \: N_{\rm H} \: v_{\rm sh} \: r_{\rm sh},
\label{eqn_shl_mfx}
\end{equation}
where $\Omega$, $C_{\rm f}$, $\mu$, $m_{\rm p}$, $N_{\rm H}$, $v_{\rm sh}$, and $r_{\rm sh}$ are the solid angle of the galactic wind ($\sim 4 \pi$), covering factor, molecular weight, proton mass, column density, shell velocity, and shell radius, respectively.
The shell model uses the observed central velocity in $v_{\rm sh}$.
The mass flux estimated from the shell model correlates positively with the SFR and stellar mass \citep{Martin05, Heckman15}.
These correlations have significantly large scatters due to the potential ambiguities of this model.
The model requires $N_{\rm H}$, which is uncertain due to the conversion from metallicity to hydrogen gas mass.
Also, $r_{\rm sh}$ requires assumptions because it is difficult to determine observationally.
Therefore, we investigate a new model that accurately estimates the mass flux from the outflow velocity.

% Chevalier & Clegg model 
Apart from the shell model, some analytical models investigated galaxy-scale outflows that extend continuously from the star-forming region ($<$ several kpc) to a distance of $\sim$100kpc.
\citet{Chevalier85}(CC85) modeled galactic-scale outflows accelerated by thermal energy injected from stars.
This model has been extended to include the gravitational potential of dark matter halo and radiative energy loss \citep{Sharma13, Bustard16, Nguyen21, Fielding22, Nguyen23}.
Some studies have also applied this model to winds from a stellar cluster with a central black hole \citep{Silich08, Silich10}.
CC85 assumed transonic outflows but fixed the locus of the transonic point (the critical point between subsonic and supersonic flows) at the edge of the star-forming region.
The locus of the transonic point should depend on the mass flux, the injected energy, and the gravitational potential.
Therefore, we re-examine this model to determine the locus of the transonic point self-consistently.

% Parker model 
Parker's theory of solar wind, which assumes a spherically symmetric steady flow in the gravitational potential of a point mass, plays a pioneering role in studies of transonic flows \citep{Parker58, Parker60, Parker65}.
This model determines the locus of the transonic point by the balance between the injected energy and the gravitational potential.
Some studies have applied the Parker model to galactic winds and assumed that the stellar mass density $\rho$ is exponential with radius $r$ \citep{Burke68, Holzer70, Johnson71}.
The transonic analysis has also been applied to winds from a stellar cluster with a central black hole \citep{Silich11, Palous13}.
Since the contribution of the dark matter halo is dominant in the gravitational potential of galaxies, \citet{Wang95} added the gravitational potential of the dark matter halo to the Parker model.
Assuming a single power law for the mass density distribution ($\rho\propto r^{-2}$), they could not obtain the locus of the transonic point.
Instead of a single power law, \citet{Tsuchiya13} assumed a double power law for the mass density distribution ($\rho\propto r^{-\alpha} (r+r_{\rm d})^{\alpha-3}$) and investigated isothermal, spherically symmetric, and steady flows.
Both numerical simulations of galaxy formation \citep{Navarro96, Moore99} and observations of local dwarf galaxies \citep{Burkert95} support the double power law.
The Tsuchiya model shows that the locus of the transonic point varies significantly with the parameters of the dark matter halo.
\citet{Igarashi14} added the gravitational potential of a super-massive black hole to this model and found that different transonic solutions can exist for the same mass density distribution of galaxies.
One solution is mainly affected by the gravitational potential of the super-massive black hole, and the transonic point locates in the inner region ($\sim$0.01 kpc).
The other solution is mainly affected by the gravitational potential of the dark matter halo, and the transonic point locates near the virial radius ($\sim$100kpc).
\citet{Igarashi17} presented that the polytropic outflow model can reproduce the observed gas density distribution more accurately than the isothermal model.
Furthermore, they proved that the transonic solution is an entropy-maximum solution independent of the structure of the gravitational potential.
It suggests that transonic acceleration is a natural consequence of astrophysical outflow systems.

% our purposes 
Previous studies have oversimplified the process of energy injection into the outflow, such as with the isothermal assumption \citep{Tsuchiya13, Igarashi14, Igarashi17}.
To accurately investigate the mass flux of transonic galactic winds, this study improves the transonic outflow model by considering the distribution of both mass and energy injections.
Our transonic outflow model correctly reveals the general properties of transonic acceleration flowing through the dark halo.
Therefore, our model can find solutions for winds from active star-forming galaxies as well as from other galaxies.
Furthermore, applying this model to the observed velocities of star-forming galaxies, we estimate mass fluxes and mass loading factors.
With this study, we clarify the impact of galactic winds on the star formation history and the chemical enrichment of the intergalactic medium.

We construct our transonic outflow model in Section 2 and show the classified transonic solutions in Section 3. 
Section 4 estimates mass fluxes and mass loading factors, and Section 5 discusses our results in other studies.
Finally, we summarize these results in Section 6.

\section{transonic galactic winds}
\subsection{Basic equations}
\label{sec_mdl_beq}

% basic equations
For transonic galactic winds, the equation of continuity, the equation of motion, and the energy equation are used:
\begin{eqnarray}
\frac{1}{r^2}\frac{d}{d r}(\rho v r^2) 
&=& \dot{\rho}_m (r), 
\label{eqn_cnt} \\
\rho v \frac{d v}{d r} 
&=& - \frac{d P}{d r} + \rho g - v \dot{\rho}_{\rm m} (r), 
\label{eqn_mot} \\
\frac{1}{r^2} \frac{d}{d r} 
\left\{\rho v r^2 \left(\frac{1}{2} v^2 + \frac{\Gamma}{\Gamma-1}\frac{P}{\rho}\right)\right\} 
&=& \rho v g + \dot{q}(r), 
\label{eqn_ene}
\end{eqnarray}
where $\rho$, $v$, $P$, and $g$ are density, velocity, pressure, and gravity, respectively. 
Assuming galactic winds of hot ionized gas, the specific heat ratio $\Gamma$ is 5/3. 
$\dot{\rho}_{\rm m}(r)$ and $\dot{q}(r)$ represent the mass injection and the energy injection, respectively. 
The energy injection $\dot{q}(r)$ represents the distribution of thermal energy sources such as supernovae (SNe). 
Because we assume that the injected mass does not have momentum, the right-hand side of equation (\ref{eqn_mot}) contains the additional term ($- v \dot{\rho}_{\rm m}$) suggesting the braking effect by the injected mass. 

% other definitions
We can write $g$ as 
\begin{eqnarray}
g(r) 
\equiv - \frac{d \phi(r)}{d r}
= - \frac{G M_{\rm gal}(r)}{r^2}
\label{eqn_gra_pot}
\end{eqnarray}
where $\phi$ and $M_{\rm gal}$ are the gravitational potential and the galactic mass inside the radius $r$, respectively. 
Here, we ignore the self-gravity of the gaseous medium. 
The sound speed $c_{\rm s}$ is defined as 
\begin{eqnarray}
c_{\rm s} 
\equiv \sqrt{\frac{\partial P}{\partial \rho} }
= \sqrt{\Gamma \frac{P}{\rho}}.
\label{eqn_sou_spe}
\end{eqnarray}
Additionally, to simplify the formula, we define the mass flux $\dot{m}(r)$ and the energy flux $\dot{e}(r)$ as 
\begin{eqnarray}
\dot{m}(r) &\equiv& 4 \pi \rho v r^2, 
\label{eqn_mfx} \\
\dot{e}(r) &\equiv& \left( \frac{1}{2}v^2 + \frac{c_s^2}{\Gamma-1} + \phi \right) \dot{m}(r). 
\label{eqn_efx}
\end{eqnarray}
The terms on the right-hand side of equation (\ref{eqn_efx}) represent the kinetic energy, the enthalpy, and the gravitational potential energy, respectively. 
Integrating equations (\ref{eqn_cnt}) and (\ref{eqn_ene}), we obtain $\dot{m}(r)$ and $\dot{e}(r)$ as 
\begin{eqnarray}
\dot{m}(r)
&=& \int_0^r 4 \pi r^2 \dot{\rho}_{\rm m}(r) dr, 
\label{eqn_flx_mas} \\
\dot{e}(r)
&=& \int_0^r 4 \pi r^2 \dot{q}(r) dr 
+  \int_0^r 4 \pi r^2 \dot{\rho}_{\rm m}(r) \phi(r) dr. 
\label{eqn_flx_ene}
\end{eqnarray}

% Mach number equation
Using equations (\ref{eqn_cnt}), (\ref{eqn_mot}), and (\ref{eqn_ene}), the differential equation is derived as 
\begin{eqnarray}
&& \frac{\mathcal{M}^2 - 1}{\mathcal{M}^2 \left\{ \left( \Gamma - 1  \right) \mathcal{M}^2 + 2 \right\}} \frac{d \mathcal{M}^2}{d r} \nonumber\\
&& \quad = \frac{2}{r} 
- \frac{\Gamma+1}{2\left( \Gamma-1 \right)} \frac{\dot{m}}{\dot{e} - \phi \dot{m}} \frac{d \phi}{d r} 
- \frac{\Gamma \mathcal{M}^2 + 1}{2} \frac{\dot{e} - 2 \phi \dot{m}}{\dot{e} - \phi \dot{m}} \frac{1}{\dot{m}} \frac{d \dot{m}}{d r} \nonumber\\
&& \qquad - \frac{\Gamma \mathcal{M}^2 + 1}{2} \frac{1}{\dot{e} - \phi \dot{m}} \frac{d \dot{e}}{d r}, 
\label{eqn_mac}
\end{eqnarray}
where $\mathcal{M} (\equiv v/c_{\rm s})$ represents the Mach number. 
The detailed derivation of equation (\ref{eqn_mac}) is summarized in Appendix \ref{app_mac}. 
When $\mathcal{M}=1$ at the transonic point, the right-hand side of equation (\ref{eqn_mac}) should be 0 to form a regular solution. 
As discussed in the Parker model, the transonic solution can be obtained from the transonic point. 
Therefore, from the right-hand side of equation (\ref{eqn_mac}) with $\mathcal{M}=1$, we define the function 
\begin{eqnarray}
N(r) 
&& = \frac{2}{r} 
- \frac{\Gamma+1}{2\left( \Gamma-1 \right)} \frac{\dot{m}}{\dot{e} - \phi \dot{m}} \frac{d \phi}{d r} 
- \frac{\Gamma + 1}{2} \frac{\dot{e} - 2 \phi \dot{m}}{\dot{e} - \phi \dot{m}} \frac{1}{\dot{m}} \frac{d \dot{m}}{d r} \nonumber\\
&&\qquad - \frac{\Gamma + 1}{2} \frac{1}{\dot{e} - \phi \dot{m}} \frac{d \dot{e}}{d r}, 
\label{eqn_n}
\end{eqnarray}
to investigate the locus of critical points, such as transonic points. 
We regularize equation (\ref{eqn_mac}) by $r$ such that $N(r)=0$.

\subsection{Model of mass and energy injection}
\label{sec_mdl_flx}

% mass loading 
The interstellar medium is entrained by the hot gas of galactic winds and this process is called as mass loading. 
The mass loading effect increases the mass fluxes ejected from the star-forming region. 
To measure quantitatively this effect, the mass loading factor $\eta$ is often used in observational studies. 
The value of $\eta$ is defined as 
\begin{eqnarray}
\eta \equiv \frac{\dot{m}_{\rm tot}}{\dot{m}_{\rm SFR}}, 
\label{eqn_eta}
\end{eqnarray}
where $\dot{m}_{\rm tot}$ and $\dot{m}_{\rm SFR}$ are defined as the total mass flux and the amount of SFR, respectively. 
From equation (\ref{eqn_flx_mas}), $\dot{m}_{\rm tot}$ can be defined as 
\begin{eqnarray}
\dot{m}_{\rm tot}
\equiv \int_0^{\infty} 4 \pi r^2 \dot{\rho}_{\rm m}(r) dr. 
\label{eqn_flx_mas_inf}
\end{eqnarray}
We assume constant $\eta$ independent of $r$ in this model. 
%equations (\ref{eqn_eta}) and (\ref{eqn_flx_mas_inf}) indicate that $\dot{\rho}_{\rm m}(r)$ is proportional to $\eta$. 
Additionally, the mass loading rate $\lambda$ is defined as 
\begin{eqnarray}
\lambda \equiv \frac{\dot{m}_{\rm tot}}{\dot{m}_{\rm SN}},
\label{eqn_lmd}
\end{eqnarray}
where $\dot{m}_{\rm SN}$ is the ejected mass from SNeII. 
Equation (\ref{eqn_lmd}) shows that the mass loading effect causes the mass flux to become $\lambda$ times larger than $\dot{m}_{\rm SN}$. 
If the mass loading effect does not exist, $\lambda = 1$. 
Because $\dot{m}_{\rm SN}$ is proportional to $\dot{m}_{\rm SFR}$, $\dot{m}_{\rm SN}$ is defined as 
\begin{eqnarray}
\dot{m}_{\rm SN} \equiv R_{\rm f} \ \dot{m}_{\rm SFR}, 
\label{eqn_flx_sne}
\end{eqnarray}
where $R_{\rm f}$ is the return fraction. 
The return fraction is the ratio of the ejected mass to the formed stellar mass. 
Therefore, $\eta$ relates to $\lambda$ as 
\begin{eqnarray}
\eta = R_{\rm f} \ \lambda. 
\label{eqn_elm}
\end{eqnarray}
By equation (\ref{eqn_elm}), $\eta$ can be estimated from $\lambda$. 
Because the return fraction $R_{\rm f}$ is smaller than unity, $\lambda$ is always larger than $\eta$. 
The value of $\eta$ represents the practical increase rate of mass flux from SFR, while that of $\lambda$ represents the pure increase rate of mass flux from the injected mass. 
Therefore, $\lambda$ is a comprehensible expression to measure the mass loading effect, compared to $\eta$. 
We focus on $\lambda$ in the following section. 

% power
Similar to mass fluxes, the mass loading effect causes powers (works per unit time) to become $\lambda$ times larger than powers without the mass loading effect ($\lambda = 1$). 
Because we assume two gravitational potentials (dark matter halo and stars), the total power consists of two components. 
The powers without mass loading effect are defined as 
\begin{eqnarray} 
\dot{w}_{\rm dmh} 
&\equiv& \frac{1}{\lambda} 
\int_0^{\infty} 4 \pi r^2 \dot{\rho}_{\rm m}(r) \left| \phi_{\rm dmh}(r) \right| dr, 
\label{eqn_pow_dmh} \\
\dot{w}_{\rm stl} 
&\equiv& \frac{1}{\lambda} 
\int_0^{\infty} 4 \pi r^2 \dot{\rho}_{\rm m}(r) \left| \phi_{\rm stl}(r) \right| dr, 
\label{eqn_pow_stl}
\end{eqnarray}
where $\phi_{\rm dmh}$ and $\phi_{\rm stl}$ correspond to the gravitational potentials of the dark matter halo and stars, respectively. 
Because $\dot{\rho}_{\rm m}(r)$ contains $\lambda$ and cancels $1/\lambda$ in equations (\ref{eqn_pow_dmh}) and (\ref{eqn_pow_stl}), the amount of $\lambda$ does not influence both $\dot{w}_{\rm dmh}$ and $\dot{w}_{\rm stl}$. 
The total power becomes $\lambda(\dot{w}_{\rm dmh} + \dot{w}_{\rm stl})$. 
Additionally, we define the injected energy flux from SNe as 
\begin{eqnarray}
\dot{e}_{\rm SN} 
\equiv \int_0^{\infty} 4 \pi r^2 \dot{q}(r) dr. 
\label{eqn_def_esn}
\end{eqnarray}
This is not influenced by the mass loading effect. 
Substituting $\dot{e}_{\rm SN}$, $\dot{w}_{\rm dmh}$, and $\dot{w}_{\rm stl}$ into equation (\ref{eqn_flx_ene}), the total energy flux becomes 
\begin{eqnarray}
\dot{e}_{\rm tot} 
&\equiv& \int_0^{\infty} 4 \pi r^2 \dot{q}(r) dr 
+  \int_0^{\infty} 4 \pi r^2 \dot{\rho}_{\rm m}(r) \phi(r) dr \nonumber\\ 
&=& \dot{e}_{\rm SN} - \lambda (\dot{w}_{\rm dmh} + \dot{w}_{\rm stl}).
\label{eqn_flx_ene_inf}
\end{eqnarray}
The second term on the right-hand side represents the decrease of energy flux by the gravitational potential. 
In Section \ref{sec_mdl_nde}, we use $\dot{m}_{\rm SN}$ and $\dot{e}_{\rm SN}$ for the non-dimensional expression.

\subsection{Requirement of escaping galactic winds}
\label{sec_res_trm}

% escaping condition 
This section examines the requirement for galactic winds to escape from galaxies.
The escaping galactic winds need to reach an infinite distance ($r\rightarrow\infty$).
In this case, the sum of the kinetic energy and the enthalpy must be positive at an infinite distance.
The limit of the energy flux becomes
\begin{eqnarray}
&& \lim_{r\rightarrow\infty} 
\left( \frac{1}{2}v^2 + \frac{c_{\rm s}^2}{\Gamma-1} \right) \dot{m}(r) \nonumber\\
&&\quad = \lim_{r\rightarrow\infty} 
\left\{\dot{e}(r)-\phi(r)\dot{m}(r) \right\} 
\qquad (\because {\rm equation} \: (\ref{eqn_efx}) \:) \nonumber\\
&&\quad = \dot{e}_{\rm tot} 
\qquad (\because \phi(\infty) = 0 \:) \nonumber\\
&&\quad = \dot{e}_{\rm SN} - \lambda (\dot{w}_{\rm dmh} + \dot{w}_{\rm stl}).  
\qquad (\because {\rm equation} \: (\ref{eqn_flx_ene_inf}) \:)
\label{eqn_ten_lim}
\end{eqnarray}
When the right-hand side of equation (\ref{eqn_ten_lim}) becomes positive, the galactic wind can reach an infinite distance.
Therefore, equation (\ref{eqn_ten_lim}) implies a necessary and sufficient condition for escaping galactic winds:
\begin{eqnarray}
0 \leq 
\dot{e}_{\rm SN} - \lambda (\dot{w}_{\rm dmh} + \dot{w}_{\rm stl}).
\label{eqn_ten_lim_ine}
\end{eqnarray}
Additionally, for escaping winds, the upper limit of $\lambda$ 
%(or $\eta$) 
becomes 
\begin{eqnarray}
\lambda_{\rm esc} 
\equiv \frac{ \dot{e}_{\rm SN} }{ \dot{w}_{\rm dmh} + \dot{w}_{\rm stl} }.
\label{eqn_lmx}
\end{eqnarray}
When $\lambda$ becomes larger than $\lambda_{\rm esc}$, the galactic wind does not reach an infinite distance, which means that it is too heavy to escape from the galaxy.
If the galactic mass increases or the energy injection decreases, $\dot{w}_{\rm dmh}$ (and $\dot{w}_{\rm stl}$) becomes larger than $\dot{e}_{\rm SN}$.
In this case, $\lambda_{\rm esc}$ becomes small and it is difficult for galactic winds to escape.

% terminal velocity 
Here, we assume that $c_{\rm s}$ becomes 0 at $r=\infty$.
If $\lambda\leq\lambda_{\rm esc}$, the terminal velocity $v_{\rm term} $ can be defined from equation (\ref{eqn_efx}) as 
\begin{eqnarray}
v_{\rm term} 
&\equiv& \sqrt{ 2 \frac{\dot{e}_{\rm tot}}{\dot{m}_{\rm tot}} } \nonumber\\
&=& \sqrt{ 2 \frac{ \dot{e}_{\rm SN} - \lambda (\dot{w}_{\rm dmh} + \dot{w}_{\rm stl}) }{\lambda\dot{m}_{\rm SN}}} \quad (\because {\rm equations} \: (\ref{eqn_lmd}), \: (\ref{eqn_flx_ene_inf})\: ) \nonumber\\
&=& \sqrt{\frac{2\dot{e}_{\rm SN}}{\lambda\dot{m}_{\rm SN}}\left\{1-\frac{\lambda}{\lambda_{\rm esc}}\right\}}. 
\quad (\because {\rm equation} \:(\ref{eqn_lmx})\: )
\label{eqn_trm}
\end{eqnarray}
With small $\lambda$, small $\dot{m}_{\rm SN}$ or large $\dot{e}_{\rm SN}$, $v_{\rm term}$ becomes large.
If $\lambda=\lambda_{\rm esc}$, $v_{\rm term}$ becomes 0.

% no gravity limit 
When the galaxy mass is small enough to be negligible, $\lambda_{\rm esc}$ becomes sufficiently larger than $\lambda$ ($\lambda \ll \lambda_{\rm esc}$).
In this case, $v_{\rm term}$ becomes
\begin{eqnarray}
v_{\rm term} 
\sim \sqrt{\frac{2\dot{e}_{\rm SN}}{\lambda\dot{m}_{\rm SN}}} 
\propto \lambda^{-\frac{1}{2}}.
\label{eqn_trm_lim}
\end{eqnarray}
Therefore, equation (\ref{eqn_trm_lim}) represents the maximum velocity without the galactic gravity, corresponding to $\lambda$ with the specific mass and energy depositions ($\dot{m}_{\rm SN}$ and $\dot{e}_{\rm SN}$).

% escape velocity 
Additionally, the escape velocity $v_{\rm esc}$ can be defined as 
\begin{eqnarray}
v_{\rm esc}(r) \equiv \sqrt{2|\phi(r)|}.
\label{eqn_esc_vel}
\end{eqnarray}
In the vicinity of the galactic centre ($r\sim 0$), the wind velocity does not exceed $v_{\rm esc}$ because galactic winds start with a small velocity as detailed in subsection \ref{sec_res_cls}.
At $r=\infty$, the wind velocity must exceed $v_{\rm esc}$ due to $\phi(\infty)=0$.

\section{classification of transonic solutions for star-forming galaxies}
\subsection{Mass distributions}
\label{sec_mdl_msd}

% dark halo mass and stellar mass 
In the proposed model of this study, we assume the dark matter halo and stars for the mass distribution of galaxies. 
For the mass density of the dark matter halo, numerical simulations based on the cold dark matter (CDM) scenario suggest a double-power-law distribution. 
The Navarro–Frenk–White (NFW) model is the most-used model in the CDM scenario \citep{Navarro96}, with a double-power-law distribution as 
\begin{eqnarray}
\rho_{\rm dmh}(r; \widetilde{\rho}_{\rm dmh},r_{\rm dmh}) 
= \frac{\widetilde{\rho}_{\rm dmh} r_{\rm dmh}^3}{r (r+r_{\rm dmh})^2}, 
\label{eqn_dis_dmh}
\end{eqnarray}
where $\widetilde{\rho}_{\rm dmh}$ and $r_{\rm dmh}$ are the scale density and the scale radius, respectively. 
For the stellar mass, this paper assumes the empirical distribution function of the bulge, which is the so-called Hernquist model, 
\begin{eqnarray}
\rho_{\rm stl}(r; r_{\rm stl},M_{\rm stl}^{\rm tot}) 
= \frac{M_{\rm stl}^{\rm tot}}{2\pi} \frac{r_{\rm stl}}{r} \frac{1}{(r+r_{\rm stl})^3}, \\
\quad \left(r_{\rm stl} \equiv \frac{r_{\rm efct}}{1+\sqrt{2}}\right) \nonumber
\label{eqn_dis_stl}
\end{eqnarray}
where $M_{\rm stl}^{\rm tot}$, $r_{\rm stl}$ and $r_{\rm efct}$ are the total mass, the scale radius, and the half-mass radius of the stellar components, respectively \citep{Hernquist90}. 

% upper limit at virial radius
Because the dark matter halo mass $M_{\rm dmh}(r)$ from equation (\ref{eqn_dis_dmh}) does not have a finite upper limit ($M_{\rm dmh}(\infty)\rightarrow\infty$), we assume the breaking point at the virial radius as 
\begin{eqnarray}
M_{\rm gal}(r) = 
\left\{
\begin{array}{ll}
M_{\rm stl}(r) + M_{\rm dmh}(r) & (r < r_{\rm vir}), \nonumber\\
M_{\rm stl}(r_{\rm vir}) + M_{\rm dmh}(r_{\rm vir}) & (r > r_{\rm vir}). \nonumber
\end{array}
\right.
\end{eqnarray} 
where $r_{\rm vir}$ is the virial radius. 
Therefore, $\rho_{\rm dmh}$(r) and $\rho_{\rm stl}$(r) become 0 in $r>r_{\rm vir}$. 
In this assumption, the mass and energy injections, $\dot{\rho}_{\rm m}(r)$, and $\dot{q}(r)$ are also assumed to vanish in $r>r_{\rm vir}$. 

% integrated mass 
Integrating equation (\ref{eqn_dis_dmh}), the dark halo mass inside a sphere of the radius $r$ becomes 
\begin{eqnarray}
M_{\rm dmh}(r;M_{\rm dmh}^{\rm vir},r_{\rm dmh},r_{\rm vir}) 
&\equiv& \int_0^{r} 4 \pi r^2 \rho_{\rm dmh}(r;\widetilde{\rho}_{\rm dmh},r_{\rm dmh}) dr \nonumber\\
&=& M_{\rm dmh}^{\rm vir} \mathcal{F}_{\rm dmh}(r;r_{\rm dmh},r_{\rm vir}), 
\label{eqn_mas_dmh}
\end{eqnarray}
where $M_{\rm dmh}^{\rm vir}$ is the virial mass of dark matter halo. 
$\mathcal{F}_{\rm dmh}(r;r_{\rm dmh})$ is the non-dimensional function defined in Appendix \ref{app_nfc}. 
Integrating equation (\ref{eqn_dis_stl}), the stellar mass inside the sphere of radius $r$ is derived as 
\begin{eqnarray}
M_{\rm stl}(r;M_{\rm stl}^{\rm vir},r_{\rm stl},r_{\rm vir}) 
&\equiv& \int_0^r 4 \pi r^2 \rho_{\rm stl}(r;M_{\rm stl}^{\rm tot},r_{\rm stl}) dr \nonumber\\
&=& M_{\rm stl}^{\rm vir} \mathcal{F}_{\rm stl}(r;r_{\rm stl},r_{\rm vir}), 
\label{eqn_mas_stl}
\end{eqnarray}
where $M_{\rm stl}^{\rm vir}$ is the stellar mass inside the virial radius. 
Due to $r_{\rm stl}\ll r_{\rm vir}$, $M_{\rm stl}^{\rm vir}\sim M_{\rm stl}^{\rm tot}$. 
$\mathcal{F}_{\rm stl}(r;r_{\rm stl},r_{\rm vir})$ is the non-dimensional function defined in Appendix \ref{app_nfc}. 
The gravitational potentials can be estimated from equations (\ref{eqn_mas_dmh}) and (\ref{eqn_mas_stl}) as
\begin{eqnarray}
&& \phi_{\rm dmh} (r;M_{\rm dmh}^{\rm vir},r_{\rm dmh},r_{\rm vir}) \nonumber\\
&& \qquad \equiv \int \frac{G M_{\rm dmh}(r)}{r} dr \nonumber\\
&& \qquad = -\frac{G M_{\rm dmh}^{\rm vir}}{r_{\rm vir}} 
\mathcal{F}_{\rm \phi dmh} (r;r_{\rm dmh},r_{\rm vir}), 
\label{eqn_pot_dmh} 
\end{eqnarray}
\begin{eqnarray}
\phi_{\rm stl} (r;M_{\rm stl}^{\rm vir},r_{\rm stl},r_{\rm vir})
&\equiv& \int \frac{G M_{\rm stl}(r)}{r} dr \nonumber\\
&=& -\frac{G M_{\rm stl}^{\rm vir}}{r_{\rm vir}} 
\mathcal{F}_{\rm \phi stl} (r;r_{\rm stl},r_{\rm vir}), 
\label{eqn_pot_stl}
\end{eqnarray}
where $\mathcal{F}_{\rm \phi dmh}(r; r_{\rm dmh}, r_{\rm vir})$ and $\mathcal{F}_{\rm \phi stl}(r; r_{\rm stl}, r_{\rm vir})$ are the non-dimensional functions defined in Appendix \ref{app_nfc}.

\subsection{Mass and energy injections}
\label{sec_mdl_inj}

% mass injection (or SFR) distribution 
In Section \ref{sec_mdl_flx}, we assumed that the mass flux increases $\lambda$ times larger than $\dot{m}_{\rm SN}$ due to the mass loading effect.
With $\lambda$ (and $\eta$), mass injection distribution can be determined as
\begin{eqnarray}
\dot{\rho}_{\rm m} (r) 
&=& \eta \: \dot{\rho}_{\rm sfr}(r) \nonumber\\
&=& \lambda \: R_{\rm f} \: \dot{\rho}_{\rm sfr}(r), 
\label{eqn_inj_mas}
\end{eqnarray}
where $\dot{\rho}_{\rm sfr}(r)$ is the spatcial distribution of SFR.
In this study, because we assume the constant mass loading, $\lambda$ (and $\eta$) is independent of the radius $r$.

% energy injection distribution 
For the source of the energy injection, we assume SNeII as the dominant energy source because this study focuses on the galactic winds in star-forming galaxies later.
The ejected momentum from SNeII will be symmetric to the plane and thermalized through shock waves.
This study assumes that the momentum contained in the ejected mass from SNeII is injected in the galactic winds after the thermalization entirely proceeds.
Therefore, the galactic winds are purely thermally driven neglecting the momentum injection from the SNeII. 
In the thermalization process, some of the energy is lost in radiative cooling. 
In this assumption, the source distribution of the energy injection is defined as 
\begin{eqnarray}
\dot{q}(r) 
\equiv \epsilon_{\rm rem} \: e_{\rm SN} \: f_{\rm SN} \: \dot{\rho}_{\rm sfr}(r), 
\label{eqn_inj_ene}
\end{eqnarray}
where $e_{\rm SN}$, $f_{\rm SN}$, and $\epsilon_{\rm rem}$ are the total energy from one SNII ($=10^{51}$erg), the SNII frequency per new stellar mass, and the efficiency of heating (the fraction of the total energy transferred into thermal energy after radiative cooling), respectively. 

% sfr distribution 
The distribution of the star-forming region depends on the species of galaxies. 
The star-forming activity of local starbursts like M82 and NGC253 concentrates in the central zones of their host galaxies \citep{O'Connell95, Satyapal97, Smith01, McCrady03, Ulvestad97, Forbes00}. 
Due to the gravitational instability, the gas accretion is induced and compressed to the center. 
This induces the star-forming activity concentrated in the center. 
In star-forming galaxies like clumpy galaxies, the star formation activity locates in young stellar clusters off the center \citep{Elmegreen05, Elmegreen09, Forster11, Genzel11, Wuyts12, Murata14}. 
In merging galaxies, the star-forming activity can spread widely in the galaxy disk \citep{Michiyama20}. 
The rapid perturbation of the velocity field results in a dense gas and induces the star-forming activity in their disks \citep{Bournaud11}. 
In any case, the star-forming activity distribution varies among galaxies. 
For simplicity, we assume that the star-forming activity spreads identically to the stellar distribution: 
\begin{eqnarray}
\dot{\rho}_{\rm sfr}(r) 
\equiv \frac{\dot{m}_{\rm SFR}}{M_{\rm stl}^{\rm vir}} \rho_{\rm stl}(r;r_{\rm stl}). 
\label{eqn_sfd}
\end{eqnarray}
When $r_{\rm stl}$ is small, the star-forming activity is concentrated in the central region. 
With large $r_{\rm stl}$, the star-forming activity becomes widely spread and diffused. 
The gravitational potential of stars can also be changed by $r_{\rm stl}$, but a dark matter halo dominates the gravitational potential. 
Therefore, $r_{\rm stl}$ mainly affects the injection distribution and does not strongly affect the global structure of the gravitational potential.

\subsection{Fluxes with mass loading}
\label{sec_mdl_pwr}

% parameters 
Substituting equations (\ref{eqn_inj_ene}) and (\ref{eqn_sfd}) into equation (\ref{eqn_def_esn}), we obtain the expression of $\dot{e}_{\rm SN}$ as
\begin{eqnarray*}
\dot{e}_{\rm SN} 
= \epsilon_{\rm SN} \: f_{\rm SN} \: e_{\rm SN} \: \dot{m}_{\rm SFR}.
\end{eqnarray*}
Additionally, substituting equations (\ref{eqn_inj_mas}) and (\ref{eqn_sfd}) into equations (\ref{eqn_pow_dmh}) and (\ref{eqn_pow_stl}), we obtain $\dot{w}_{\rm dmh}$ and $\dot{w}_{\rm stl}$ as 
\begin{eqnarray}
\dot{w}_{\rm dmh} 
&=& \dot{m}_{\rm SN} \frac{G M_{\rm dmh}^{\rm vir}}{r_{\rm vir}} 
\mathcal{F}_{\rm w dmh} (r_{\rm vir};r_{\rm dmh}, r_{\rm stl},r_{\rm vir}), 
\label{eqn_pow_dmh_inn} \\
\dot{w}_{\rm stl} 
&=& \dot{m}_{\rm SN} \frac{G M_{\rm stl}^{\rm vir}}{r_{\rm vir}} 
\mathcal{F}_{\rm w stl} (r_{\rm vir}; r_{\rm stl},r_{\rm vir}), 
\label{eqn_pow_stl_inn}
\end{eqnarray}
where $\mathcal{F}_{\rm w dmh}(r;r_{\rm dmh}, r_{\rm stl},r_{\rm vir})$ and $\mathcal{F}_{\rm w stl}(r; r_{\rm stl},r_{\rm vir})$ are the non-dimensional functions defined in Appendix \ref{app_nfc}. 

% mass & energy fluxes
Substituting equations (\ref{eqn_inj_mas}) and (\ref{eqn_sfd}) into equation (\ref{eqn_flx_mas}), the mass flux at the radius $r$ is derived as
\begin{eqnarray}
\dot{m}(r) 
= \lambda \: \dot{m}_{\rm SN} \: \mathcal{F}_{\rm stl}(r;r_{\rm stl},r_{\rm vir}). 
\label{eqn_mfx_non} 
\end{eqnarray}
%Because $\mathcal{F}_{\rm stl}(r;r_{\rm stl},r_{\rm vir})=1$ in $r>r_{\rm vir}$, the total mass flux $\dot{m}(\infty)$ is $\lambda$ times larger than $\dot{m}_{\rm SN}$ as defined in equation (\ref{eqn_lmd}). 
Additionally, substituting equations (\ref{eqn_inj_ene}) and (\ref{eqn_sfd}) into equation (\ref{eqn_flx_ene}), we obtain the energy flux at the radius $r$ as 
\begin{eqnarray}
\dot{e}(r) 
&=& \dot{e}_{\rm SN} \: \mathcal{F}_{\rm stl}(r;r_{\rm stl},r_{\rm vir}) \nonumber\\
&& - \lambda \: \dot{w}_{\rm dmh} \: \frac{\mathcal{F}_{\rm w dmh}(r;r_{\rm dmh},r_{\rm stl},r_{\rm vir})}{\mathcal{F}_{\rm w dmh}(r_{\rm vir};r_{\rm dmh},r_{\rm stl},r_{\rm vir})} \nonumber\\
&& - \lambda \: \dot{w}_{\rm stl} \: \frac{\mathcal{F}_{\rm w stl}(r;r_{\rm stl},r_{\rm vir})}{\mathcal{F}_{\rm w stl}(r_{\rm vir};r_{\rm stl},r_{\rm vir})}. 
\label{eqn_efx_non}
\end{eqnarray}
Equation (\ref{eqn_efx_non}) indicates that the total energy flux is the residue after subtracting the powers from the ejected energy from stars. 
The mass loading effect increases the powers by a factor $\lambda$ and decreases the total energy flux.

\subsection{Normalized expressions}
\label{sec_mdl_nde}

% non-dimensional parameters 
To simplify the analysis, the non-dimensional expressions are defined as follows:
\begin{eqnarray}
&&x \equiv \frac{r}{r_{\rm dmh}}, \\
&&x_{\rm stl} \equiv \frac{r_{\rm stl}}{r_{\rm dmh}}, 
\quad \left( x_{\rm stl} = \frac{x_{\rm efct}}{1+\sqrt{2}} \right)\\
&&c \equiv \frac{r_{\rm vir}}{r_{\rm dmh}}, \\
&&\dot{w}_{\rm dmh,n} \equiv \frac{\dot{w}_{\rm dmh}}{\dot{e}_{\rm SN}}, \\
&&\dot{w}_{\rm stl,n} \equiv \frac{\dot{w}_{\rm stl}}{\dot{e}_{\rm SN}}, \\
&&\dot{m}_{\rm n}(x) 
\equiv \frac{\dot{m}(r)}{\dot{m}_{\rm SN}} 
= \lambda \mathcal{F}_{\rm stl}(x), \\
&&\dot{e}_{\rm n}(x) 
\equiv \frac{\dot{e}(r)}{\dot{e}_{\rm SN}} 
= \mathcal{F}_{\rm stl}(x) 
- \lambda \: \frac{\dot{w}_{\rm dmh,n}}{\mathcal{F}_{\rm w dmh}(c;x_{\rm stl},c)} \mathcal{F}_{\rm w dmh}(x;x_{\rm stl},c) \nonumber\\
&&\qquad\qquad\qquad\qquad - \lambda \: \frac{\dot{w}_{\rm stl,n}}{\mathcal{F}_{\rm w stl}(c;x_{\rm stl},c)} 
\mathcal{F}_{\rm w stl}(x;x_{\rm stl},c), 
\end{eqnarray}
and 
\begin{eqnarray}
&&\phi_{\rm n}(x) 
\equiv \frac{\phi(r)}{\dot{e}_{\rm SN}/\dot{m}_{\rm SN}} 
= - \frac{\dot{w}_{\rm dmh,n}}{\mathcal{F}_{\rm w dmh}(c;x_{\rm stl},c)} 
\mathcal{F}_{\rm \phi dmh}(x;c) \nonumber\\
&&\qquad\qquad\qquad\qquad\qquad - \frac{\dot{w}_{\rm stl,n}}{\mathcal{F}_{\rm w stl}(c;x_{\rm stl},c)} 
\mathcal{F}_{\rm \phi stl}(x;x_{\rm stl},c), 
\end{eqnarray}
where the subscript $\mathrm{n}$ implies non-dimensional. 
equation (\ref{eqn_mac}) becomes 
\begin{eqnarray}
&& \frac{\mathcal{M}^2 - 1}{\mathcal{M}^2 \left\{ \left( \Gamma - 1  \right) \mathcal{M}^2 + 2 \right\}} \frac{d \mathcal{M}^2}{d x} \nonumber\\
&&\qquad = \frac{2}{x} 
- \frac{\Gamma+1}{2\left( \Gamma-1 \right)} \frac{\dot{m}_{\rm n}}{\dot{e}_{\rm n} - \phi_{\rm n} \dot{m}_{\rm n}} \frac{d \phi_{\rm n}}{d x} \nonumber\\
&&\qquad \quad - \frac{\Gamma \mathcal{M}^2 + 1}{2} \frac{\dot{e}_{\rm n} - 2 \phi_{\rm n} \dot{m}_{\rm n}}{\dot{e}_{\rm n} - \phi_{\rm n} \dot{m}_{\rm n}} \frac{1}{\dot{m}_{\rm n}} \frac{d \dot{m}_{\rm n}}{d x} \nonumber\\
&&\qquad \quad - \frac{\Gamma \mathcal{M}^2 + 1}{2} \frac{1}{\dot{e}_{\rm n} - \phi_{\rm n} \dot{m}_{\rm n}} \frac{d \dot{e}_{\rm n}}{d x}. 
\label{eqn_mac_non}
\end{eqnarray}
Similar to equation (\ref{eqn_n}), we define $\widetilde{N}(x)$ from the right-hand side of equation (\ref{eqn_mac_non}) as 
\begin{eqnarray}
\widetilde{N}(x)
&& = \frac{2}{x} 
- \frac{\Gamma+1}{2\left( \Gamma-1 \right)} \frac{\dot{m}_{\rm n}}{\dot{e}_{\rm n} - \phi_{\rm n} \dot{m}_{\rm n}} \frac{d \phi_{\rm n}}{d x} \nonumber\\
&&\qquad - \frac{\Gamma + 1}{2} \frac{\dot{e}_{\rm n} - 2 \phi_{\rm n} \dot{m}_{\rm n}}{\dot{e}_{\rm n} - \phi_{\rm n} \dot{m}_{\rm n}} \frac{1}{\dot{m}_{\rm n}} \frac{d \dot{m}_{\rm n}}{d x} \nonumber\\
&&\qquad - \frac{\Gamma + 1}{2} \frac{1}{\dot{e}_{\rm n} - \phi_{\rm n} \dot{m}_{\rm n}} \frac{d \dot{e}_{\rm n}}{d x}, 
\end{eqnarray}
to investigate critical points ($\widetilde{N}(x)=0$).

\subsection{Classification of all types}
\label{sec_res_cls}

% critical points 
In this model, the types of critical points are the saddle type (transonic point), the spiral type (spiral point) and the cliff edge (terminal point).
Our model can have only these three types.
The emergence of these critical points depends on how the total energy of the system evolves. Our model generates critical solutions classified as spiral types due to non-adiabatic effects associated with mass and energy injections. However, when non-adiabatic effects are weak, the solutions asymptotically approach saddle Type transonic solutions, represented by Parker's solution. Furthermore, in cases where the total energy is negative and the system bounded, a terminal point emerges.
If a transonic point exists, there can be a transonic solution through the locus of that point.
If a spiral point exists, the solution curve bends significantly in the vicinity of that point and forms a spiral structure.
If a terminal point exists, the solution curves cannot extend beyond a radius determined for each curve.
All solution curves cannot extend beyond the locus of that point.

\begin{figure*}
  \begin{center}
   \includegraphics[width=1.0\linewidth]{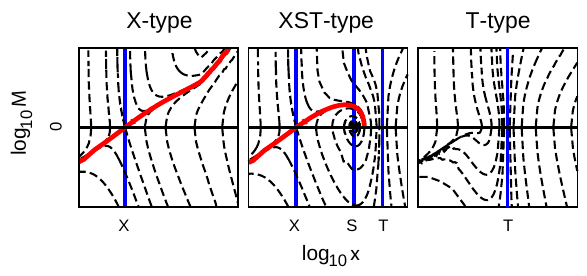}
   \caption{
Examples of Mach number diagram for three solution types.
The vertical axis represents the Mach number, and the horizontal axis shows the non-dimensional radius.
The solid red lines and black dashed lines represent transonic outflow solutions and other solutions.
Blue lines represent the locus of critical points: transonic points, spiral points, or terminal points.
The X type has one transonic point. In this type, the transonic galactic wind can escape from galaxies.
The XST type has one transonic point with one spiral point and one terminal point, where the transonic outflow solution cannot extend to an infinite distance.
The T type does not have transonic points.
In the XST type and T type, Mach number profiles break in the vicinity of the terminal point and %velocities (and densities) 
solutions of outflow cannot extend beyond the terminal point.
   }
   \label{img_pat}
  \end{center}
\end{figure*}

% soltuion types 
This section defines all types of galactic winds as a combination of critical points corresponding to a set of parameters.
We classify the combination of critical points into three types: X, XST, and T.
For example, Figure \ref{img_pat} shows the Mach number diagrams for these solution types.
Equation (\ref{eqn_mac_non}) gives the solution curve for the Mach number corresponding to the integration constant.
The red line indicates the transonic solution for outflows.
The black dashed lines show subsonic and supersonic solutions.
The blue lines represent the locus of critical points; transonic, spiral, or terminal points.
% The X Type solution has only a transonic point.
The transonic solution starts at the center ($r=0$) and continues to infinity ($r=\infty$).
The transonic galactic wind can escape from the galaxy with the terminal velocity defined by equation (\ref{eqn_trm}).
The XST Type solution has a transonic point, a spiral point, and a terminal point.
The transonic solution starts at the center, but the inclination of the transonic solution curve becomes infinite at a finite radius between the spiral point and the terminal point.
Therefore, such a solution breaks at that radius, and the transonic galactic wind cannot escape from the galaxy.
Since this contradicts the steady assumption, we discuss the XST Type solution in Section \ref{sec_dis_ass}.
The T Type solution has only a terminal point.
Since there is no transonic point, there can be no transonic solution.
The medium ejected from SNe cannot escape from the galaxy.

\begin{figure*}
  \begin{center}
    \includegraphics[width=1.00\linewidth]{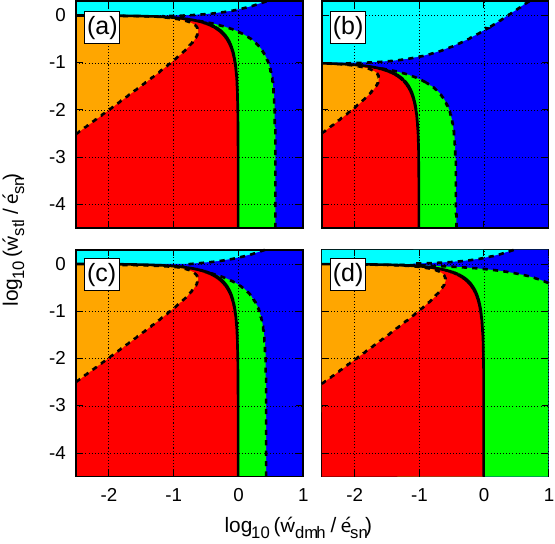}
   \caption{
Classification of solution types.
The four diagrams have different parameters; (a): ($\lambda$,$c$,$x_{\rm efct}$)=(1,10,0.1), (b): ($\lambda$,$c$,$x_{\rm efct}$)=(10,10,0.1), (c): ($\lambda$,$c$,$x_{\rm efct}$)=(1,2,0.1), and (d): ($\lambda$,$c$,$x_{\rm efct}$)=(1,10,0.01), respectively
The red region represents the X Type solution with only one transonic point in the gravitational potential of the dark matter halo mass.
The orange region represents the X Type solution with only one transonic point in the gravitational potential of stellar mass.
The green region represents the XST Type solution with a transonic point, a spiral point and a terminal point.
The blue region represents the T Type solution with only one terminal point in the gravitational potential of the dark matter halo mass.
The cyan region represents the T Type solution with only one terminal point in the gravitational potential of the stellar mass.
The black lines show the borders of these regions.
    }
   \label{img_cls}
  \end{center}
\end{figure*}

% solution sub Types 
Fixing $\lambda$, $x_{\rm stl}$ and $c$, figure \ref{img_cls} shows the results of solution classification.
The red and orange areas indicate the X type in the parameter space ($\dot{w}_{\rm stl,n}$, $\dot{w}_{\rm dmh,n}$), respectively.
Similarly, the green area represents the XST type, and the blue and cyan areas represent the T type.
The X type is divided into sub-types: X$_{\rm dmh}$ (red area) and X$_{\rm stl}$ (orange area).
The transonic point of X$_{\rm dmh}$ type is mainly determined by the gravitational potential of the dark matter halo mass, while that of X$_{\rm stl}$ type by the gravitational potential of stellar mass.
Also, we divide the T type into sub-types: T$_{\rm dmh}$ (blue area) and T$_{\rm stl}$ (cyan area).
The T$_{\rm dmh}$ type has a terminal point in the gravitational potential of the dark matter halo, that of the T$_{\rm stl}$ type locates in the gravitational potential of the stellar mass.

% escape solution to parameters 
As shown in figure \ref{img_cls}, when $\dot{w}_{\rm dmh,n}$ and $\dot{w}_{\rm stl,n}$ decrease, the solution becomes X type.
This indicates that transonic galactic winds can escape from the gravitational potential well when the injected energy becomes larger than the gravitational potential energy.
In figure \ref{img_cls}, the solid black line between X and the other types (XST and T) corresponds to the boundary of the escape requirement, $\lambda=\lambda_{\rm esc}$.
Therefore, with large $\lambda$, the region of X Type and XST Type solutions shrinks, and that of T Type solutions expands in the parameter space.
This implies that the mass loading effect suppresses the formation of transonic galactic winds.
Figure \ref{img_cls} also shows that the concentration $c$ does not change the region of X Type, as the virial radius has a negligible effect inside the halo scale.
Also, the expanse of injections, $x_{\rm stl}$ does not affect the region of the X type because $x_{\rm stl}$ does not affect the total amount of the injected energy.
We note that $x_{\rm stl}$ has little impact on the gravitational potential because the dark halo dominates the gravitational potential.
Nevertheless, the region of XST Type erodes that of T Type with small $x_{\rm stl}$.
This indicates that the centrally concentrated injections can form transonic winds that cannot escape.
Therefore, $x_{\rm stl}$ needs to be taken into account to determine the possibility of transonic galactic winds.

\begin{figure*}
 \begin{center}
\includegraphics[width=1.0\linewidth]{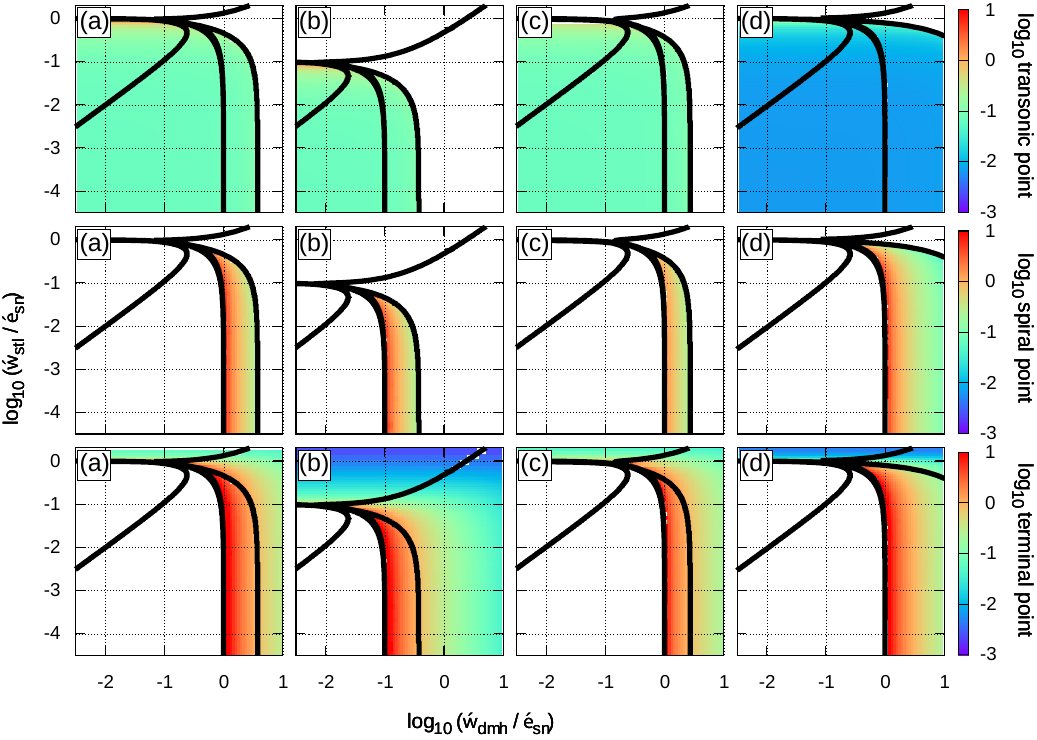}
  \caption{
Non-dimensional locus of critical points (top row: transonic points, middle row: spiral points, bottom row: terminal points).
Indexes (a)-(d) are the same as in figure\ref{img_cls}.
The right color bar represents the non-dimensional locus of points.}
  \label{img_cls_cpt}
 \end{center}
\end{figure*}

% critical points to parameters 
Figure \ref{img_cls_cpt} shows the locus of the critical point non-dimensionalized by $r_{\rm dmh}$.
As shown in the top row of figure \ref{img_cls_cpt}, the locus of the transonic point is mainly affected by $\dot{w}_{\rm stl,n}$, $x_{\rm stl}$, and $\lambda$.
As $\dot{w}_{\rm stl,n}$, $x_{\rm stl}$, or $\lambda$ increases, the transonic point moves outward.
On the other hand, $\dot{w}_{\rm dmh,n}$ and $c$ have little effect on the locus of the transonic point.
Therefore, the stellar mass and mass loading affect the locus of the transonic point, while the dark halo mass has little effect.
As shown in the middle row of figure \ref{img_cls_cpt}, the locus of the spiral point is mainly affected by $\dot{w}_{\rm dmh,n}$, $c$, and $\lambda$.
As $\dot{w}_{\rm dmh,n}$ or $\lambda$ increases, the spiral point moves inward.
As $c$ increases, the spiral point moves outward.
On the other hand, $\dot{w}_{\rm stl,n}$ and $x_{\rm stl}$ have little effect.
Therefore, the dark halo mass and mass loading affect the locus of the spiral point, while the stellar mass has little effect.
As shown in the bottom row of figure \ref{img_cls_cpt}, the locus of the terminal point of T$_{\rm stl}$ type is affected by $\dot{w}_{\rm stl,n}$, $x_{\rm stl}$ and $\lambda$, that of T$_{\rm dmh}$ type and XST type is affected by $\dot{w}_{\rm dmh,n}$; $c$ and $\lambda$.
In the XST type, the locus of the terminal point can increase to almost the virial radius.


\section{mass fluxes in star-forming galaxies}
\label{mass fluxes in star-forming galaxies}
\subsection{Parameters in outflow model}
\label{sec_mdl_prm}

% injections 
To estimate the mass flux in Section \ref{sec_res_mlr}, we use several empirical laws suggested by observations and predictions deduced from numerical experiments.
First, we determine the parameters required for mass and energy injection.
The return fraction $R_{\rm f}$ depends on the initial mass function (IMF) \citep{Salpeter55, Chabrier03, Kroupa01, Kroupa02}.
Typically, 30-40\% of the newly formed stellar mass returns to the interstellar medium.
This paper adopts $R_{\rm f} =$0.35.
For the Kroupa IMF, $f_{\rm SN}=$1.86$\times 10^{-2}M_{\odot}^{-1}$.
Assuming that the energy from SNeII has an injection rate of 10\% after radiative cooling, $\epsilon_{\rm rem} = 0.1$ \citep{Strickland09, Sashida13}.

% dark halo mass 
Next, we estimate the dark matter halo mass from the observed stellar mass. 
The abundance matching technique (matching observed galaxy distributions to dark halo distributions calculated from numerical experiments) leads to the relation between the dark matter halo mass and the stellar mass \citep{Behroozi10, Behroozi13}, 
\begin{eqnarray}
&& \log_{10} M_{\rm stl}^{\rm tot} = \log_{10} (\epsilon(z) M_{\rm l}(z)) \nonumber\\
&& \qquad\qquad\qquad + f\left(\log_{10}\left(\frac{M_{\rm dmh}^{\rm vir}}{M_{\rm l}(z)}\right)\right) - f(0), \\
&& f(x) = - \log_{10} (10^{\alpha(z) x}+1) + \delta(z) \frac{( \log_{10} \{1+ \exp(x)\} )^{\gamma(z)}}{1+\exp(10^{-x})}, \nonumber
\end{eqnarray}
where $M_{\rm l}(z)$, $\epsilon(z)$, $\alpha(z)$, $\delta(z)$, and $\gamma(z)$ are parameters specified by redshift $z$.

% scale radius 
Finally, $r_{\rm vir}$, $r_{\rm dmh}$, and $r_{\rm efct}$ are determined from $M_{\rm dmh}^{\rm vir}$ and $M_{\rm stl}^{\rm tot}$. 
From theoretical studies \citep{Bryan98, Bullock01}, $r_{\rm vir}$ can be defined by $M_{\rm dmh}^{\rm vir}$ and $z$. 
For this definition, we use the result by WMAP (\citet{Dunkley09}; $h=0719^{+0.026}_{-0.027}$, $\Omega_{\rm \Lambda} = 0.742\pm 0.030$, $\Omega_{\rm m} + \Omega_{\rm \Lambda} = 1$). 
\citet{Munoz-Cuartas11} found that the concentration parameter $c$ has the theoretical relation to $M_{\rm dmh}^{\rm vir}$, 
\begin{eqnarray}
\log_{10} c = a(z) \log_{10} (M_{\rm dmh}^{\rm vir}/[h^{-1}M_{\odot}]) + b(z)
\end{eqnarray}
where $a(z)$ and $b(z)$ are determined by $z$. 
%Using these relations, $r_{\rm vir}$ and $r_{\rm dmh}$ can be determined by $M_{\rm dmh}^{\rm vir}$ and $z$. 
%$r_{\rm efct}$ can be estimated from the observed relation, 
%\begin{align}
%r_{\rm efct} &= r_{\rm 0} \left( \frac{L_{\rm UV}}{L_{\rm 0}} \right)^{0.27\pm 0.01} \\
%&\quad r_{\rm 0} = B_{\rm z} (1 + z)^{\beta_{\rm z}} \nonumber \\
%&\quad \log_{10} \left( \frac{L_{\rm UV}}{L_{\rm 0}} \right) = %\frac{M_{\rm UV} - M_{\rm 0}}{-2.5} \nonumber\\
%&\quad M_{\rm UV} = \frac{\log_{10} M_{\rm stl}^{\rm tot} - p_{\rm 1}(z)}{p_{\rm 2}(z)}, \nonumber
%\end{align}
%where $B_{\rm z}$, $\beta_{\rm z}$, $p_{\rm 1}(z)$, and $p_{\rm 2}(z)$ are the parameters determined by $z$ \citep{Shibuya15}. 
%$M_{\rm UV}$ and $L_{\rm UV}$ represent the UV magnitude and luminosity, respectively. 
%The specific luminosity $L_{\rm 0}$ corresponds to $M_{\rm 0}=$-21. 
%The specific radius $r_{\rm 0}$ represents the effective radius at a luminosity of $L_{\rm 0}$ at $z\sim$3. 
From the observed relation, $r_{\rm efct}$ can be estimated by $M_{\rm stl}^{\rm tot}$ and $z$ \citep{Shibuya15}.
In addition, we use the definition of \citet{Mo02} for the circular velocity $v_{\rm circ}$.

% summary
As a result, $M_{\rm dmh}^{\rm vir}$, $r_{\rm vir}$, $r_{\rm dmh}$, $r_{\rm efct}$, and $v_{\rm circ}$ can be determined by the given $M_{\rm stl}^{\rm tot}$ and $z$.
Therefore, our model can give velocity, density, temperature, and pressure distributions from the parameters $z$, $M_{\rm stl}^{\rm tot}$, $\dot{m}_{\rm SFR}$, $\dot{m}$ ($\lambda$ and $\eta$).
Estimated parameters for local galaxies are summarized in Table \ref{tab_est_loc} and for high-redshift galaxies in Table \ref{tab_est_hiz} in the Appendix 4.
Therefore, our model can give velocity, density, temperature, and pressure distributions from the parameters $z$, $M_{\rm stl}^{\rm tot}$, $\dot{m}_{\rm SFR}$, $\dot{m}$ ($\lambda$ and $\eta$).
Because $z$, $M_{\rm stl}^{\rm tot}$ and $\dot{m}_{\rm SFR}$ are observable parameters, our model can predict $\dot{m}$ ($\lambda$ and $\eta$) from observed velocities.
In Section \ref{sec_res_mfx}, we apply the transonic outflow model to local and high-redshift star-forming galaxies.

\subsection{Influence of mass flux to the transonic galactic winds}
\label{sec_res_mfx}

This section estimates the mass flux from the observed velocity with the transonic outflow model. 
Additionally, we derive $\lambda$ and $\eta$ from the estimated mass flux. 

\subsubsection{Relation between mass flux and velocity profiles}
\label{sec_res_mfx_1}

\begin{figure}
 \begin{center}
    \includegraphics[width=\linewidth]{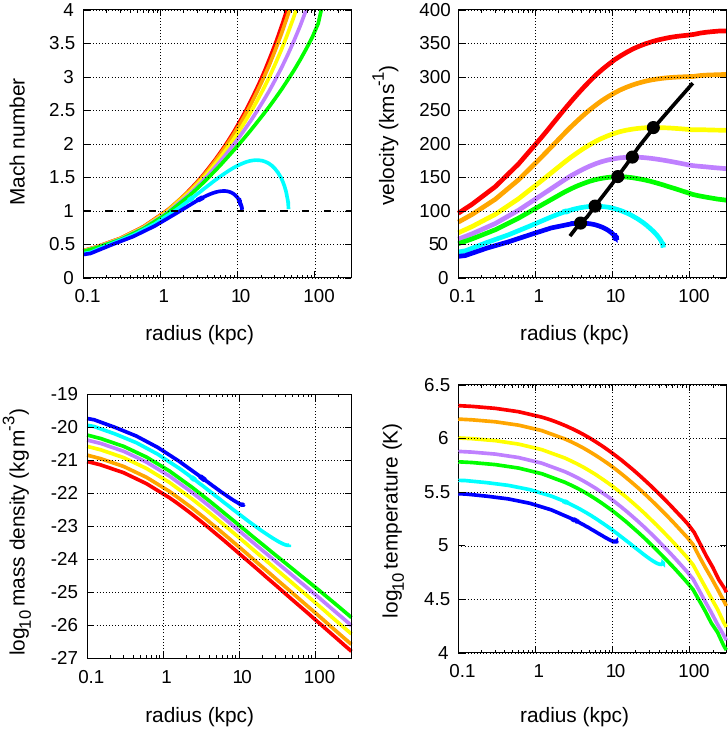}
  \caption{
Predicted Mach number, velocity, mass density, and temperature in a galaxy ($z$, $\log_{10}M_{\rm stl} $, $\log_{10}M_{\rm dmh}$, SFR)=(0.1, 8.6, 11.0,  and 10), respectively. 
The colors represent different $\dot{m}$ (and $\lambda$): red, $\dot{m}$=10.5($\lambda$=3); orange, 14.0(4); yellow, 21.0(6); purple, 28.0(8); green, 35.0(10); cyan, 52.5(15); and blue, 70.0(20), respectively.
The black line shows the locus of maximum velocity in the velocity graph. 
  }
  \label{img_vel}
 \end{center}
\end{figure}

% velocity profile 
Figure \ref{img_vel} shows the Mach number, velocity, number density, and temperature profiles for various mass fluxes.
The line color represents the mass flux.
Section \ref{sec_res_cls} shows that the type of solution varies with the mass flux.
In figure \ref{img_vel}, for large galaxies ($\log_{\rm 10} M_{\rm dmh} = 11.0$) with small mass fluxes ($\dot{m} \lesssim 50$ $M_{\rm \odot}$ ${\rm yr}^{-1}$), the solution is X Type, and the outflow line extends to the infinite distance.
For moderate mass fluxes ($\dot{m} \sim 50-70$ $M_{\rm \odot}$ ${\rm yr}^{-1}$), the solution type becomes XST Type.
For large mass fluxes ($\dot{m} \gtrsim 70$ $M_{\rm \odot}$ ${\rm yr}^{-1}$), it becomes T Type and no transonic solution exists.

% SFR influence to velocity 
Here, we note that the velocity is independent of the SFR in our model.
In Section \ref{sec_mdl_beq}, we define $\dot{m}$ and $\dot{e}$ as in equations (\ref{eqn_flx_mas}) and (\ref{eqn_flx_ene}).
Since $\dot{\rho}_{\rm m}$ and $\dot{q}$ have the same distribution proportional to SFR, $\dot{m}$ and $\dot{e}$ are also proportional to SFR.
Therefore, the energy per mass $\dot{e}/\dot{m}$ is independent of SFR.
Furthermore, $\mathcal{M}$ is independent of SFR because $\mathcal{M}$ depends on non-dimensional parameters and SFR is canceled out in non-dimensional parameters.
Equation (\ref{eqn_efx}) determines the velocity from $\dot{e}/\dot{m}$ and $\mathcal{M}$.
As a result, the velocity is independent of SFR.
SFR affects the distribution of other quantities, such as density and pressure.
Unlike SFR, $\lambda$ increases $\dot{\rho}_{\rm m}$ but does not affect $\dot{q}$.
Therefore, $\lambda$ changes the velocity.
This independence of SFR and velocity differs from other wind models.
Other models make different assumptions than our model and include physical processes such as radiative cooling.
Therefore, other models produce a velocity distribution that depends on the amount of SFR.

% solution patterns 
Figure \ref{img_vel} shows that the outflow of the XST Type solution ($\dot{m} \sim 50-70$ $M_{\rm \odot}$ ${\rm yr}^{-1}$) accelerates in the star-forming region and decelerates due to the gravitational potential of the dark matter halo.
Furthermore, in the XST Type solution, the outflow velocity profile shows the maximum velocity at a specific locus.
As the mass flux decreases, the maximum velocity increases and the specific locus moves outward.
For X Type solutions with moderate mass fluxes ($\dot{m} \sim 20-35 $ $M_{\rm \odot}$ ${\rm yr}^{-1}$), the maximum velocity is greater than the terminal velocity given by equation (\ref{eqn_trm}).
Also, for X Type solutions with small mass fluxes ($\dot{m} \lesssim 20 $ $M_{\rm \odot}$ ${\rm yr}^{-1}$), the outflow does not decelerate.
Therefore, the maximum velocity is equivalent to the terminal velocity.
As a result, we classify X Type solutions into two sub-types depending on the maximum velocity relative to the terminal velocity: (1) solutions where the maximum velocity is greater than the terminal velocity, and (2) solutions with maximum velocity equivalent to the terminal velocity.
If the specific location of the maximum velocity is outside the virial radius, we fix the maximum velocity to the terminal velocity.

\subsubsection{Relation between mass flux and solution type}

As mentioned in Section \ref{sec_mdl_prm}, our model includes a set of parameters ($z$, $M_{\rm stl}^{\rm tot}$, $\dot{m}_{\rm SFR}$, $\lambda$).
Since the velocity distribution is independent of the SFR as noted in sub-section \ref{sec_res_mfx_1}, the parameters ($z$, $M_{\rm stl}^{\rm tot}$, $\lambda$) are needed to determine the maximum velocity $v_{\rm max}$.
In this section, we examine the relation between ($z$, $M_{\rm stl}^{\rm tot}$, $\lambda$) and $v_{\rm max}$.

\begin{figure*}
 \begin{center}
  \includegraphics[width=\linewidth]{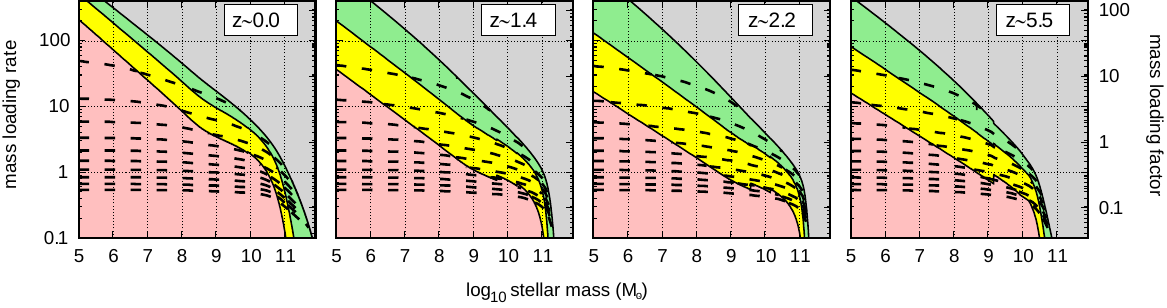}
    \includegraphics[width=\linewidth]{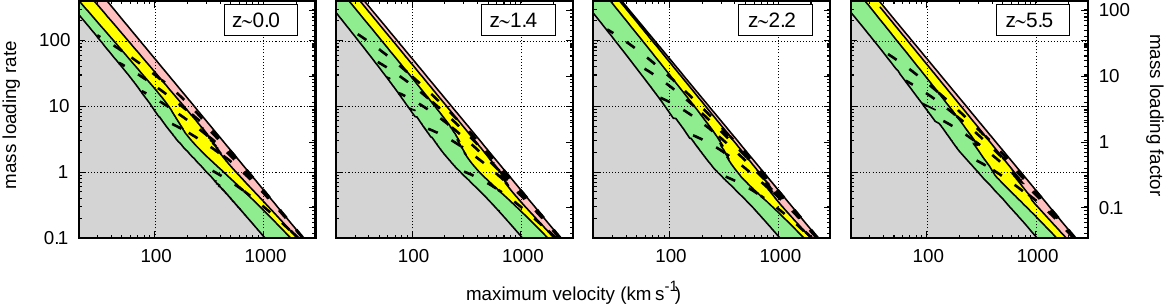} 
  \caption{
Top: solution types in stellar mass and mass loading rate (mass loading factor) parameter space.
The diagrams have different redshifts $z$.
The pink area represents an X Type solution where the maximum velocity is consistent with the terminal velocity.
The yellow region shows the X Type solution where the maximum velocity is larger than the terminal velocity.
The green region represents the XST Type solution.
The grey region shows the forbidden region (T Type region).
The dashed lines show different velocities ($v_{\rm max}=$100-1000 ${\rm km}$ ${\rm s}^{-1}$).
Bottom: solution types in maximum velocity and mass loading rate (mass loading factor) parameter space.
The black dashed lines show different stellar masses ($M_{\rm stl}^{\rm tot}=10^{7-11} M_{\odot}$).}
  \label{img_mst_vmx_mlr}
 \end{center}
\end{figure*}

% mass loading vs velocity 
Diagrams for ($v_{\rm max}$, $\lambda$) or ($M_{\rm stl}^{\rm tot}$, $\lambda$) for different values of $z$ are shown at figure \ref{img_mst_vmx_mlr}.
The background color indicates the type of solution.
The grey area represents the T type for which no transonic outflow exists.
This region represents the forbidden region of transonic winds.
The green region indicates the XST type.
In this region, the transonic flow forms a maximum velocity and stops in the gravitational potential of the dark matter halo.
The yellow and pink regions show the X type.
In the yellow region, the transonic flow decelerates in the gravitational potential of the dark matter halo and its maximum velocity is greater than the terminal velocity. 
This solution possibly forms shock waves as it decelerates.
In the pink region, the transonic flow does not decelerate, and therefore its maximum and terminal velocities are the same value.
Although these two solutions are outflows that can escape from the galaxy, they differ in the relation between the maximum and terminal velocities.
Here, we consider them separately as theoretically different solutions.

% border lines 
For a given $\dot{e}_{\rm SN} / \dot{m}_{\rm SN}$, $\lambda$ has an upper limit to a given maximum velocity.
This upper limit is the case without the gravity of the galaxy as already given in equation (\ref{eqn_trm_lim}).
Our model does not treat outflow processes beyond this upper limit.
In the bottom row of figure \ref{img_mst_vmx_mlr}, the white region locates beyond this upper limit.
The black solid line between the white and pink regions represents the upper limit by equation (\ref{eqn_trm_lim}).
Using the parameters given in Section \ref{sec_mdl_prm}, equation (\ref{eqn_trm_lim}) becomes $v_{\rm term} \sim \sqrt{2\dot{e}_{\rm SN} / \lambda\dot{m}_{\rm SN}}\sim$ 731 $\lambda^{-1/2}$ km s$^{-1}$.
The black solid line between the yellow and green regions represents the transonic outflow where the terminal velocity becomes 0.
Therefore, this line derives from $\lambda=\lambda_{\rm esc}$.
We derive the black solid line between the pink and yellow regions numerically and that between the green and grey regions.

% mass loading limit 
The top row of figure \ref{img_mst_vmx_mlr} shows diagrams for ($M_{\rm stl}^{\rm tot}$, $\lambda$) for different $z$.
The black dashed lines represent $v_{\rm max}$ from 100 to 1000 ${\rm km}$ ${\rm s}^{-1}$.
These diagrams show that the upper limit of $\lambda$ decreases for $M_{\rm stl}^{\rm tot}$.
This indicates that the gravitational potential determines the upper limit of the mass loading effect.
Even with large $\lambda$, transonic galactic winds can exist with small $M_{\rm stl}^{\rm tot}$.
For example, if $M_{\rm stl}^{\rm tot}\sim$ 10$^{6-7}M_{\rm \odot}$, $\lambda$ can increase up to tens.
This indicates that transonic galactic winds effectively remove the interstellar medium from small galaxies.
At high $z$, the upper limit of $\lambda$ decreases for $M_{\rm stl}^{\rm tot}$ because the stellar-to-halo mass ratio varies with $z$.
For small galaxies, $\lambda$ can still increase to tens.
If $M_{\rm stl}^{\rm tot}$ is more massive than a specific value ($\sim$10$^{11}M_{\rm \odot}$), $\lambda$ cannot be larger than unity.
At high $z$, this specific value of $M_{\rm stl}^{\rm tot}$ decreases to $\sim$10$^{10}M_{\rm \odot}$.
This indicates that transonic galactic winds do not effectively remove the interstellar medium from massive galaxies ($M_{\rm stl}^{\rm tot}$ $\gtrsim$ 10$^{10-11}M_{\rm \odot}$).

% tight correlation 
Next, the bottom row of figure \ref{img_mst_vmx_mlr} shows diagrams for ($v_{\rm max}$, $\lambda$) for different $z$.
The black dashed lines represent $M_{\rm stl}^{\rm tot}$ from 10$^7$ to 10$^{11} M_{\odot}$.
These lines extend in a narrow area.
This tight correlation indicates that $v_{\rm max}$ correlates strongly negatively with $\lambda$ (and $\eta$).
The velocity distribution strongly depends on $\lambda$ because $\lambda$ increases the mass flux and strengthens the braking effect of the gravitational potential.
This tight correlation exists at high $z$ as well as at low $z$.

\begin{figure*}
 \begin{center}
\includegraphics[width=\linewidth]{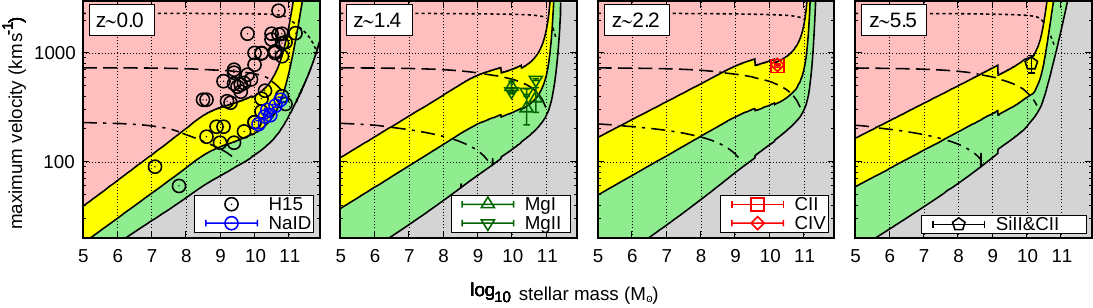}
  \caption{
Solution types of galactic outflows in the parameter space (stellar mass and maximum velocity).
The background colors show solution types in the same manner as in figure \ref{img_mst_vmx_mlr}.
The black lines represent different mass loading ratios (dotted:0.1, dashed:1, dot-dashed:10).
The black circles are observation data from \citet{Heckman15, Heckman16}.
The colored dots are observation data from \citet{Sugahara17, Sugahara19} using different tracers.}
  \label{img_mst_vmx}
 \end{center}
\end{figure*}

% velocity vs stellar mass 
Finally, figure \ref{img_mst_vmx} shows diagrams of ($M_{\rm stl}^{\rm tot}$, $v_{\rm max}$) with different $z$.
The background color and black solid lines are the same as in Figure \ref{img_mst_vmx_mlr}.
The black dashed lines indicate $\lambda$ values of 0.1, 1, and 10.
Since the escape velocity of large galaxies is greater than that of small galaxies, the forbidden region (green region) extends widely with large $M_{\rm stl}^{\rm tot}$.
In the green and yellow regions, the black dashed lines show that $v_{\rm max}$ correlates negatively with $M_{\rm stl}^{\rm tot}$.
This indicates that the gravitational potential decreases velocities.
In the pink region, these lines extend horizontally because the small gravitational potential does not strongly affect velocities.

\subsection{Massfluxes in star-forming galaxies}
\label{sec_res_mlr}

% maximum velocity assumption 
The outflow velocity, such as $v_{\rm max}$, is deduced from the observed line width.
In actual galaxies, flowing gas with various velocities broadens the observed line width.
Therefore, there are various manners of defining outflow velocity.
The central velocity (or average velocity) is defined as the 50\% point of the cumulative velocity distribution (CVD: starting from the red side of the line).
The flowing absorbers spread over a wide range of velocities.
Therefore, the central velocity represents the average velocity of the absorbers.
The maximum velocity is defined as the 90\% point of the CVD; $v_{90}$ \citep{Rupke05a,Rupke05b,Chisholm15,Arribas14} or the 98\% point of the CVD; $v_{98}$ \citep{Heckman16}.
The maximum velocity corresponds to the velocity of the absorber most accelerated by the hot gas flow.
In this study, both $v_{90}$ and $v_{98}$ are used as $v_{\rm max}$.

% absorption lines and maximum velocity assumption (upper limit) 
The absorbers that form the observed absorption lines, such as NaID, are warm ionized gases distributed with the cold gas in the hot gas flow.
The ram pressure accelerates the absorbers in the hot gas flow, but hydrodynamic instabilities can destroy them \citep{Scannapieco17}.
This instability depends on the velocity gap between these gases and the hot gas flow.
It is still unclear whether they are completely destroyed before the hot gas flow fully accelerates them.
Additionally, if the hot gas temperature decreases far from the star-forming region, absorbers may form from the hot gas due to radiative cooling \citep{Thompson16}.
%These absorbers already have the velocity of the hot gas flow.
These several possible acceleration mechanisms exist for the absorbers forming absorption lines.
This study does not consider these mechanisms in detail.
For simplicity, we estimate the mass flux, assuming that the most accelerated absorber has a velocity equivalent to that of the hot gas flow.
If the hot gas flow does not sufficiently accelerate the absorbers, the observed maximum velocity will be lower than the velocity of the hot gas flow.
In this case, the predicted mass flux will be larger than the true value.
Therefore, our predicted value corresponds to the upper limit of the mass flux.

% solution types for observed galaxies 
We add the observed parameters of the actual galaxies to the diagrams in figure \ref{img_mst_vmx}.
Tables \ref{tab_prm_loc} and \ref{tab_prm_hiz} summarize the parameters adopted in this study.
All observed parameters are located in the X-type (pink, yellow) or XST-type (green) regions for local star-forming galaxies.
They are not in the forbidden regions (T-type, grey).
Therefore, we can estimate the mass flux and $\lambda$ (and $\eta$) for local star-forming galaxies.
Also, all observed parameters locate in the X-type or XST-type regions for high-redshift star-forming galaxies.
Therefore, we can estimate the mass fluxes.

\begin{figure*}
 \begin{center}
\includegraphics[width=\linewidth]{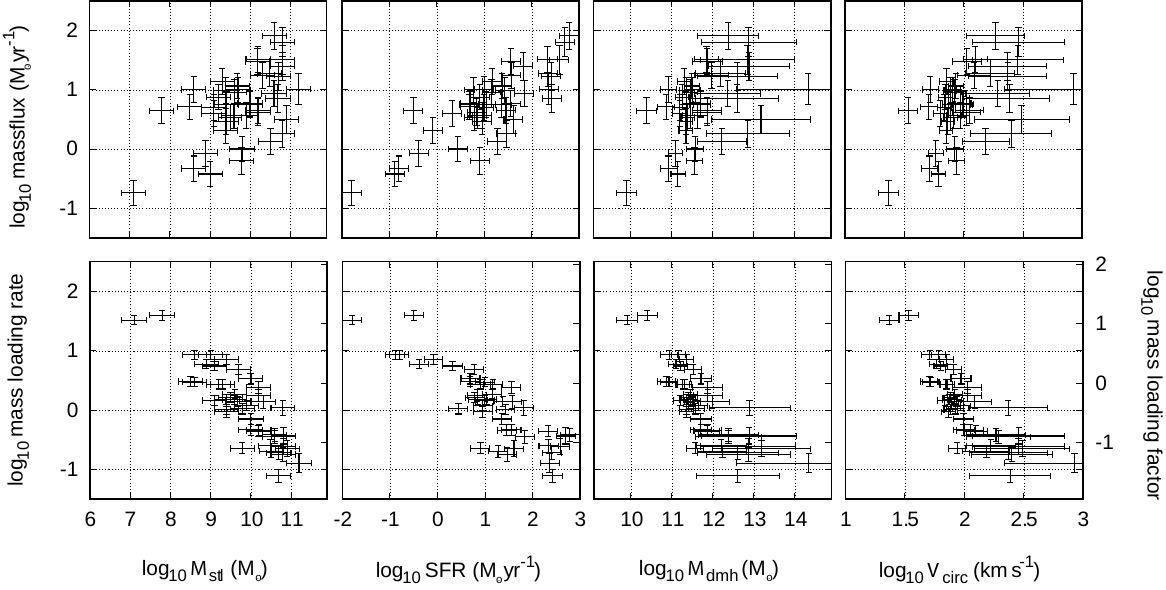}
  \caption{
Mass fluxes and mass loading ratio (mass loading factors) are estimated by the transonic outflow model in local galaxies using the data from \citet{Heckman15, Heckman16}. 
  }
  \label{img_mfx_loc}
 \end{center}
 \begin{center}
\includegraphics[width=\linewidth]{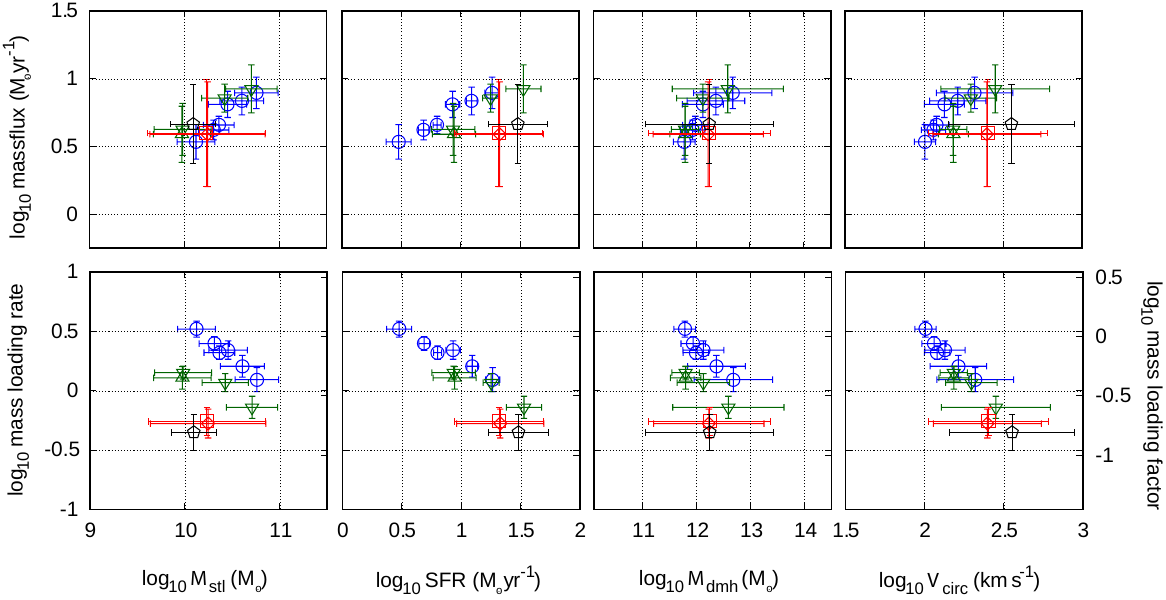}
  \caption{
Mass fluxes and the mass loading ratio (mass loading factors) are estimated by our transonic outflow model in high-redshift galaxies using the data from \citet{Sugahara17,Sugahara19}.
The open symbols show the results of our transonic outflow model: blue circles, $z\sim 0.0$ (NaID); green triangles, $z\sim 1.4$ (MgI); green inverted triangles, $z\sim 1.4$ (MgII); red squares, $z\sim 2.2$ (CII); red diamonds, $z\sim 2.2$ (CIV); and black pentagons, $z\sim 5.5$ (SiII and CII). 
  }
  \label{img_mfx_hiz}
 \end{center}
\end{figure*}

% mass flux of local galaxies 
Figure \ref{img_mfx_loc} shows the results for local star-forming galaxies, which are summarized in Table \ref{tab_mfx_loc}.
The circular velocity is estimated from the halo mass, as described in Section \ref{sec_mdl_prm}.
As shown in the top row of figure \ref{img_mfx_loc}, the mass fluxes correlate positively with stellar mass, halo mass (circular velocity), and SFR (and redshift).
This indicates that massive galaxies have a large mass flux compared to less massive galaxies.

% mass loading of local galaxies 
Next, we estimate the $\lambda$ (and $\eta$) value for local star-forming galaxies.
As shown in the bottom row of Figure \ref{img_mfx_loc}, $\lambda$ correlates negatively with stellar mass, halo mass (circular velocity), and SFR (and redshift).
For less massive galaxies ($M_{\rm stl}\lesssim 10^{9-10}M_{\rm \odot},$ $M_{\rm dmh}\lesssim 10^{11-12}M_{\rm \odot}$), $\lambda$ is larger than unity.
This indicates that the mass flux of less massive galaxies is larger than the mass flux injected from SNe due to the entrainment of the interstellar medium.
Galactic winds can effectively entrain the interstellar medium.
Therefore, in less massive galaxies, the gravitational potential cannot strongly suppress the acceleration process of the galactic wind.
This result implies that galactic winds significantly suppress star formation in less massive galaxies.
In contrast, for massive galaxies ($M_{\rm stl}\gtrsim 10^{9-10}M_{\rm \odot},$ $M_{\rm dmh}\gtrsim 10^{11-12}M_{\rm \odot}$), $\lambda$ is smaller than unity.
Therefore, due to the strong gravity in massive galaxies, the mass flux ejected from the galaxy is likely smaller than the mass flux injected from the SNe.
In these galaxies, non-negligible masses remain in the galaxy.
Massive galaxies cannot effectively eject the interstellar medium.

% mass loading and metal enrichment 
Furthermore, the magnitude of the mass loading affects metal enrichment in intergalactic space.
Our results show that the mass loading ratio depends on the halo mass.
Since $\lambda$ is smaller in massive galaxies, interstellar medium with low metallicity cannot be efficiently ejected by galactic winds.
Therefore, the metallicity of the mass flux may increase in massive galaxies.
The number of galaxies also affects the metal enrichment in intergalactic space.
The number of massive galaxies is smaller than that of less massive galaxies \citep{Lilly95, Lin99, Cole01, Bell03, Wolf03}.
Therefore, the metallicity of massive galaxies is expected to be high, but we do not determine the impact on metal enrichment in intergalactic space.

% high-z galaxies 
Finally, we estimate the mass flux and $\lambda$ (and $\eta$) in high-z star-forming galaxies ($z\sim$1.4, 2.2, 5.5).
The results are shown in figure \ref{img_mfx_hiz} and summarized in Table \ref{tab_mfx_hiz}.
As shown in the top row of figure \ref{img_mfx_hiz}, at $z\sim$1.4, the mass flux has a similar correlation to that of local star-forming galaxies ($z\sim$0).
At $z\sim$2.2 and 5.5, the mass flux has significant uncertainty.
As shown in the bottom row of figure \ref{img_mfx_hiz}, $\lambda$ for high-z galaxies ($z\sim$ 1.4, 2.2, 5.5) is slightly smaller than that for local star-forming galaxies ($z\sim$ 0).
Furthermore, $\lambda$ correlates with the circular velocity independently of $z$.
This indicates that the efficiency of mass loading depends on the structure of the gravitational potential (and hence on the circular velocity).

\subsection{Estimation formula for mass flux}
\label{sec_res_frm}

% fitting functions 
In Section \ref{sec_res_mfx}, we have shown that our transonic wind model can estimate $\lambda$ from $M_{\rm stl}^{\rm tot}$ and $v_{\rm max}$.
In this section, we show the equations that approximate $\lambda$ to $M_{\rm stl}^{\rm tot}$ and $v_{\rm max}$.
We assume a function with a constant core and power-law wing for the fitting.
This function will be like a double power law.
Also, we assume that the range of $v_{\rm max}$ has a lower limit $v_{\rm max,m}$.
The lower limit $v_{\rm max,m}$ is the boundary between XST Type and T Type regions as shown in Figure \ref{img_mst_vmx_mlr}. 
For practical evaluation, the upper limit of $v_{\rm max}$ is fixed at 100 times $v_{\rm circ}$.
The circular velocity $v_{\rm circ}$ can be estimated from $M_{\rm stl}^{\rm tot}$ as described in Section \ref{sec_mdl_prm}.
For $v_{\rm max,m} < v_{\rm max} < 100 v_{\rm circ}$, the fitting formula to $\eta$ (or $\lambda$) is assumed as 
\begin{eqnarray}
\eta(z,v_{\rm max},M_{\rm stl}^{\rm tot})
&=& R_{\rm f} \lambda(z,v_{\rm max},M_{\rm stl}^{\rm tot}) \nonumber\\
&=& A_{\rm m} 
\left\{ 
\frac{v_{\rm max}+v_{\rm m}}{100\ {\rm km}\ {\rm s}^{-1}} 
\right\}^{n_{\rm m}} 
\label{eqn_apx_vmx}
\end{eqnarray}
where $n_{\rm m}$, $A_{\rm m}$, $v_{\rm m}$ and $v_{\rm max,m}$ are parameters determined by $M_{\rm stl}^{\rm tot}$ with a fixed $z$. 
As $v_{\rm max}$ decreases, $\lambda$ approaches constant.
The transition (breakpoint) from the core constant to the power-law wing is indicated by $v_{\rm m}$.
We use the chi-squared method for approximation.

% fitting results 
We summarize the results of the fitting at $z\sim 0$ in Table \ref{tab_res_frm}.
The value of equation (\ref{eqn_apx_vmx}) differs from the exact value of our model by about $\pm 3-4\%$.
The stellar mass $M_{\rm stl}^{\rm tot}$ strongly affects $v_{\rm m}$.
The estimate of $v_{\rm term}$ by equation (\ref{eqn_trm_lim}) shows that $\lambda$ is proportional to $v_{\rm max}^{-2}$ for large $v_{\rm max}$.
The power exponent $n_{\rm m}$ is almost $-2$, but $\lambda$ is constant at $v_{\rm max}$ smaller than $v_{\rm m}$.
Thisindicates that the mass loading increases the braking effect of the gravitational potential and that $\lambda$ is affected by the gravitational potential.

% other fitting 
As $v_{\rm max}/100 < v_{\rm circ}< v_{\rm circ,s}$, equation (\ref{eqn_apx_vmx}) can be rewritten as
\begin{eqnarray}
\eta(z,v_{\rm max},M_{\rm stl}^{\rm tot})
&=& R_{\rm f}\lambda(z,v_{\rm max},M_{\rm stl}^{\rm tot}) \nonumber\\
&=& A_{\rm s} 
\left\{ 
\frac{M_{\rm stl}^{\rm tot} + M_{\rm s}}{10^{7} M_{\rm \odot}} 
\right\}^{n_{\rm s}} 
\end{eqnarray}
where $n_{\rm s}$, $A_{\rm s}$, $M_{\rm s}$, and $v_{\rm circ,s}$ are parameters determined by $v_{\rm max}$ with fixed $z$.
Next, we obtain approximate expressions for $\lambda$ and $v_{\rm circ}$ to investigate the relation between mass loading and gravitational potential.
As $v_{\rm max}/100 < v_{\rm circ} < v_{\rm circ,c}$, equation (\ref{eqn_apx_vmx}) can be rewritten as
\begin{eqnarray}
\eta(z,v_{\rm max},v_{\rm circ})
&=& R_{\rm f} \lambda(z,v_{\rm max},v_{\rm circ}) \nonumber\\
&=& A_{\rm c} 
\left\{ 
\frac{v_{\rm circ} + v_{\rm c}}{100\ {\rm km}\ {\rm s}^{-1}} 
\right\}^{n_{\rm c}} 
\end{eqnarray}
where $n_{\rm c}$, $A_{\rm c}$, $v_{\rm c}$, and $v_{\rm circ,c}$ are parameters determined by $v_{\rm max}$ with fixed $z$.
For small $v_{\rm max}$, the power exponent $n_{\rm c}$ is almost $-2$.
In this case, $\lambda$ depends strongly on $v_{\rm circ}$ because the gravitational potential suppresses the acceleration of the galactic wind.
For large $v_{\rm max}$, the power exponent $n_{\rm c}$ is almost $0$.
In this case, $\lambda$ does not depend much on $v_{\rm circ}$ because the energy injection sufficiently exceeds the gravitational potential energy.

\section{discussion}
\subsection{Comparison to the shell model}
\label{sec_dis_shm}

% shell model 
This section compares our model with the shell model \citep{Desjacques04, Ferrara06}.
Hot gas ($\sim$10$^{6-8}$K) spreads widely from the star-forming region and creates a shock front.
In the hot gas, hydrodynamic instabilities destroy the cold gas ($\sim$10$^{2-4}$K).
The expanding hot gas sweeps up the surviving cold gas.
The compressed cold gas forms a thin shell at the shock front.
As a result, an expanding dense shell surrounds the dilute hot gas.
The shell model considers this expanding gas as the mass flux of the galactic wind.
The mass flux is defined as
\begin{eqnarray}
\dot{m} = \Omega \: C_{\rm f} \: \mu \: m_{\rm p} \: N_{\rm H} \: v_{\rm sh}\: r_{\rm sh},
\end{eqnarray}
where $\Omega$, $C_{\rm f}$, $\mu$, $m_{\rm p}$, $N_{\rm H}$, $v_{\rm sh}$, and $r_{\rm sh}$ are the solid angle, covering factor, molecular weight, proton mass, column density, shell velocity, and shell radius, respectively, which are all constants.
From the metallicity of the outflow gas, $N_{\rm H}$ is estimated.
There are unavoidable uncertainties in estimating the metallicity and converting it to column density.
In addition, most previous studies use the central velocity as $v_{\rm sh}$.
Since absorbers have a variety of velocities, the central velocity corresponds to the average of those velocities.
The shell model does not consider the velocity profile.
It is also difficult to determine $r_{\rm sh}$ by observation.
The shell model assumes several times the radius of the star-forming region, such as the UV half-light radius.
Therefore, the shell model leads to inevitable uncertainties in estimating the mass flux.

\begin{figure*}
 \begin{center}
  \includegraphics[width=\linewidth]
  {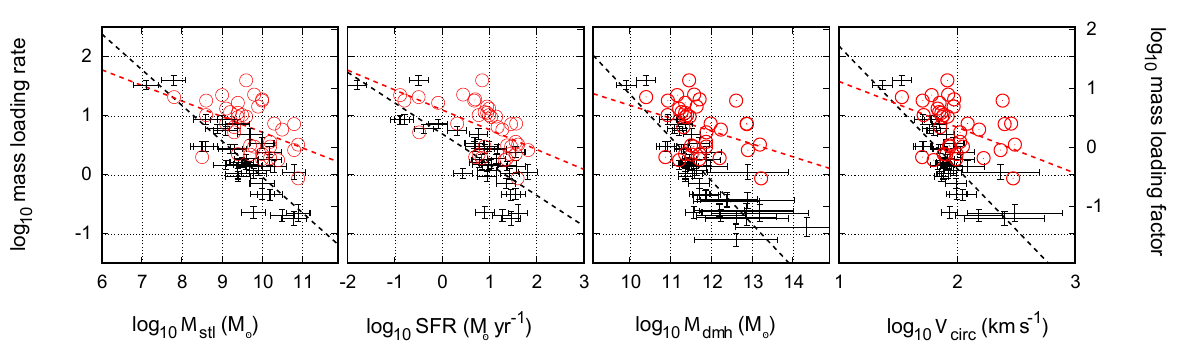}
  \caption{
Estimated mass loading ratio (mass loading factors) by the transonic outflow model and shell outflow model in local galaxies. 
The black crosses represent the results of our transonic outflow model. 
The red circles represent the results of the shell model \citep{Heckman15}. 
The black dashed line and the red dashed line show the average lines. 
  }
  \label{img_mlr_shm_loc}
 \end{center}
\end{figure*}

% comparison to shell model 
Figure \ref{img_mlr_shm_loc} compares the results of the two models.
The black dots and red circles represent the $\lambda$ values of our model and the shell model \citep{Heckman15}, respectively.
We find that the $\lambda$ value of our model correlates clearly with stellar mass, SFR, halo mass, and circular velocity, while that of the shell model correlates less clearly.
For large galaxies ($M_{\rm dmh} \gtrsim 10^{11-12} M_{\odot}$), the $\lambda$ of our model is smaller than that of the shell model, and for small galaxies ($M_{\rm dmh} \lesssim 10^{11-12} M_{\odot}$), the two models have nearly equal values.
The chi-squared fitting yields a correlation of $\eta$ ($=R_{\rm f} \lambda$) for the transonic wind model as
\begin{eqnarray}
\log_{10} \eta 
&= 5.549 - 0.604 \times \log_{10} M_{\rm stl}^{\rm tot}  \nonumber\\
&= 0.237 - 0.521 \times \log_{10} \dot{m}_{\rm SFR}  \nonumber\\
&= 8.227 - 0.732 \times \log_{10} M_{\rm dmh}  \nonumber\\
&= 3.798 - 2.074 \times \log_{10} v_{\rm circ} 
\end{eqnarray}
and the shell model as
\begin{eqnarray}
\log_{10} \eta 
&= 2.903 - 0.264 \times \log_{10} M_{\rm stl}^{\rm tot}  \nonumber\\
&= 0.642 - 0.336 \times \log_{10} \dot{m}_{\rm SFR} \nonumber\\
&= 2.899 - 0.218 \times \log_{10} M_{\rm dmh}  \nonumber\\
&= 1.900 - 0.777 \times \log_{10} v_{\rm circ} 
\end{eqnarray}
These results would suggest that uncertainties in the shell model prevent accurate estimation of the mass flux.
Our model can estimate the mass flux more accurately than the shell model.
However, as described below, the different outflow images in the two models may also affect this result.

% difference of two models 
Our transonic outflow model focuses on galaxy-scale hot flows, while the shell model focuses on an expanding thin shell filled with hot gas.
Over time, the expanding thin shell becomes unstable due to the Rayleigh-Taylor instability \citep{Ferrara06}.
The hot gas can pass through the fragmented shell.
Therefore, a galaxy-scale hot flow spreads inside and outside the fragmented shell.
The hot gas escapes from the galaxy at terminal velocity, entraining fragments.
Meanwhile, surviving fragments cannot escape due to gravitational potential.
Hence, theoretically, only a fraction of the temporarily moving gas is the mass flux escaping from the galaxy.
The escape mass flux depends on the injected energy and gravitational potential.
Also, they determine the maximum velocity.
Our transonic outflow model estimates the escape mass flux from the maximum velocity.
In contrast, the shell model estimates the temporary moving mass.
As a result, the mass flux of the shell model can be larger than the escape mass flux.
In other words, even for the same galaxy, the mass flux of our model could be smaller than that of the shell model.
This conjecture is consistent with the results shown in Figure \ref{img_mlr_shm_loc}.
For large galaxies, the gravitational potential strongly influences the galactic wind; therefore, the mass fluxes shown by the two models tend to be very different.
The two mass fluxes tend to be almost the same value for smaller galaxies.
It should be noted that analyses of mass loading and entrainment based on similar theoretical depictions have recently been investigated by \citet{Nguyen21} and \citet{Fielding22}.

\begin{figure*}
 \begin{center}
  \includegraphics[width=\linewidth]{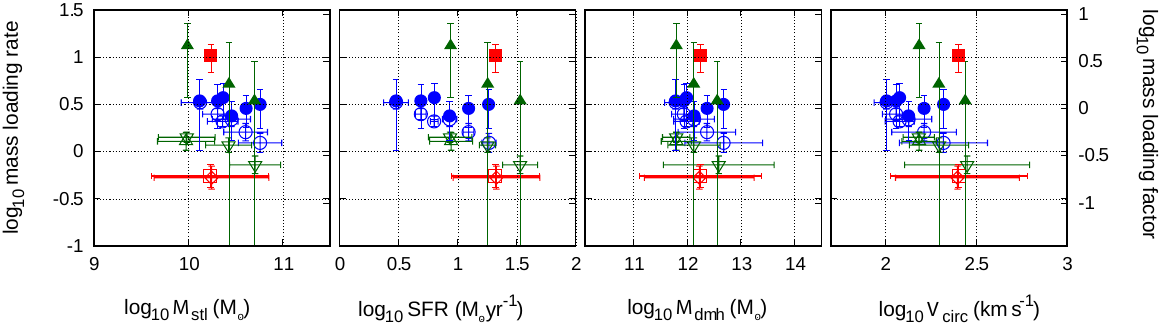}
  \caption{
Estimated mass loading ratio (mass loading factors) by the transonic outflow model and shell outflow model in high-redshift galaxies. 
The open symbols show the results of our transonic outflow model; blue circles: $z\sim 0.0$ (NaID), green triangles: $z\sim 1.4$ (MgI), green inverted triangles: $z\sim 1.4$ (MgII), red squares: $z\sim 2.2$ (CII), and red diamonds: $z\sim 2.2$ (CIV), respectively. 
The filled symbols show the results of the shell model \citep{Sugahara17}; blue circles: $z\sim 0.0 $ (NaID), green triangles: $z\sim 1.4$ (MgI), and red squares: $z\sim 2.2$ (CII), respectively. 
  }
  \label{img_mlr_shm_hiz}
 \end{center}
\end{figure*}

Figure \ref{img_mlr_shm_hiz} shows a comparison of our model with shell models in high-z star-forming galaxies.
In section \ref{sec_res_mlr}, we find that the $\lambda$ of our model correlates negatively with $z$ in the given range of stellar masses ($\sim 10^{10-11} M_{\odot}$) and halo mass.
Independent of $z$, $\lambda$ value correlates to the circular velocity.
On the other hand, as shown in Figure \ref{img_mlr_shm_hiz}, $\lambda$ in the shell model correlates positively with $z$.
The large gas fraction in high-z galaxies allows a large mass of swept-up components to move around star-forming regions.
Therefore, the swept-up shell can be more massive.
As shown in figure \ref{img_mlr_shm_hiz}, the $\lambda$ value of the shell model is particularly large in high-z galaxies.
However, as noted above, some of the swept-up mass cannot escape because of the gravitational potential.
Therefore, the gap between the swept-up mass and the escaping mass may increase in high-z galaxies.
As shown in figure \ref{img_mlr_shm_hiz}, the $\lambda$ value in our model is not larger, even in high-z galaxies, compared to the shell model.
Therefore, the $\lambda$ gap between the two models is large in high-z galaxies.

% modified shell outflow model 
Several studies \citep{Chisholm16, Chisholm17} have modified the shell model using the velocity profile of CC85.
Because CC85 assumes explosion-like galactic winds and ignores the gravitational potential, its velocity and density profiles differ from those of our model.
%\citet{Chevalier85}.
%Because \citet{Chevalier85} assumes explosion-like galactic winds and ignores the gravitational potential, its velocity and density profiles differ from those of our model.
Nevertheless, the $\lambda$ value of this modified shell model correlates clearly with stellar mass, as in our model.
The correlation has an intermediate slope $(-0.43\pm0.07)$ between our $(-0.604)$ and that of the unmodified shell model $(-0.264)$.
This modified shell model does not explicitly include the gravitational potential, but it determines the velocity and density distributions from the observed spectra.
The acceleration process of the galactic wind determines the observed spectrum.
Since the acceleration process is affected by the gravitational potential, this modified shell model should implicitly include the effect of the gravitational potential.
This effect may be responsible for the intermediate slope.


\subsection{Comparison to other transonic outflow models}
\label{sec_dis_tom}

% transonic solutions with no gravity 
Previous studies also used spherically symmetric and steady assumptions for transonic analysis \citep{Tsuchiya13, Igarashi14, Igarashi17}.
In their results, transonic points appear when the magnitude of gravity is sufficiently large.
They described this property as an analogy for the de Laval nozzle.
In the case of the de Laval nozzle, the transonic point appears when the nozzle constriction, the so-called throat, is sufficiently narrow.
Thermal energy converts to kinetic energy through the nozzle.
The gravitational potential effectively forms this throat.
Therefore, their model requires sufficiently strong gravity to form the transonic point.
In contrast, as shown in Section \ref{sec_res_cls}, our new model allows transonic points to appear even with weak gravity.
Also, other studies have obtained transonic solutions of the stellar cluster winds even without the gravitational potential \citep{Silich11, Palous13}.
Our model assumes that a mass injected at no velocity has the same velocity as the galactic wind at the injection point.
This momentum dilution reduces the kinetic energy per unit mass.
Therefore, the continuous mass injection causes a braking effect similar to the effect of gravity.
The right-hand side of the equation (\ref{eqn_mot}) also shows that mass injection has a braking effect similar to that of gravity.
If the mass injection has a continuous distribution, the energy injection also needs a continuous distribution.
If the energy injection has no continuous distribution, the flow will have all energy injected at the starting point.
It is unnatural to have much energy despite slight mass flux.
Therefore, the energy injection also needs a continuous distribution.
As a result, the transonic point appears due to the mass and energy injections, even with weak gravity.
Furthermore, it indicates that transonic points can form if the model has factors corresponding to the de Laval nozzle.

% slow-accelerating outflows 
Previous studies used isothermal or polytropic assumptions \citep{Tsuchiya13, Igarashi14, Igarashi17}.
In their models, the transonic point can exist in the large magnitude of gravity.
In that case, the transonic outflow would have low velocity, and the transonic point would locate well beyond the scale radius of the halo.
They resolved the observational problem by this slowly accelerating transonic outflow (see \citet{Igarashi14, Igarashi17}).
However, as they argued, this slowly accelerating transonic outflow requires an additional heat source outside the stellar region.
Our model assumes only energy injection from supernovae in star-forming galaxies.
There are no additional heat sources at a distance in this model.
Therefore, in this model, the transonic outflows disappear in the large magnitude of the gravity and slowly accelerating transonic outflows do not appear.
We will examine the slowly accelerating transonic outflows in future work.

% comparing to CC85 model
It is important to note that the CC85 model fixes the transonic point manually at 200 pc, and the velocity distribution inside and outside the transonic point is estimated separately without a junction analysis at the transonic point. The estimated $\lambda$ value of the CC85 model is greater than that of our model for the same observed galaxies. For some star-forming galaxies, the $\lambda$ value of the CC85 model is located in the forbidden region of our model. So far, the studies, \citet{Bustard16} have  extended the CC85 model by including the effects of gravity and non-uniform mass and energy injections, as well as radiative cooling, which is not incorporated in this study, although the critical point remains fixed at 200 pc. Recently, \citet{Nguyen23} also examined CC85, integrating the effects of non-uniform mass and energy injections. In these studies, the obtained velocities are shown to be faster than those of CC85, which is consistent with the present study. These results indicate that the transonic process, including gravitational potentials and the spatial distribution of the mass and energy injections, is crucial for correctly estimating mass fluxes. In Appendix \ref{app_ccm}, we report a detailed comparison of our transonic outflow model with CC85. 

%\subsection{Comparison to the theoretical prediction}
\subsection{Predictions of the galaxy formation studies}
\label{sec_dis_thp}

\begin{figure*}
 \begin{center}
  \includegraphics[width=\linewidth]{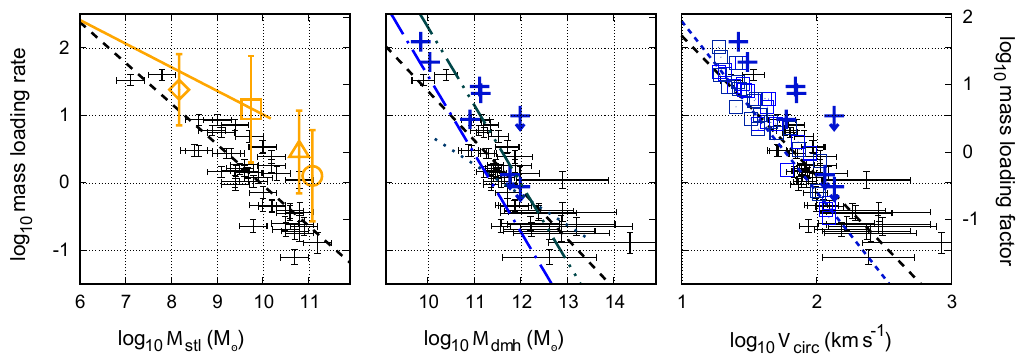}
  \caption{
Estimated mass loading ratio by the transonic outflow model and other theoretical studies in local galaxies. 
The black crosses represent the results of this transonic outflow model, while the black dashed line represents the average line. 
Left: The orange line shows the result of \citet{Muratov15} with $z\sim 0.0-4.0$ and the orange symbols show those of \citet{Hopkins12}: diamonds, SMC; triangles, MW; squares, Sbc; and circles, HiZ. 
Middle: The dotted line represents the results of \citet{Barai15} with $z\sim 0.8$, while the one-dot-dashed line and the two-dot-dashed line represent those of \citet{Mitra15} with $z\sim 0.0$ and $z\sim 1.0$, respectively. The blue crosses show the results of \citet{Muratov15} with $z\sim 0.0-0.5$. 
Right: The squares represent the results of  \citet{Christensen16} with $z\sim 0.0$ and $z\sim 0.5$ and the dotted line shows the average line of \citet{Christensen16}. 
  }
  \label{img_mlr_thp_loc}
 \end{center}
 \begin{center}
  \includegraphics[width=\linewidth]{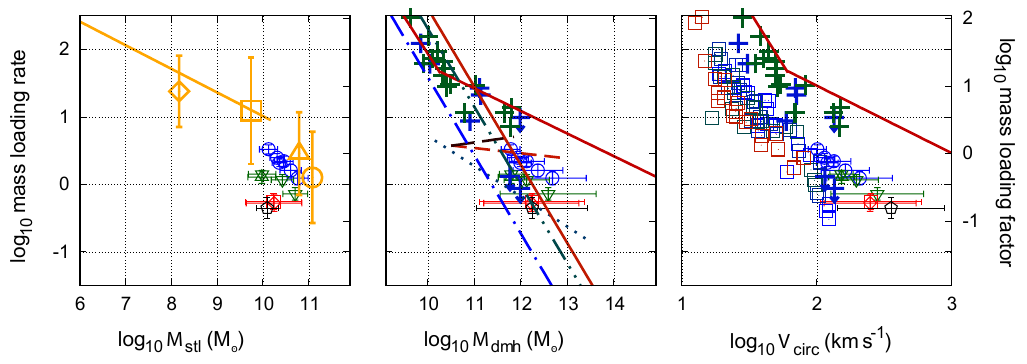}
  \caption{
Estimated mass loading ratio by the transonic outflow model and other theoretical studies in high-redshift galaxies. 
The open symbols show the results of our transonic outflow model: blue circles, $z\sim 0.0$ (NaID); green triangles, $z\sim 1.4$ (MgI); green inverted triangles, $z\sim 1.4$ (MgII); red squares, $z\sim 2.2$ (CII); red diamonds, $z\sim 2.2$ (CIV); and black pentagons, $z\sim 5.5$ (SiII\&CII).
Left: The orange line shows the results of \citet{Muratov15} with $z\sim 0.0-4.0$ and the orange symbols show those of \citet{Hopkins12}: diamonds, SMC; triangles, 
 MW; squares, Sbc; and circles, HiZ. 
Middle: The dashed lines represent the results of \citet{Barai15}: short-dashed, $z\sim 0.80$; middle-dashed, $z\sim 2.02$; and long-dashed, $z\sim 4.96$. 
The one-dot-dashed line, the two-dot-dashed line, and the three-dot-dashed line represent the results of \citet{Mitra15} with $z\sim 0.0$, $z\sim 1.0$, and $z\sim 2.0$, respectively. 
The blue crosses and the green crosses show the results of \citet{Muratov15} with $z\sim 0.0-0.5$ and $z\sim 0.5-2.0$, respectively. 
The solid line shows the results of \citet{Muratov15} with $z\sim 2.0-4.0$. 
Right: The squares represent the results of \citet{Christensen16} with $z\sim 0.0$, $0.5$, $1.0$, and $2.0$. 
  }
  \label{img_mlr_thp_hiz}
 \end{center}
\end{figure*}

% comparing to numerical predictions 
Numerical experiments \citep{Muratov15, Barai15, Christensen16,Nelson19} and semi-analytical studies \citep{Mitra15} predicted correlations between the mass loading rate $\lambda$ and other physical quantities.
These theoretical studies of galaxy formation should be able to reproduce our $\lambda$ results.
Figure \ref{img_mlr_thp_loc} compares our $\lambda$ value for local star-forming galaxies with the predictions of theoretical studies.
The $\lambda$ value of \citet{Mitra15} falls within the range of our results but is slightly smaller than our average at $M_{\rm dmh}\gtrsim 10^{11}M_{\rm \odot}$.
The $\lambda$ value of \citet{Muratov15} also falls within the range of our results but is slightly larger than our average by $M_{\rm dmh}\sim 10^{10-12}M_{\rm \odot}$.
The $\lambda$ value of \citet{Barai15} is smaller than our results for $M_{\rm dmh}\lesssim 10^{11}M_{\rm \odot}$, but about equal for $M_{\rm dmh}\gtrsim 10^{11}M_{\rm \odot}$.
The $\lambda$ value of \citet{Christensen16} is approximately equivalent to our results.
Therefore, theoretical studies of galaxy formation can roughly reproduce our results.

% high-z case 
Figure \ref{img_mlr_thp_hiz} compares our results with the predictions of theoretical studies for high-redshift star-forming galaxies.
Several theoretical studies \citep{Muratov15, Barai15, Mitra15} predicted that $\lambda$ increases with redshift, but in the same range of halo mass, our estimated $\lambda$ decreases with redshift.
Therefore, the gap between our $\lambda$ value and the predictions of theoretical studies increases at high redshifts.
Some theoretical studies of galaxy formation do not well reproduce $\lambda$ at high redshifts.

% christensen 2016
In the right panel of figure \ref{img_mlr_thp_hiz}, we compare our $\lambda$ and circular velocity relation with the numerical experiment of \citep{Christensen16}.
Their numerical experiments reproduced observational relations such as the Tully-Fisher relation, the stellar-to-halo mass ratio, and the mass-metallicity relation.
They defined the mass flux as the subtracted mass flowing in and out of the galaxy throughout 1 Gyr.
Therefore, their mass flux represents the effective amount of the escape mass flux, not the temporary amount of swept-up mass.
Our model estimates the escape mass flux by the observed maximum velocity.
Therefore, their definition of mass flux is the same as ours.
As a result, we expect their experimental results to be similar to ours.
Also, as mentioned in Section \ref{sec_res_mlr}, our estimated $\lambda$ correlates tightly with the circular velocity, regardless of the redshift.
Their experiments also implied this trend.
At high redshifts, their predictions overlap slightly with our results, and our results seem to be an extension of their predictions.
Numerical experiments on galaxy formation have not been able to resolve the transonic acceleration process of galactic winds rigorously.
Nevertheless, they have succeeded in roughly reproducing the mass flux of the galactic winds.

\subsection{The availability of our assumptions}
\label{sec_dis_ass}

Our transonic outflow model uses adiabatic, spherically symmetric, and steady assumptions.
This section examines the availability of these assumptions.

% spherical assumption 
First, we consider the spherically symmetric assumption. Naturally, spherical symmetry is not always a perfect approximation in a realistic galaxy environment. For example, a ring-like structure has been reported by \citet{Nakai87} and \citet{Weiss01}, and galactic wind models for such a case have been considered by \citet{Nguyen22}. It should be added that \citet{Fielding18} have also investigated more complex models. However, we can approximate the flow locally as spherically symmetric under several requirements described below.
The streamlines need to be linearly straight, with no interaction between the streamlines.
Also, the cross-sectional area of the streamline needs to increase outwards with the square of $r$.
The absence of stream interaction conserves the mass flux as in equation (\ref{eqn_mfx}).
Even if the total flow does not satisfy the conditions, some of it can.
In that part, we can approximate the flow locally as spherically symmetric.
For example, in the local star-forming galaxy M82, observations show that the bipolar structure of filaments and hot gas starts from the galactic plane and spreads to several kiloparsecs in the rotational axis direction \citep{Fabbiano88, Strickland00, Ohyama02}.
This indicates that the wind extends linearly in the rotational axis direction, increasing the cross-sectional area of the streamline.
If this wind conserves the mass flux, we can approximate the acceleration process around the rotational axis as spherically symmetric.
Additionally, in Section \ref{sec_res_mlr}, we use the maximum velocity observed in star-forming galaxies for the mass flux estimation.
Assuming that this maximum velocity is the velocity around the rotational axis, the spherically symmetric assumption is available for the estimation.
In this context, precisely speaking, the approximation we used should be called the 'conical streamline approximation', which, as mentioned above, consists in a local space where each streamlines is a straight line and the cross-section of the flux tube is proportional to the square of its distance from the center. Therefore, it should be emphasized again that the results of this study do not require the above conditions to be fulfilled throughout the entire space; they can be applicable even in local environments as long as they are fulfilled within such subspace.

\begin{figure}
 \begin{center}
  \includegraphics[width=\linewidth]{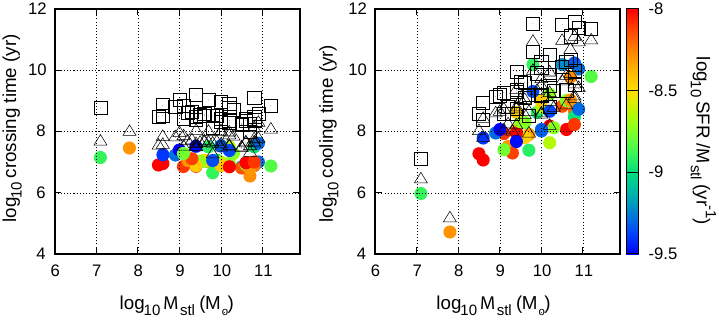}
  \caption{
Crossing times and cooling times.
left: crossing time defined by equation (\ref{eqn_tim_crs}).
right: cooling time defined by equation (\ref{eqn_tim_col}) assuming primordial metallicity.
The filled circles are timescales evaluated at the transonic point.
The open triangles represent those at the scale radius of the dark matter halo.
The open squares represent those at the virial radius.
The color of the circles represents specific SFR (the ratio of SFR to stellar mass).}
  \label{img_tim_crs}
 \end{center}
\end{figure}

% steady assumption (crossing time) 
Next, we consider the steady approximation.
The steady assumption is available if the crossing time is shorter than the star formation timescale.
The local crossing time is defined as
\begin{eqnarray}
\Delta t_{\rm cross} \equiv \frac{\Delta r}{ v(r)}
\label{eqn_tcs_def}
\end{eqnarray}
where $\Delta r$ denotes the radius difference.
Using equation (\ref{eqn_tcs_def}), we define the crossing time as 
\begin{eqnarray}
t_{\rm cross} \equiv \int_0^{r_{\rm sp}} \frac{dr}{v(r)} 
\label{eqn_tim_crs}
\end{eqnarray}
where $r_{\rm sp}$ denotes the specific radius for determining the timescales.
The left part of figure \ref{img_tim_crs} shows $t_{\rm cross}$ with various $r_{\rm sp}$ such as transonic points, scale radii and virial radii.
If $r_{\rm sp}$ is the transonic point, $t_{\rm cross}\sim$ 10 Myr.
If $r_{\rm sp}$ is the scale radius, $t_{\rm cross}\sim$ several tens of Myr.
If the star formation timescale is several tens of Myr, $t_{\rm cross}$ value is approximately equal to the star formation timescale.
Therefore, the steady assumption preserves sufficient availability for many observed star-forming galaxies beyond the transonic point.
If $r_{\rm sp}$ is the virial radius, $t_{\rm cross}$ exceeds 100 Myr.
Since this $t_{\rm cross}$ is larger than the star formation timescale, the steady assumption breaks down near the virial radius.
Nevertheless, the outflow velocity has already increased significantly near the scale radius.
For small $\lambda$, the velocity almost saturates beyond the scale radius and becomes close to the terminal velocity.
For large $\lambda$, the maximum velocity is determined well inside the virial radius because the gravitational potential decreases the velocity to the terminal velocity in the outer region.
Therefore, a breakdown of the steady assumption near the virial radius does not affect the maximum velocity and the mass flux estimation in Section \ref{sec_res_mfx}.

\begin{figure}
 \begin{center}
  \includegraphics[width=0.70\linewidth]{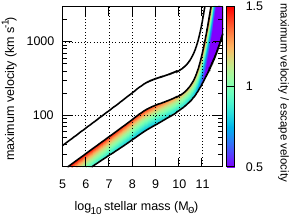}
  \caption{
The ratio of maximum velocity to escape velocity for XST-type solutions.
Escape velocities are evaluated at the virial radius.}
  \label{img_mst_vmx_esc}
 \end{center}
\end{figure}

% XST-type solution 
As mentioned in section \ref{sec_res_cls}, XST-type winds cannot escape from the gravitational potential well.
This is inconsistent with the steady assumption, but XST-type winds have several possibilities.
As shown in Figure \ref{img_cls_cpt}, the terminal point of the XST type can reach as far as the vicinity of the virial point.
In this case, the XST-type wind can practically extend into intergalactic space and the galactic wind can almost escape from the galaxy.
For another possibility, when XST-type winds pass the transonic point, a supersonic outflow can transition to a subsonic inflow due to the shock.
%Such a case corresponds to a galactic fountain \citep{Shapiro76}.
Furthermore, the XST-type solution can be an unsteady flow.
Because our model does not cover these possibilities, we will discuss them in future work.
In addition, XST-type winds have intriguing potential as an observational property.
Figure \ref{img_mst_vmx_esc} shows that the maximum velocity of XST-type winds can be greater than the escape velocity evaluated at the virial radius using equation (\ref{eqn_esc_vel}).
Therefore, it is possible to observe XST-type galactic winds with a maximum velocity higher than the escape velocity, although they cannot escape the gravitational potential.

\begin{figure}
 \begin{center}
  \includegraphics[width=\linewidth]{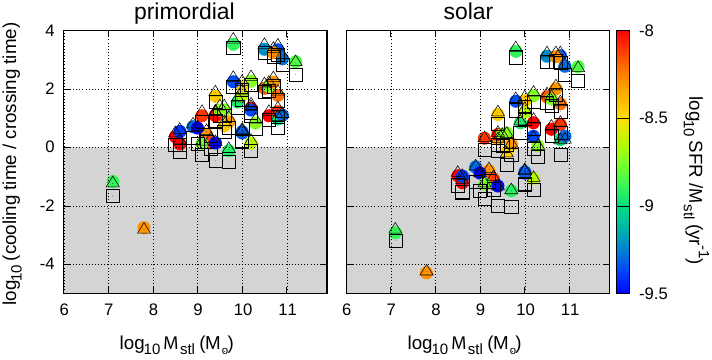}
  \caption{
Comparison of the crossing times to the cooling times.
Dots are used in the same manner as in figure \ref{img_tim_crs}.
The grey region shows the area where the cooling time becomes shorter than the crossing time.
We assume primordial (left) and solar (right) metallicity.}
  \label{img_tim_col}
 \end{center}
\end{figure}

% adiabatic assumption (cooling time) 
Finally, we discuss the availability of the adiabatic assumption.
There are theoretical studies that have examined the effect of radiative cooling in the classical CC85
%Chevalier \& Clegg 
model \citep{Silich04, Tenorio-Tagle07, Lochhaas21} and Parker model \citep{Everett07}.
Because of the complexity of radiative cooling effects, we discuss only the cooling time and do not examine it in detail.
The adiabatic assumption could be available when the crossing time is shorter than the cooling time.
We define the cooling time as
\begin{eqnarray}
t_{\rm cool} \equiv \frac{U(r_{\rm sp})}{\Lambda(r_{\rm sp})},
\label{eqn_tim_col}
\end{eqnarray}
where $U(r)$ and $\Lambda(r)$ are the total thermal energy and the total energy loss, respectively.
We define $U(r)$ and $\Lambda(r)$ as
\begin{eqnarray}
U(r) 
&\equiv \int_0^r 4 \pi r^2 \Delta U(r) n(r) dr, \\
\Lambda(r) 
&\equiv \int_0^r 4 \pi r^2 \Lambda_{\rm N}(T) \{n(r)\}^2 dr,
\end{eqnarray}
where $\Delta U(r)$, $\Lambda_{\rm N}(T)$, and $n(r)$ are the thermal energy of a single particle, cooling function, and number density, respectively.
We define $\Delta U(r)$ and $n(r)$ as
\begin{eqnarray}
\Delta U(r) 
\equiv \frac{3}{2} k_{\rm B} T(r), \\
n(r) 
\equiv \frac{\rho(r)}{(1/2)m_{\rm h}},
\end{eqnarray}
where $k_{\rm B}$ and $m_{\rm h}$ are the Boltzmann constant and proton mass, respectively.
The metallicity varies with $\Lambda_{\rm N}(T)$.
High metallicity increases $\Lambda_{\rm N}(T)$ and shortens the cooling time.
As shown in Section \ref{sec_res_mlr}, large galaxies have small $\lambda$.
Therefore, they can have high metallicity, which derives from the outflow gas enriched with SN products.
On the other hand, because small galaxies have large $\lambda$, the low metallicity derives from the entrainment of low-metallicity interstellar gas.
Indeed, the observed gas-phase metallicity is low in less massive star-forming galaxies and high in more massive galaxies \citep{Andrews13}.
Therefore, we assume both primordial and solar metallicity, adopting $\Lambda_{\rm N}(T)$ from \citet{Sutherland93}.
We note that $\Lambda_{\rm N}(T)$ changes significantly with temperature as well as metallicity.
The right panel of figure \ref{img_tim_crs} shows $t_{\rm cool}$ for various $r_{\rm sp}$, assuming primordial metallicity.
It shows that $t_{\rm cool}$ has a broader range of values than $t_{\rm cross}$.
Independent of $r_{\rm sp}$, $t_{\rm cool}$ correlates positively with stellar mass.
With a larger specific SFR, $t_{\rm cool}$ becomes slightly shorter.
Figure \ref{img_tim_col} compares $t_{\rm cross}$ with $t_{\rm cool}$.
For larger galaxies ($M_{\rm stl}^{\rm tot}$ $\sim$ $10^{10-11}M_{\odot}$), we find that $t_{\rm cross}$ is shorter than $t_{\rm cool}$ in both cases of primordial and solar metallicity.
Therefore, the adiabatic assumption is available for larger galaxies.
For middle-mass galaxies ($M_{\rm stl}^{\rm tot}$ $\sim$ $10^{8-10}M_{\odot}$), $t_{\rm cool}$ is longer than $t_{\rm cross}$ in the case of primordial metallicity and shorter in the case of solar metallicity.
Since $\lambda$ value is larger than unity for these galaxies, the metallicity may decrease.
Therefore, the adiabatic assumption may be available for middle-mass galaxies.
On the other hand, for smaller galaxies ($M_{\rm stl}^{\rm tot}$ $\sim$ 10$^{7-8}M_{\odot}$), $t_{\rm cool}$ is shorter than $t_{\rm cross}$ in both cases of primordial and solar metallicity.
The larger $\lambda$ of these galaxies reduces the metallicity, but it also strongly affects the gas density and temperature.
As a result, large $\lambda$ significantly decreases $t_{\rm cool}$.
Hence, the adiabatic assumption is available, except for small galaxies.

%We define the local cooling time as
%\begin{align}
%\Delta t_{\rm cool} \equiv \frac{\Delta U(r)}{\Lambda_{\rm N}(T)n(r)}, 
%\end{align}
%where $\Delta U(r)$, $\Lambda_{\rm N}(T)$, and $n(r)$ are the thermal energy of a single particle, cooling function, and number density, respectively.

\begin{figure}
 \begin{center}
  \includegraphics[width=0.70\linewidth]{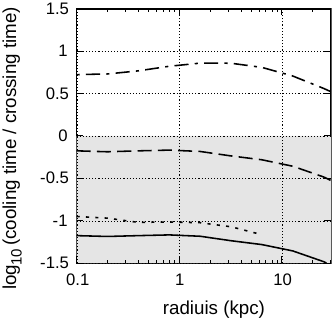}
  \caption{
Ratio of cooling time (in primordial metallicity) and crossing time in a small galaxy.
For example, we use IZw18 ($\log_{10}M_{\rm stl} = $7.1, $\log_{10}\dot{m}_{\rm SFR}=$-1.796, $\lambda$=33.5).
The crossing time and the cooling time are defined by equations (\ref{eqn_tim_crs}) and (\ref{eqn_tim_col}), respectively.
The solid line shows IZw18.
The dotted line represents the case of a large stellar mass (10 times the original).
The parameters of the gravitational potential are revised, corresponding to the large stellar mass.
The dashed line represents the case of a small SFR (one-tenth of the original).
The dot-dashed line represents the case of a small mass loading rate (one-third of the original).}
  \label{img_tim_loc}
 \end{center}
\end{figure}

% cooling in small galaxies 
As discussed above, for smaller galaxies, gas density and temperature (i.e. $\lambda$ and SFR) can affect $t_{\rm cool}$ more than metallicity.
For example, figure \ref{img_tim_loc} shows the ratio of $t_{\rm cross}$ and $t_{\rm cool}$ versus radius.
It shows that $t_{\rm cool}$ is globally shorter than $t_{\rm cross}$ for small galaxies ($M_{\rm stl}^{\rm tot}$ $\sim$ 10$^7M_{\rm \odot}$).
Even for small galaxies, $t_{\rm cool}$ can be longer than $t_{\rm cross}$ for small $\lambda$.
Since $\lambda$ affects gas density and temperature, larger $\lambda$ in smaller galaxies shortens $t_{\rm cool}$.
The amount of SFR also affects $t_{\rm cool}$, but not as much as $\lambda$.
The gravitational potential has little effect on $t_{\rm cool}$.
Therefore, the availability of the adiabatic assumption depends mainly on $\lambda$.
Also, figure \ref{img_tim_loc} shows that the ratio of $t_{\rm cool}$ and $t_{\rm cross}$ decreases outwards.
Several theoretical studies have shown that radiative cooling leads to the formation of cold components in hot gas flow \citep{Thompson16, Schneider18, Zhang18}.
The formation of cold components may affect the evolution of small star-forming galaxies.

%figure \ref{img_tim_loc} also shows that near the centre of small galaxies ($\leq$1 kpc), the cooling time becomes significantly shorter ($\leq$ 1 Myr).
%As discussed in Section \ref{sec_dis_shm}, radiative cooling can form a cool component in hot gas.
%There are several theoretical studies of this formation \citep{Thompson16, Schneider18, Zhang18}.
%These studies explain the formation process of fast-moving cool gas observed outside the star-forming region ($\geq $1 kpc) of star-forming galaxies.
%Although outflows are not yet fully accelerated near the centre ($\leq $1 kpc), the formation of a cold component may also affect the absorption lines of small star-forming galaxies.

\begin{figure}
 \begin{center}
  \includegraphics[width=\linewidth]{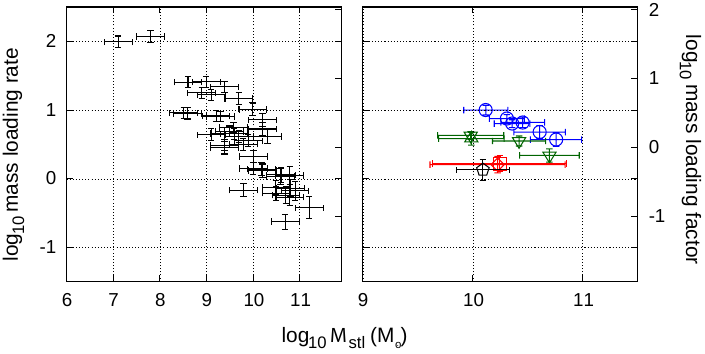}
  \caption{
The mass loading rate is estimated with high efficiency ($\epsilon_{\rm SN}$=0.3). 
The symbols used in the same manner as in figures \ref{img_mfx_loc} and \ref{img_mfx_hiz}. }
  \label{img_mlr_e03}
 \end{center}
\end{figure}

% energy efficiency 
In Section \ref{sec_mdl_prm} we fixed the SNeII energy injection rate to 10\% ($\epsilon_{\rm SN}$ = 0.1).
Since radiative cooling in SNeII is still unclear, here we test the mass flux estimation for the high-efficiency case (30\%, $\epsilon_{\rm SN}$ = 0.3).
The estimation method is the same as in Section \ref{sec_res_mlr}.
The higher efficiency accelerates the galactic wind and increases the mass flux corresponding to the same observed velocity.
Therefore, as shown in figure \ref{img_mlr_e03}, $\lambda$ is several times larger than in the 10\% case.
In Section \ref{sec_res_mlr}, we find that $\lambda$ correlates negatively with the galaxy mass.
This correlation is the same for the high-efficiency case.
Therefore, we find that $\lambda$ correlates negatively with the galaxy mass, independent of efficiency.

\section{conclusion}
% classification 
We constructed a transonic model that extends the Parker theory of the solar wind to investigate the acceleration process of the galactic wind considering a widely spread dark matter gravitational potential, energy/momentum and mass injection along the streamline.
The model adopts spherically symmetric and steady assumptions.
We consider the dark matter halo and stellar mass as the galactic gravitational potential.
The existence of transonic galactic winds depends on the balance between galactic gravity, injected energy from SNe, and the amount of mass flux.
A transonic acceleration process is possible if the injected energy per mass flux is greater than the gravitational potential energy.
As shown in the left panel of figure \ref{img_pat}, this galactic wind starts from the center and accelerates from subsonic to supersonic speeds.
This galactic wind can escape from the galactic gravitational potential.
If the injected energy is sufficiently larger than the gravitational potential energy, the maximum velocity will be the terminal velocity.
If the injection energy is close to the gravitational potential energy, the galactic wind decelerates outside the star-forming region due to the gravitational potential of the dark matter halo.
In this case, the maximum velocity is greater than the terminal velocity.
There is also a transonic acceleration process if the injection energy is slightly less than the gravitational potential energy.
This galactic wind cannot escape due to the gravitational potential, as shown in the central panel in figure \ref{img_pat}.
If the injection energy is sufficiently smaller than the gravitational potential energy, there is no transonic acceleration process, as shown in the right panel of figure \ref{img_pat}.

% mass flux 
Mass fluxes are essential for estimating the impact of galactic winds on galaxy evolution.
We apply the transonic outflow model to the observed outflow velocities to estimate the mass flux of individual star-forming galaxies.
As shown in figure \ref{img_mfx_loc}, we find that the mass flux correlates positively with stellar mass, dark matter halo mass, circular velocity, and SFR.
We then divide the mass flux by the SFR to obtain the mass loading rate (and mass loading factor).
The mass loading rate indicates the degree of the mass loading effect in which the hot gas flow entrains the interstellar medium and increases the mass flux.
The mass loading rate correlates negatively with stellar mass, dark halo mass, circular velocity, and SFR.
For small galaxies (stellar mass $\sim 10^{7-8} M_{\odot}$), the mass loading rate is higher than unity.
This indicates that small galaxies can efficiently eject interstellar medium into intergalactic space.
For large galaxies (stellar mass $\sim 10^{10-11} M_{\odot}$), the mass loading rate is lower than unity.
It indicates that large galaxies have mass fluxes smaller than the injected mass by SNe.
As shown in figure \ref{img_mfx_hiz}, the results for high-z star-forming galaxies show a similar trend.
Therefore, the gravitational well of a massive dark matter halo may effectively increase stellar mass in galaxy evolution.

% comparing to shell model 
The shell model is also often used to estimate mass fluxes, but this model requires ambiguous parameters such as column density.
Column density has uncertainty due to the conversion between metallicity and column density.
Our transonic outflow model does not require ambiguous parameters for mass flux estimation.
Also, the shell model does not explicitly include the gravitational potential.
Figures \ref{img_mlr_shm_loc} and \ref{img_mlr_shm_hiz} compare the results of our transonic outflow model and the shell model.
The mass loading rates of the two models are close for small galaxies but differ significantly for large galaxies.
Our mass flux (or mass loading rate) correlates clearly with stellar mass, dark halo mass, circular velocity, and SFR.
On the other hand, the shell model has no apparent correlation due to the large scatter.
These results suggest that correct estimation of the mass flux requires consideration of the gravitational potential and the outflow profile.

% comparing to theoretical prediction 
We compare our results with the theoretical predictions of previous studies.
These theoretical studies compute the galaxy formation history to reproduce observed relations such as the stellar-to-halo mass ratio, the mass-metallicity relation, and the Tully-Fisher relation.
As shown in Figures \ref{img_mlr_thp_loc} and \ref{img_mlr_thp_hiz}, their predictions roughly reproduce our results, although they do not resolve transonic acceleration processes.
This suggests that their numerical experiment roughly traces the galaxy formation history.

% comparing to CC85 model 
%In addition, we compare the transonic outflow model with the classical model of the galactic wind by \citet{Chevalier85}. Their model manually fixed the transonic point and injected mass and energy into the subsonic region. Their model neglected the gravitational potential. Therefore, our model is a realistic refinement of their model. Figures \ref{img_mlr_ccm_loc} and \ref{img_mlr_ccm_hiz} compare the estimated results of our transonic outflow model and their model. The mass flux of their model increases more than that of our model, especially for large mass loading rates. This result indicates that the gravitational potential is necessary for the mass flux estimation because the locus of the transonic point is affected by the gravitational potential.

\begin{ack}
We thank the anonymous referee for their helpful comments and suggestions, and M. Ouchi and Y. Sugahara for kindly providing the information in Table \ref{tab_prm_hiz}. 
This work was supported in part by JSPS KAKENHI Grant Number JP20K04022, and by the Multidisciplinary Cooperative Research Program in CCS, University of Tsukuba.

\end{ack}

\appendix
\clearpage

\section{The Mach number equation}
\label{app_mac}

Transforming equations (\ref{eqn_mfx}) and (\ref{eqn_efx}) with $\mathcal{M}\equiv v/c_{\rm s}$, we obtain
\begin{eqnarray}
\rho c_{\rm s} 
&=& \frac{\dot{m}}{4\pi \mathcal{M} r^2},
\label{eqn_mfx_app}\\
c_{\rm s}^2 
&=& \left(\frac{\dot{e}}{\dot{m}} - \phi\right) \frac{2(\Gamma-1)}{(\Gamma-1)\mathcal{M}^2+2}.
\label{eqn_efx_app} 
\end{eqnarray}
From equation (\ref{eqn_mot}), we obtain
\begin{eqnarray}
\rho c_{\rm s} \mathcal{M} \frac{d}{dr} (c_{\rm s} \mathcal{M})
&=& -\frac{dP}{dr} 
- \rho \frac{d\phi}{dr} - v \dot{\rho}_{\rm m} 
\quad (\because {\rm equation} (\ref{eqn_gra_pot}) \: ) \nonumber\\
&=& -\frac{d}{dr}\left( \frac{\rho c_{\rm s}^2}{\Gamma} \right) 
- \rho \frac{d\phi}{dr} - v \dot{\rho}_{\rm m} \nonumber\\
&&\qquad\qquad\qquad\qquad\qquad
(\because {\rm equation} (\ref{eqn_sou_spe}) \: ) \nonumber\\
&=& -\frac{d}{dr}\left( \frac{\rho c_{\rm s}^2}{\Gamma} \right) 
- \rho \frac{d\phi}{dr} - \frac{c_{\rm s} \mathcal{M} }{4\pi r^2} \frac{d\dot{m}}{dr} \nonumber\\
&&\qquad\qquad\qquad\qquad\qquad
(\because {\rm equation} (\ref{eqn_flx_mas}) \: ) \nonumber\\
\frac{\mathcal{M}^2}{2} \frac{dc_{\rm s}^2}{dr} + \frac{c_{\rm s}^2}{2} \frac{d\mathcal{M}^2}{dr} 
&=& -\frac{1}{2\Gamma} \frac{dc_{\rm s}^2}{dr} - \frac{c_{\rm s}^2}{\Gamma \rho c_{\rm s}} \frac{d (\rho c_{\rm s})}{dr} \nonumber\\
&&\quad - \frac{d\phi}{dr} - \frac{c_{\rm s}^2}{\rho c_{\rm s}} \frac{\mathcal{M}}{4 \pi r^2} \frac{d\dot{m}}{dr}.
\label{eqn_mot_app} 
\end{eqnarray}
Substituting equations (\ref{eqn_mfx_app}) and (\ref{eqn_efx_app}) into equation (\ref{eqn_mot_app}) and eliminating $\rho c_{\rm s}$ and $c_{\rm s}^2$, we obtain equation (\ref{eqn_mac}).
First, we determine the transonic point from $N(r)=0$.
Next, integrating equation(\ref{eqn_mac}) using the fact that $\mathcal{M}=1$ at the transonic point, we obtain the distribution of $\mathcal{M}$ with respect to $r$.
It is careful to note that the transonic solution is a double solution of the inflow and outflow.
We obtain other physical quantities from the $\mathcal{M}$ distribution.
For integration, we use an accurate method such as the Runge-Kutta method.

\clearpage

\section{The non-dimensional function} 
\label{app_nfc}

In equations (\ref{eqn_mas_dmh}) and (\ref{eqn_mas_stl}), the following normalized expressions are used at $r > r_{\rm vir}$: 
\begin{eqnarray}
\mathcal{F}_{\rm dmh}(r; r_{\rm dmh}, r_{\rm vir}) 
= \mathcal{F}_{\rm stl}(r; r_{\rm stl}, r_{\rm vir}) 
= 1,
\end{eqnarray}
where $r_{\rm stl}$, $r_{\rm dmh}$, and $r_{\rm vir}$ represent the effective radius, scale radius of the dark halo, and virial radius, respectively. 
In $r < r_{\rm vir}$, both functions become 
\begin{eqnarray}
&&\mathcal{F}_{\rm dmh}(r; r_{\rm dmh}, r_{\rm vir}) \nonumber\\
&&\quad \equiv \left\{ \int_0^r 4 \pi r^2 \rho_{\rm dmh} (r) dr \right\}
\left\{ \int_0^{r_{\rm vir}} 4 \pi r^2 \rho_{\rm dmh} (r) dr \right\}^{-1} \nonumber\\
&&\quad = \left\{ \int_0^r \frac{rdr}{(r+r_{\rm dmh})^2} \right\}
\left\{ \int_0^{r_{\rm vir}} \frac{rdr}{(r+r_{\rm dmh})^2} \right\}^{-1} \nonumber\\
&&\quad = \left\{ \log \left( \frac{r+r_{\rm dmh}}{r_{\rm dmh}} \right) - \frac{r}{r+r_{\rm dmh}} \right\} \nonumber\\
&&\quad \qquad \times 
\left\{ \log \left( \frac{r_{\rm vir}+r_{\rm dmh}}{r_{\rm dmh}} \right) - \frac{r_{\rm vir}}{r_{\rm vir}+r_{\rm dmh}} \right\}^{-1} \nonumber\\
&&\quad = \left\{ \log \left( x+1 \right) - \frac{x}{x+1} \right\} 
\left\{ \log \left( c+1 \right) - \frac{c}{c+1} \right\}^{-1}, 
\end{eqnarray}
\begin{eqnarray}
&&\mathcal{F}_{\rm stl}(r; r_{\rm stl}, r_{\rm vir}) \nonumber\\
&&\quad \equiv \left\{ \int_0^r 4 \pi r^2 \rho_{\rm stl} (r) dr \right\}
\left\{ \int_0^{r_{\rm vir}} 4 \pi r^2 \rho_{\rm stl} (r) dr \right\}^{-1} \nonumber\\
&&\quad = \left\{ \int_0^r \frac{2 r_{\rm stl} r dr}{(r + r_{\rm stl})^3} \right\}
\left\{ \int_0^{r_{\rm vir}} \frac{2 r_{\rm stl} r dr}{(r  +r_{\rm stl})^3}  \right\}^{-1} \nonumber\\
&&\quad = \left\{ \frac{r^2}{(r + r_{\rm stl})^2} \right\}
\left\{\frac{r_{\rm vir}^2}{(r_{\rm vir} + r_{\rm stl})^2} \right\}^{-1} \nonumber\\
&&\quad = \left\{ \frac{x^2}{(x + x_{\rm stl})^2} \right\}
\left\{\frac{c^2}{(c + x_{\rm stl})^2} \right\}^{-1}, 
\end{eqnarray}
In equations (\ref{eqn_pot_dmh}) and (\ref{eqn_pot_stl}), the following normalized expressions in $r>r_{\rm vir}$ are estimated as 
\begin{eqnarray}
\mathcal{F}_{\rm \phi dmh}(r;r_{\rm dmh},r_{\rm vir}) 
= \mathcal{F}_{\rm \phi stl}(r;r_{\rm stl},r_{\rm vir}) 
= \frac{r_{\rm vir}}{r} 
\label{eqn_pot_inn} 
\end{eqnarray}
where the integration constants are determined in order with $\phi(\infty)=0$. 
In $r < r_{\rm vir}$, both functions become 
\begin{eqnarray}
&& \mathcal{F}_{\rm \phi dmh}(r; r_{\rm dmh}, r_{\rm vir}) \nonumber\\
&&= \frac{r_{\rm vir}}{r} 
\left\{ \log\left( \frac{r+r_{\rm dmh}}{r_{\rm dmh}} \right) - \frac{r}{r_{\rm vir}+r_{\rm dmh}} \right\} \nonumber\\
&&\qquad \times \left\{ \log \left( \frac{r_{\rm vir}+r_{\rm dmh}}{r_{\rm dmh}} \right) - \frac{r_{\rm vir}}{r_{\rm vir}+r_{\rm dmh}} \right\}^{-1} \nonumber\\
&&= \frac{c}{x} 
\left\{ \log\left( x+1 \right) - \frac{x}{c+1} \right\} \nonumber\\
&&\qquad \times \left\{ \log \left( c+1 \right) - \frac{c}{c+1} \right\}^{-1}, 
\label{eqn_pot_dmh_inn} 
\end{eqnarray}
\begin{eqnarray}
&& \mathcal{F}_{\rm \phi stl}(r; r_{\rm stl}, r_{\rm vir}) \nonumber\\
&&= \frac{r_{\rm vir}}{r} 
\left\{ \frac{r}{r+r_{\rm stl}}-\frac{r_{\rm stl} r}{(r_{\rm vir}+r_{\rm stl})^2}\right\} 
\left\{ \frac{r_{\rm vir}^2}{(r_{\rm vir} + r_{\rm stl})^2} \right\}^{-1} \nonumber\\
&&= \frac{c}{x} 
\left\{ \frac{x}{x+x_{\rm stl}}-\frac{x_{\rm stl} x}{(c+x_{\rm stl})^2}\right\} 
\left\{ \frac{c^2}{(c + x_{\rm stl})^2} \right\}^{-1}, 
\label{eqn_pot_stl_inn}
\end{eqnarray}
where the integration constants are determined to be consistent with equation (\ref{eqn_pot_inn}) at $r=r_{\rm vir}$. 
In equations (\ref{eqn_pow_dmh_inn}) and (\ref{eqn_pow_stl_inn}), the following normalized expressions are used in $r < r_{\rm vir}$: 
\begin{eqnarray}
&& \mathcal{F}_{\rm w dmh}(r; r_{\rm stl}, r_{\rm dmh}, r_{\rm vir}) \nonumber\\
&&\equiv \left\{ \frac{r_{\rm vir}^2}{(r_{\rm vir} + r_{\rm stl})^2} \right\}^{-1}
\int_0^r \frac{2 r_{\rm stl} r}{(r + r_{\rm stl})^3} \mathcal{F}_{\rm \phi dmh}(r;r_{\rm dmh},r_{\rm vir}) dr \nonumber
\end{eqnarray}
\begin{eqnarray}
&&= \left\{ \frac{r_{\rm vir}^2}{(r_{\rm vir} + r_{\rm stl})^2} \right\}^{-1}
\left\{ \log \left( \frac{r_{\rm vir}+r_{\rm dmh}}{r_{\rm dmh}} \right) - \frac{r_{\rm vir}}{r_{\rm vir}+r_{\rm dmh}} \right\}^{-1} \nonumber\\
&&\qquad \times \int_0^r \frac{2 r_{\rm stl} r}{(r + r_{\rm stl})^3} \frac{r_{\rm vir}}{r} \left\{ \log\left(\frac{r + r_{\rm dmh}}{r_{\rm dmh}}\right) - \frac{r}{r_{\rm vir} + r_{\rm dmh}} \right\} dr \nonumber
\end{eqnarray}
\begin{eqnarray}
&&= \left\{ \frac{r_{\rm vir}^2}{(r_{\rm vir} + r_{\rm stl})^2} \right\}^{-1}
\left\{ \log \left( \frac{r_{\rm vir}+r_{\rm dmh}}{r_{\rm dmh}} \right) - \frac{r_{\rm vir}}{r_{\rm vir}+r_{\rm dmh}} \right\}^{-1} \nonumber\\
&&\qquad \times \left\{ r_{\rm stl} r_{\rm vir} \frac{(r + r_{\rm stl})^2 - (r_{\rm dmh} -r_{\rm stl})^2}{(r_{\rm dmh} -r_{\rm stl})^2 (r + r_{\rm stl})^2}  \log\left( \frac{r+r_{\rm dmh}}{r_{\rm dmh}} \right) \right. \nonumber\\
&&\qquad \qquad - \frac{r_{\rm stl}r_{\rm vir}}{(r_{\rm dmh} - r_{\rm stl})^2} \log\left( \frac{r+r_{\rm stl}}{r_{\rm stl}} \right) \nonumber\\
&&\qquad \qquad \left. + r_{\rm vir} r \frac{ (r_{\rm vir}+r_{\rm dmh})(r+r_{\rm stl}) - (r_{\rm dmh} - r_{\rm stl}) r }{(r_{\rm dmh} - r_{\rm stl})(r_{\rm vir}+r_{\rm dmh})(r + r_{\rm stl})^2} \right\} \nonumber
\end{eqnarray}
\begin{eqnarray}
&&= \left\{ \frac{c^2}{(c + x_{\rm stl})^2} \right\}^{-1}
\left\{ \log \left( c+1 \right) - \frac{c}{c+1} \right\}^{-1} \nonumber\\
&&\qquad \times \left\{ c x_{\rm stl} \frac{(x + x_{\rm stl})^2 - (1 -x_{\rm stl})^2}{(1 -x_{\rm stl})^2 (x + x_{\rm stl})^2}  \log\left( x+1 \right) \right. \nonumber\\
&&\qquad \qquad - \frac{c x_{\rm stl}}{(1 - x_{\rm stl})^2} \log\left( \frac{x+x_{\rm stl}}{x_{\rm stl}} \right) \nonumber\\
&&\qquad \qquad \left. + c x \frac{ (c+1)(x+x_{\rm stl}) - (1 - x_{\rm stl}) x }{(1 - x_{\rm stl})(c+1)(x + x_{\rm stl})^2} \right\}, 
\end{eqnarray}
\begin{eqnarray}
&&\mathcal{F}_{\rm w stl}(r; r_{\rm stl}, r_{\rm vir}) \nonumber\\
&&\equiv \left\{ \frac{r_{\rm vir}^2}{(r_{\rm vir} + r_{\rm stl})^2} \right\}^{-1}
\int_0^r \frac{2 r_{\rm stl} r}{(r + r_{\rm stl})^3} \mathcal{F}_{\rm \phi stl}(r;r_{\rm stl},r_{\rm vir}) dr \nonumber
\end{eqnarray}
\begin{eqnarray}
&&= \left\{ \frac{r_{\rm vir}^2}{(r_{\rm vir} + r_{\rm stl})^2} \right\}^{-2} \nonumber\\
&&\qquad \times\int_0^r \frac{2 r_{\rm stl} r}{(r + r_{\rm stl})^3} \frac{r_{\rm vir}}{r} \left\{ \frac{r}{r + r_{\rm stl}} - \frac{r_{\rm stl} r}{(r_{\rm vir} + r_{\rm stl})^2} \right\} dr \nonumber\\
&&= \left\{ \frac{r_{\rm vir}^2}{(r_{\rm vir} + r_{\rm stl})^2} \right\}^{-2} \nonumber\\
&&\qquad \times r_{\rm vir} r^2 \frac{(r_{\rm vir} + r_{\rm stl})^2 (r + 3 r_{\rm stl}) - 3r_{\rm stl}^2 (r+r_{\rm stl})}{3 r_{\rm stl} (r_{\rm vir} + r_{\rm stl})^2 (r+r_{\rm stl})^3} \nonumber\\ 
&&= \left\{ \frac{c^2}{(c + x_{\rm stl})^2} \right\}^{-2} \nonumber\\
&&\qquad \times c x^2 \frac{(c + x_{\rm stl})^2 (x + 3 x_{\rm stl}) - 3x_{\rm stl}^2 (x+x_{\rm stl})}{3 x_{\rm stl} (c + x_{\rm stl})^2 (x+x_{\rm stl})^3} . 
\end{eqnarray}
In $r>r_{\rm vir}$, both functions become 
\begin{eqnarray}
\mathcal{F}_{\rm w dmh}(r; r_{\rm stl}, r_{\rm dmh}, r_{\rm vir}) 
&&= \mathcal{F}_{\rm w dmh}(r_{\rm vir}; r_{\rm stl}, r_{\rm dmh}, r_{\rm vir}), \\
\mathcal{F}_{\rm w stl}(r; r_{\rm stl}, r_{\rm vir}) 
&&= \mathcal{F}_{\rm w stl}(r_{\rm vir}; r_{\rm stl}, r_{\rm vir}). 
\end{eqnarray}

\clearpage

\section{Comparison to Chevalier \& Clegg (1985) model} 
\label{app_ccm}

% CC85 model & M82 
CC85 used a simple outflow model (the CC85 model) to study the galactic wind observed in the local star-forming galaxy M82.
The CC85 model ignores the gravitational potential of the galaxy.
Also, they manually fix the transonic point and assume injections of mass and energy inside the transonic point.
Therefore, the CC85 model does not rigorously examine the transonic acceleration process.
Our transonic outflow model includes the distribution of gravitational potential and injections.
Therefore, our model corresponds to a refinement of the CC85 model.
In this section, we apply both models to the observations of M82 and compare the results to determine the effects of gravitational potential and injections.
\citet{Strickland97} found that the CC85 model cannot reproduce the emission observed in M82 due to the sharp decrease in the density profile of CC85.
They concluded that shock-heated clouds comprise the majority of the emission \citep{Strickland00}. 
In contrast to the CC85 model, our model has a slowly decreasing density distribution.
Therefore, the outflowing gas could affect the emission as much as the heated clouds.

\begin{figure}
 \begin{center}
  \includegraphics[width=0.99\columnwidth]{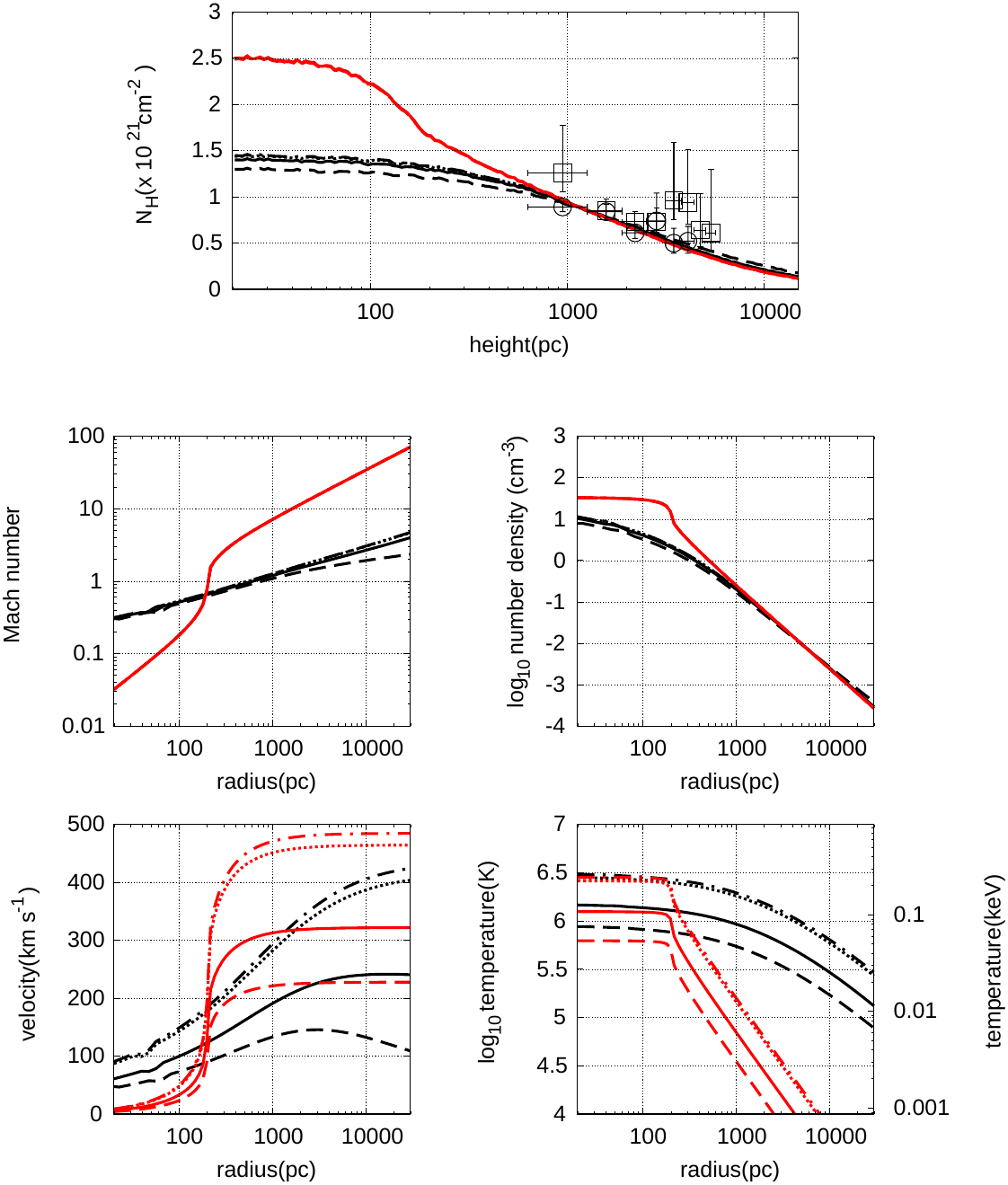}
  \caption{
The top panel represents the fitting results of the observed column density distribution of M82 with the given SFR and the residual rate $\epsilon_{\rm rem}$ of SN energy after radiative cooling. 
Black lines represent the transonic outflow model results: solid, (SFR,$\epsilon_{rem}$)=(6.8,0.1); dotted, (6.8,0.3); dashed, (2.4,0.1); and dotted-dashed, (23.1,0.1). 
Red lines represent the results of the CC85 model; solid: (SFR,$\epsilon_{rem}$)=(6.8,0.1), dotted: (6.8,0.3), dashed: (2.4,0.1) and dotted-dashed: (23.1,0.1).
The open squares and open circles show the north and south sides of M82 \citep{Strickland97}, respectively. 
The detailed results are summarized in Table \ref{tab_m82_mfx}. 
The other four panels show the estimated Mach number, velocity, number density, and temperature. 
  }
  \label{img_m82}
 \end{center}
\end{figure}

% column density fitting 
M82 is a well-studied star-forming galaxy in the local universe.
X-ray observations have revealed a column density distribution of hot gas along the minor axis \citep{Strickland97}.
We fit two outflow models to this profile to predict the mass flux (and $\lambda$).
For this fitting, our model requires stellar and dark halo mass distributions.
We can estimate the stellar mass distribution of M82 from infrared observations \citep{Jarrett03, Dale07, Bell03} and the dark halo mass distribution from fitting the observed rotation curves \citep{Sofue98, Greco12}.
Table \ref{tab_m82_prm} summarizes the parameters used for M82.
Due to the large uncertainty of the SFR \citep{Schober17}, we investigate the following three cases: average SFR (6.8 
 $M_{\odot}$ yr$^{-1}$), low SFR (2.4 $M_{\odot}$ yr$^{-1}$), high SFR (23.1 $M_{\odot}$ yr$^{-1}$).
In addition, \citet{Strickland09} examined the X-ray properties of M82 and found the possibility of a high thermalization efficiency ($\epsilon_{\rm rem}>$0.1).
Therefore, we investigate two cases with low and high thermalization efficiency ($\epsilon_{\rm rem}=$0.1 and 0.3).
Since the observed column densities were extracted from a series of rectangular regions of 0.1$^\circ$ width and 0.01$^\circ$ height along the northern and southern minor axes \citep{Strickland97}, we calculate the column densities of these regions for fitting.
Figure \ref{img_m82} shows the estimated column density of rectangular regions of 0.1$^\circ$ width with lines. 
Table \ref{tab_m82_mfx} summarizes the detailed results of the fitting. 

% result 
Figure \ref{img_m82} shows that both models approximately reproduce the column density in the observed region ($\sim$1-5kpc).
The mass fluxes are different in each case.
The CC85 model assumes concentrated injections of mass and energy inside the transonic point (200pc), whereas our model considers the distribution of the injections.
Therefore, inside the observed region ($\sim$1-5kpc), the CC85 model shows a large column density compared to our model.
Both models show almost the same column density outside of the observed region.
This may be due to reduced injection and adiabatic expansion of the supersonic flow.
The middle and bottom rows of figure \ref{img_m82} show the estimated Mach number, velocity, number density, and temperature.
Because the CC85 model manually fixes the transonic point at 200 pc, these quantities change abruptly around 200 pc.
In particular, the temperature decreases rapidly outside of 200 pc because the outflow adiabatically expands outside of the fixed transonic point in the CC85 model.
As a result, the Mach number is much higher than our model.
On the other hand, the transonic outflow model passes smoothly through the correctly estimated transonic point, resulting in a slow change in velocity, number density, and temperature.
Therefore, our model has a large density and high temperature, which may contribute to the observed emission.
This result is different from the CC85 model.

% velocity 
As shown in figure \ref{img_m82}, the predicted velocities of both models reach up to $\sim$400-500 km s$^{-1}$ with large SFR or thermalization efficiency.
In observational studies, the CO emission line determines the velocity of molecular gas \citep{Walter02, Leroy15}.
After correcting for the opening angle of the flow, the observed maximum velocity is $\sim$230 km s$^{-1}$.
The H$\alpha$ line determines the velocity of the ionized gas \citep{Shopbell98, Yoshida11}.
This velocity is approximately $\sim$600 km s$^{-1}$.
These observational results are roughly consistent with our results.

\begin{figure*}
 \begin{center}
  \includegraphics[width=\linewidth]{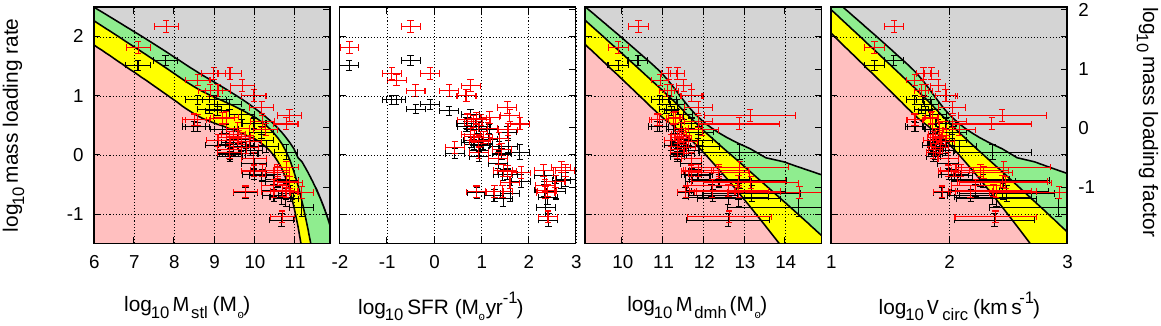}
  \caption{
Estimated mass loading ratio by the transonic outflow model and CC85 model in local galaxies.
The black and red crosses represent our model and the CC85 model. 
The background colors correspond to the solution types in the same manner as in figure \ref{img_mst_vmx_mlr}. 
  }
  \label{img_mlr_ccm_loc}
 \end{center}
\end{figure*}

\begin{figure*}
 \begin{center}
  \includegraphics[width=\linewidth]{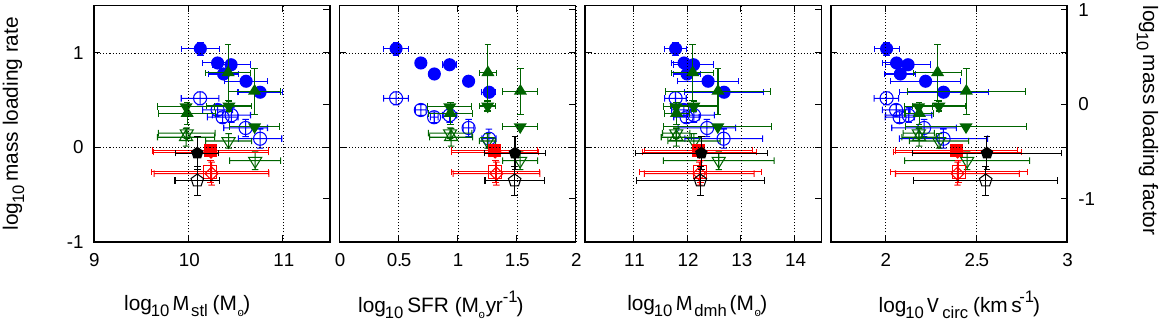}
  \caption{
Estimated mass loading ratio of the transonic outflow model and CC85 model in high-redshift galaxies. 
The open and filled symbols represent the results of the transonic outflow and CC85 models, respectively. 
  }
  \label{img_mlr_ccm_hiz}
 \end{center}
\end{figure*}

% CC8 massflux 
Next, we compare the mass flux estimate from the CC85 model with that from the transonic outflow model described in Section \ref{sec_res_mfx}.
Assuming that the terminal velocity of the CC85 model corresponds to the observed maximum velocity, the estimated mass flux correlates positively with stellar mass, halo mass (circular velocity), and SFR.
The estimated $\lambda$ correlates negatively with these quantities.
These trends are the same as the transonic flow model, but the CC85 model shows a larger scatter.
Despite the similar halo mass, the differences are several times larger when $\lambda$ is large.
In our model, large $\lambda$ strongly causes the gravitational potential to decelerate the acceleration process.
On the other hand, there is no gravitational potential in the CC85 model.
As a result, the CC85 model overestimates $\lambda$ more than our model.
In figure \ref{img_mlr_ccm_loc}, these galaxies with larger $\lambda$ distribute in the green and yellow regions defined in Section \ref{sec_res_mfx}.
Some of the galaxies are in the forbidden region (grey).
When using our model, they do not distribute in the forbidden region.
This indicates that the gravitational potential is essential to correctly estimate the mass flux (and $\lambda$) in the case of galaxies with large $\lambda$.
In other words, the CC85 model is not a good approximation for large $\lambda$.
On the other hand, the two models have almost identical values for the same halo mass range and smaller $\lambda$.
These galaxies distribute in the pink region defined in Section \ref{sec_res_mfx}.
The CC85 model is a good approximation for small $\lambda$ since the gravitational potential is less affected when $\lambda$ is small.
As shown in figure \ref{img_mlr_ccm_hiz}, the $\lambda$ obtained for high-redshift star-forming galaxies also show a similar trend to those of local star-forming galaxies.


\clearpage

\section{Tables}
\label{app_tab}

%--- 1 ---
\begin{longtable}{lcccc} 
\caption{
Parameters of galaxies. 
(1) Redshift cited from NED (that of J142856+165339 is cited from SIMBAD). 
(2) Stellar mass, (3) SFR, and (4) maximum velocity of local star-forming galaxies cited from \citet{Heckman15,Heckman16}. 
The errors of stellar mass, SFR, and maximum velocity are 0.3, 0.2, and 0.05 in the logarithmic scale, respectively.}
\label{tab_prm_loc}

\multicolumn{5}{c}{} \\
      \ & (1) & (2) & (3) & (4) \\
      name & z   & $\log_{10}$ stellar mass & $\log_{10}$ SFR  & $\log_{10}$ maximum velocity \\ 
      \ & \ & ($M_{\odot}$) & ($M_{\odot}$ yr$^{-1}$) & (km s$^{-1}$) \\
\hline
\endfirsthead
      \ & (1) & (2) & (3) & (4) \\
      name & z   & $\log_{10}$ stellar mass & $\log_{10}$ SFR  & $\log_{10}$ maximum velocity \\ 
      \ & \ & ($M_{\odot}$) & ($M_{\odot}$ yr$^{-1}$) & (km s$^{-1}$) \\
\hline
\endhead
\hline
\endfoot
\endlastfoot
J002101+005248 & 0.098300$\pm$0.0003000 & 9.3 & 1.176 & 2.544 \\ 
J005527-002148 & 0.167600$\pm$0.0001000 & 9.7 & 1.380 & 2.724 \\ 
J015028+130858 & 0.147000$\pm$0.0002100 & 10.3 & 1.568 & 2.653 \\ 
J021348+125951 & 0.219200$\pm$0.0000680 & 10.5 & 1.279 & 3.176 \\ 
J080844+394852 & 0.091080$\pm$0.0009670 & 9.8 & 0.903 & 3.176 \\ 
J082354+280621 & 0.046780$\pm$0.0001200 & 8.6 & 1.000 & 2.568 \\ 
J092159+450912 & 0.234800$\pm$0.0001750 & 10.8 & 1.462 & 3.176 \\ 
J092600+442736 & 0.180700$\pm$0.0007250 & 9.1 & 1.000 & 2.740 \\ 
J093813+542825 & 0.102100$\pm$0.0005490 & 9.4 & 1.041 & 2.716 \\ 
J102548+362258 & 0.126500$\pm$0.0007310 & 9.2 & 0.903 & 2.556 \\ 
J111244+550347 & 0.131500$\pm$0.0008290 & 10.2 & 1.462 & 2.996 \\ 
J111323+293039 & 0.170500$\pm$0.0001710 & 9.6 & 0.845 & 2.708 \\ 
J114422+401221 & 0.127000$\pm$0.0000420 & 9.9 & 0.954 & 2.756 \\ 
J141454+054047 & 0.081900$\pm$0.0007510 & 8.5 & 0.699 & 2.568 \\ 
J141612+122340 & 0.123100$\pm$0.0008000 & 10.0 & 1.362 & 2.892 \\ 
J142856+165339 & 0.181670$\pm$0.0000100 & 9.6 & 1.146 & 2.643 \\ 
J142947+064334 & 0.173600$\pm$0.0007760 & 9.4 & 1.431 & 2.820 \\ 
J152141+075921 & 0.094270$\pm$0.0007310 & 9.5 & 0.778 & 2.690 \\ 
J152521+075720 & 0.075670$\pm$0.0008060 & 9.4 & 0.954 & 2.845 \\ 
J161245+081701 & 0.149100$\pm$0.0009390 & 10.0 & 1.556 & 3.000 \\ 
J210358-072802 & 0.136900$\pm$0.0001500 & 10.9 & 1.613 & 3.100 \\ 
Haro11      & 0.020600$\pm$0.0000330 & 10.2 & 1.556  & 2.462 \\ 
VV114       & 0.020070$\pm$0.0000770 & 10.8 & 1.820  & 2.602 \\ 
NGC1140     & 0.005007$\pm$0.0000040 & 9.4  & -0.081 & 2.176 \\ 
SBS0335-052 & 0.013520$\pm$0.0000070 & 7.8  & -0.495 & 1.778 \\ 
Tol0440-381 & 0.040860$\pm$0.0001500 & 10.0 & 0.699  & 2.362 \\ 
NGC1705  & 0.002112$\pm$0.0000190 & 8.6  & -0.796 & 2.230 \\ 
NGC1741  & 0.013470$\pm$0.0000180 & 9.7  & 0.778  & 2.279 \\ 
IZw18    & 0.002505$\pm$0.0000070 & 7.1  & -1.796 & 1.954 \\ 
NGC3310  & 0.003312$\pm$0.0000090 & 9.8  & 0.447  & 2.799 \\ 
Haro3    & 0.003149$\pm$0.0000170 & 8.9  & -0.387 & 2.322 \\ 
NGC3690  & 0.010410$\pm$0.0000100 & 10.9 & 1.602  & 2.531 \\ 
NGC4214  & 0.000970$\pm$0.0000100 & 9.0  & -0.886 & 2.176 \\ 
IRAS19245-4140 & 0.009453$\pm$0.0000180 & 9.1 & 0.322 & 2.322 \\ 
NGC7673  & 0.011370$\pm$0.0000030 & 10.0 & 0.681 & 2.362 \\ 
NGC7714  & 0.009333$\pm$0.0000030 & 10.2 & 0.839 & 2.580 \\ 
J0826+43 & 0.603000$\pm$0.0003400 & 10.8 & 2.580 & 3.090 \\ 
J0905+57 & 0.712100$\pm$0.0001450 & 10.7 & 2.415 & 3.389 \\ 
J0944+09 & 0.514100$\pm$0.0001800 & 10.5 & 2.342 & 3.124 \\ 
J1104+59 & 0.573200$\pm$0.0002040 & 10.6 & 1.845 & 3.017 \\ 
J1506+54 & 0.607900$\pm$0.0002360 & 10.7 & 2.398 & 3.170 \\ 
J1506+61 & 0.436800$\pm$0.0001490 & 10.2 & 2.322 & 3.000 \\ 
J1558+39 & 0.402200$\pm$0.0001240 & 10.6 & 2.785 & 3.000 \\ 
J1613+28 & 0.449400$\pm$0.0003460 & 11.2 & 2.362 & 3.182 \\ 
J1713+28 & 0.577000$\pm$0.0015290 & 10.8 & 2.699 & 2.968 \\ \hline
\end{longtable}

%--- 4 ---
\begin{table*}
\caption{
Parameters of high-redshift galaxies cited from \citet{Sugahara17}.}
  \begin{center}
    \begin{tabular}{lcccc} \hline
      \    & (1) & (2)                      & (3)              & (4)              \\
      line & redshift   & $\log_{10}$ stellar mass & $\log_{10} $SFR  & maximum velocity \\ 
      \    & \   & ($M_{\odot}$)            & ($M_{\odot}$ yr$^{-1}$) & (km s$^{-1}$)           \\ \hline
NaID & 0.0654$\pm$0.01126 & 10.120$\pm$0.2011 & 0.479$\pm$0.1056 & 221$\pm$18 \\ 
NaID & 0.0763$\pm$0.01991 & 10.310$\pm$0.1609 & 0.689$\pm$0.0452 & 261$\pm$12 \\ 
NaID & 0.0874$\pm$0.01770 & 10.370$\pm$0.1609 & 0.803$\pm$0.0332 & 299$\pm$11 \\ 
NaID & 0.1065$\pm$0.02534 & 10.460$\pm$0.2011 & 0.932$\pm$0.0613 & 267$\pm$11 \\ 
NaID & 0.1269$\pm$0.02896 & 10.610$\pm$0.2313 & 1.092$\pm$0.0452 & 327$\pm$10 \\ 
NaID & 0.1379$\pm$0.03349 & 10.760$\pm$0.2313 & 1.264$\pm$0.0593 & 373$\pm$20 \\ 
MgI & 1.3720$\pm$0.04123 & 9.990$\pm$0.3017 & 0.941$\pm$0.1830 & 486$\pm$64 \\ 
MgI & 1.3810$\pm$0.04022 & 10.430$\pm$0.2413 & 1.255$\pm$0.0684 & 309$\pm$91 \\ 
MgI & 1.3850$\pm$0.03932 & 10.700$\pm$0.2715 & 1.530$\pm$0.1498 & 382$\pm$98 \\ 
MgII & 1.3720$\pm$0.04123 & 9.990$\pm$0.3017 & 0.941$\pm$0.1830 & 445$\pm$9 \\ 
MgII & 1.3810$\pm$0.04022 & 10.430$\pm$0.2413 & 1.255$\pm$0.0684 & 442$\pm$27 \\ 
MgII & 1.3850$\pm$0.03932 & 10.700$\pm$0.2715 & 1.530$\pm$0.1498 & 569$\pm$12 \\ 
CII & 2.2690$\pm$0.17620 & 10.240$\pm$0.6134 & 1.323$\pm$0.3690 & 759$\pm$20 \\ 
CIV & 2.2690$\pm$0.17620 & 10.240$\pm$0.6134 & 1.323$\pm$0.3690 & 776$\pm$16 \\ 
SiIIandCII & 5.4890$\pm$0.17120 & 10.09$\pm$0.23 & 1.48$\pm$0.25 & 801$\pm$150 \\ \hline
    \end{tabular}
  \end{center}
\label{tab_prm_hiz}
\end{table*}

%--- 5 ---
\begin{longtable}{lcccc} 
\caption{Estimated mass flux of local galaxies in Section \ref{sec_res_mfx}. 
To estimate mass fluxes, the parameters in Tables \ref{tab_prm_loc} and \ref{tab_est_loc} are applied to the transonic outflow model as mentioned in Section \ref{sec_mdl_prm}. 
(1), (2), and (3) show the estimated mass flux, mass loading rate, (mass loading factor) and transonic point, respectively. 
Additionally, (4) represents the estimated mass flux using \citet{Chevalier85} model (CC85). }
\label{tab_mfx_loc}
\multicolumn{5}{c}{} \\
      \ & (1)  & (2) & (3) & (4) \\
      name & mass flux & mass loading rate (mass loading factor) & transonic point & mass flux (CC85) \\ 
      \ & ($M_{\odot}$ yr$^{-1}$) & \ & (kpc) & ($M_{\odot}$ yr$^{-1}$) \\
\hline
\endfirsthead
      \ & (1)  & (2) & (3) & (4) \\
      name & mass flux & mass loading rate (mass loading factor) & transonic point & mass flux (CC85) \\ 
      \ & ($M_{\odot}$ yr$^{-1}$) & \ & (kpc) & ($M_{\odot}$ yr$^{-1}$) \\
\hline
\endhead
\hline
\endfoot
\endlastfoot
J002101+005248 & 15.6$\pm$8.34 & 2.8$\pm$0.50(1.0$\pm$0.17) & 2.0$\pm$0.47 & 26.1$\pm$14.00 \\
J005527-002148 & 13.0$\pm$6.65 & 1.4$\pm$0.27(0.5$\pm$0.09) & 2.3$\pm$0.60 & 18.1$\pm$10.09 \\
J015028+130858 & 20.0$\pm$10.70 & 1.4$\pm$0.32(0.5$\pm$0.11) & 4.1$\pm$1.18 & 39.0$\pm$21.56 \\
J021348+125951 & 1.5$\pm$0.83 & 0.2$\pm$0.05(0.1$\pm$0.02) & 3.6$\pm$0.85 & 1.8$\pm$0.98 \\
J080844+394852 & 0.7$\pm$0.39 & 0.2$\pm$0.05(0.1$\pm$0.02) & 2.4$\pm$0.55 & 0.8$\pm$0.44 \\
J082354+280621 & 11.5$\pm$5.96 & 3.0$\pm$0.58(1.1$\pm$0.20) & 1.3$\pm$0.27 & 15.5$\pm$8.52 \\
J092159+450912 & 2.1$\pm$1.18 & 0.2$\pm$0.05(0.1$\pm$0.02) & 4.4$\pm$1.05 & 2.7$\pm$1.46 \\
J092600+442736 & 5.6$\pm$3.02 & 1.5$\pm$0.29(0.5$\pm$0.10) & 1.5$\pm$0.32 & 7.1$\pm$3.87 \\
J093813+542825 & 6.4$\pm$3.41 & 1.5$\pm$0.28(0.5$\pm$0.10) & 2.0$\pm$0.47 & 8.8$\pm$4.77 \\
J102548+362258 & 8.5$\pm$4.59 & 2.8$\pm$0.49(1.0$\pm$0.17) & 1.8$\pm$0.40 & 13.3$\pm$7.60 \\
J111244+550347 & 5.2$\pm$2.85 & 0.5$\pm$0.10(0.2$\pm$0.04) & 3.3$\pm$0.77 & 6.3$\pm$3.50 \\
J111323+293039 & 4.1$\pm$2.12 & 1.5$\pm$0.28(0.5$\pm$0.10) & 2.2$\pm$0.56 & 5.8$\pm$3.09 \\
J114422+401221 & 4.1$\pm$2.18 & 1.2$\pm$0.23(0.4$\pm$0.08) & 2.8$\pm$0.72 & 5.9$\pm$3.19 \\
J141454+054047 & 5.8$\pm$3.08 & 3.1$\pm$0.60(1.1$\pm$0.21) & 1.2$\pm$0.25 & 7.9$\pm$4.32 \\
J141612+122340 & 6.4$\pm$3.47 & 0.7$\pm$0.15(0.3$\pm$0.05) & 2.9$\pm$0.74 & 7.9$\pm$4.40 \\
J142856+165339 & 10.2$\pm$5.16 & 1.9$\pm$0.34(0.7$\pm$0.12) & 2.2$\pm$0.57 & 15.4$\pm$8.56 \\
J142947+064334 & 10.9$\pm$5.57 & 1.1$\pm$0.21(0.4$\pm$0.07) & 1.8$\pm$0.40 & 13.3$\pm$7.07 \\
J152141+075921 & 3.8$\pm$2.06 & 1.6$\pm$0.30(0.6$\pm$0.11) & 2.2$\pm$0.52 & 5.3$\pm$2.87 \\
J152521+075720 & 3.3$\pm$1.77 & 1.0$\pm$0.19(0.3$\pm$0.07) & 2.0$\pm$0.44 & 3.9$\pm$2.14 \\
J161245+081701 & 6.6$\pm$3.56 & 0.5$\pm$0.10(0.2$\pm$0.04) & 2.8$\pm$0.67 & 7.7$\pm$4.27 \\
J210358-072802 & 3.7$\pm$2.11 & 0.2$\pm$0.07(0.1$\pm$0.02) & 5.2$\pm$1.32 & 5.5$\pm$2.92 \\
Haro11 & 34.2$\pm$18.91 & 2.5$\pm$0.55(0.9$\pm$0.19) & 4.9$\pm$1.61 & 91.2$\pm$51.38 \\
VV114 & 28.9$\pm$16.51 & 1.1$\pm$0.33(0.4$\pm$0.11) & 7.3$\pm$2.36 & 88.6$\pm$50.31 \\
NGC1140 & 2.3$\pm$1.22 & 7.3$\pm$1.39(2.6$\pm$0.49) & 3.2$\pm$1.19 & 7.9$\pm$4.31 \\
SBS0335-052 & 5.0$\pm$2.59 & 40.5$\pm$7.87(14.2$\pm$2.75) & 1.2$\pm$0.38 & 19.0$\pm$10.63 \\
Tol0440-381 & 6.6$\pm$3.47 & 3.5$\pm$0.73(1.2$\pm$0.26) & 4.4$\pm$1.57 & 20.1$\pm$11.56 \\
NGC1705 & 0.5$\pm$0.27 & 8.7$\pm$1.62(3.0$\pm$0.57) & 1.5$\pm$0.37 & 1.2$\pm$0.65 \\
NGC1741 & 11.2$\pm$5.99 & 5.0$\pm$0.99(1.7$\pm$0.35) & 3.7$\pm$1.37 & 35.5$\pm$19.62 \\
IZw18 & 0.2$\pm$0.11 & 33.5$\pm$6.42(11.7$\pm$2.25) & 0.6$\pm$0.15 & 0.4$\pm$0.23 \\
NGC3310 & 1.1$\pm$0.58 & 1.1$\pm$0.21(0.4$\pm$0.07) & 2.9$\pm$0.70 & 1.5$\pm$0.78 \\
Haro3 & 1.0$\pm$0.51 & 6.0$\pm$1.10(2.1$\pm$0.38) & 1.8$\pm$0.45 & 2.0$\pm$1.08 \\
NGC3690 & 16.4 & 1.2(0.4) & 8.8 & 74.2$\pm$43.24 \\
NGC4214 & 0.4$\pm$0.23 & 8.6$\pm$1.60(3.0$\pm$0.56) & 2.2$\pm$0.63 & 1.2$\pm$0.66 \\
IRAS19245-4140 & 4.5$\pm$2.41 & 5.6$\pm$1.00(2.0$\pm$0.35) & 2.1$\pm$0.52 & 10.1$\pm$5.62 \\
NGC7673 & 6.4$\pm$3.39 & 3.5$\pm$0.71(1.2$\pm$0.25) & 4.6$\pm$1.75 & 18.9$\pm$10.34 \\
NGC7714 & 4.8$\pm$2.65 & 1.8$\pm$0.39(0.6$\pm$0.14) & 4.4$\pm$1.29 & 10.0$\pm$5.50 \\
J0826+43 & 37.4$\pm$22.64 & 0.3$\pm$0.07(0.1$\pm$0.03) & 3.7$\pm$0.98 & 52.5$\pm$28.72 \\
J0905+57 & 8.3$\pm$4.85 & 0.1$\pm$0.02(0.0$\pm$0.01) & 3.0$\pm$0.75 & 9.2$\pm$5.08 \\
J0944+09 & 21.7$\pm$12.15 & 0.3$\pm$0.06(0.1$\pm$0.02) & 3.0$\pm$0.74 & 26.8$\pm$14.73 \\
J1104+59 & 9.7$\pm$5.16 & 0.4$\pm$0.09(0.1$\pm$0.03) & 3.3$\pm$0.88 & 13.9$\pm$7.70 \\
J1506+54 & 19.4$\pm$10.70 & 0.2$\pm$0.05(0.1$\pm$0.02) & 3.3$\pm$0.86 & 24.7$\pm$13.52 \\
J1506+61 & 37.2$\pm$20.05 & 0.5$\pm$0.10(0.2$\pm$0.03) & 2.6$\pm$0.67 & 45.1$\pm$24.70 \\
J1558+39 & 93.3$\pm$52.39 & 0.4$\pm$0.10(0.1$\pm$0.04) & 3.7$\pm$0.96 & 129.8$\pm$72.42 \\
J1613+28 & 12.1$\pm$7.40 & 0.1$\pm$0.05(0.0$\pm$0.02) & 5.2$\pm$1.49 & 21.2$\pm$11.48 \\
J1713+28 & 73.9$\pm$42.82 & 0.4$\pm$0.11(0.1$\pm$0.04) & 4.0$\pm$1.11 & 124.0$\pm$68.66 \\
\hline
\end{longtable}

%--- 6 ---
\begin{table*}
\caption{
Estimated mass flux and mass loading rate of high-redshift galaxies as mentioned in Section \ref{sec_res_mfx}. 
(4) represents the estimated mass flux using \citet{Chevalier85} model. }
  \begin{center}
    \begin{tabular}{lcccc} \hline
      \    & (1)              & (2)                                     & (3)             & (4)              \\
      line & mass flux         & mass loading rate (mass loading factor) & transonic point & mass flux (CC85)  \\ 
      \    & ($M_{\odot}$ yr$^{-1}$) & \                                       & (kpc)           & ($M_{\odot}$ yr$^{-1}$) \\ \hline
NaID & 3.6$\pm$1.05 & 3.3$\pm$0.52(1.2$\pm$0.18) & 5.0$\pm$1.33 & 12.1$\pm$3.66 \\
NaID & 4.2$\pm$0.71 & 2.5$\pm$0.33(0.9$\pm$0.11) & 5.4$\pm$1.02 & 13.5$\pm$1.87 \\
NaID & 4.6$\pm$0.67 & 2.1$\pm$0.27(0.7$\pm$0.09) & 5.2$\pm$0.98 & 13.4$\pm$1.42 \\
NaID & 6.6$\pm$1.46 & 2.2$\pm$0.39(0.8$\pm$0.14) & 6.2$\pm$1.57 & 22.8$\pm$3.83 \\
NaID & 7.0$\pm$1.62 & 1.6$\pm$0.33(0.6$\pm$0.12) & 6.3$\pm$1.78 & 21.9$\pm$2.72 \\
NaID & 8.1$\pm$2.13 & 1.3$\pm$0.29(0.4$\pm$0.10) & 6.8$\pm$1.90 & 25.2$\pm$4.47 \\
MgI & 4.4$\pm$2.17 & 1.3$\pm$0.29(0.5$\pm$0.10) & 1.9$\pm$0.64 & 7.9$\pm$4.42 \\
MgI & 10.7 & 1.7(0.6) & 3.6 & 57.3$\pm$126.96 \\
MgI & 13.5 & 1.2(0.4) & 4.3 & 70.7$\pm$565.39 \\
MgII & 4.7$\pm$2.13 & 1.4$\pm$0.17(0.5$\pm$0.06) & 1.9$\pm$0.66 & 8.9$\pm$4.06 \\
MgII & 7.4$\pm$1.72 & 1.2$\pm$0.20(0.4$\pm$0.07) & 3.0$\pm$0.88 & 17.6$\pm$3.60 \\
MgII & 9.1$\pm$3.76 & 0.7$\pm$0.14(0.3$\pm$0.05) & 3.6$\pm$1.08 & 20.8$\pm$7.63 \\
CII & 6.0$\pm$6.51 & 0.6$\pm$0.13(0.2$\pm$0.05) & 2.0$\pm$1.34 & 9.5$\pm$9.71 \\
CIV & 5.8$\pm$6.37 & 0.5$\pm$0.13(0.2$\pm$0.04) & 2.0$\pm$1.32 & 9.3$\pm$9.36 \\
SiIIandCII & 5.5$\pm$3.08 & 0.5$\pm$0.17(0.2$\pm$0.06) & 1.2$\pm$0.35 & 11.5$\pm$8.06 \\
      \hline
    \end{tabular}
  \end{center}
\label{tab_mfx_hiz}
\end{table*}

%--- 7 ---
\begin{table*}
\caption{
Results of parameter fitting for the mass loading rate $\lambda$ in $z\sim 0$. 
See Section \ref{sec_res_frm}. }
  \begin{center}
\small
    \begin{tabular}{cccccc} \hline
      $\log_{10}M_{\rm stl}^{\rm tot}(M_{\rm \odot})$ & $n_{\rm m}$ & $\log_{10}A_{\rm m}$ & $\log_{10}v_{\rm m}({\rm km}$ ${\rm s}^{-1})$ & $v_{\rm max,m}({\rm km}$ ${\rm s}^{-1})$ & max error (\%) \\\hline
4.0 & -2.13 & 1.39 & 1.14 & 7.1 & 3.72 \\
4.5 & -2.12 & 1.39 & 1.23 & 8.9 & 3.83 \\
5.0 & -2.10 & 1.39 & 1.32 & 11.1 & 3.93 \\
5.5 & -2.09 & 1.39 & 1.41 & 14.5 & 4.01 \\
6.0 & -2.08 & 1.40 & 1.50 & 18.0 & 4.13 \\
6.5 & -2.07 & 1.39 & 1.59 & 23.2 & 4.21 \\
7.0 & -2.07 & 1.40 & 1.70 & 29.7 & 4.27 \\
7.5 & -2.06 & 1.39 & 1.79 & 37.7 & 4.14 \\
8.0 & -2.06 & 1.40 & 1.90 & 46.9 & 4.01 \\
8.5 & -2.06 & 1.41 & 2.02 & 62.1 & 3.85 \\
9.0 & -2.09 & 1.45 & 2.14 & 76.2 & 3.35 \\
9.5 & -2.14 & 1.54 & 2.29 & 95.6 & 2.84 \\
10.0 & -2.17 & 1.61 & 2.43 & 120.9 & 3.49 \\
10.5 & -2.16 & 1.62 & 2.55 & 163.8 & 3.26 \\
11.0 & -2.06 & 1.44 & 2.68 & 268.2 & 2.08 \\
11.5 & -1.98 & 1.21 & 2.84 & 567.4 & 2.29 \\
12.0 & -1.97 & 1.15 & 3.26 & 1771.2 & 3.71 \\
    \end{tabular}
    \begin{tabular}{cccccc} \hline
      $v_{\rm max}({\rm km}$ ${\rm s}^{-1})$ & $n_{\rm s}$ & $\log_{10}A_{\rm s}$ & $\log_{10}M_{\rm s}(M_{\rm \odot})$ & $v_{\rm circ,s}$ ($\log_{10}M_{\rm stl,s}^{\rm tot}$) & max error (\%) \\\hline
100 & -0.19 & 1.03 & 5.77 & 76 (9.6) & 4.64 \\
200 & -0.23 & 0.79 & 7.71 & 155 (10.7) & 10.74 \\
300 & -0.27 & 0.74 & 8.75 & 282 (11.0) & 11.93 \\
400 & -0.35 & 0.90 & 9.50 & 530 (11.2) & 10.61 \\
500 & -0.46 & 1.21 & 10.01 & 1174 (11.4) & 8.81 \\
600 & -0.58 & 1.67 & 10.39 & 1860 (11.5) & 7.11 \\
700 & -0.74 & 2.32 & 10.72 & 3081 (11.6) & 6.02 \\
800 & -0.68 & 2.00 & 10.74 & 3081 (11.6) & 4.57 \\
900 & -0.81 & 2.58 & 10.96 & 5360 (11.7) & 3.83 \\
1000 & -0.80 & 2.49 & 11.02 & 5360 (11.7) & 3.40 \\
1100 & -0.89 & 2.88 & 11.14 & 9830 (11.8) & 3.87 \\
1200 & -0.95 & 3.13 & 11.23 & 9830 (11.8) & 4.07 \\
1300 & -0.89 & 2.82 & 11.22 & 19063 (11.9) & 3.87 \\
1400 & -1.00 & 3.31 & 11.33 & 19063 (11.9) & 4.39 \\
1500 & -1.00 & 3.27 & 11.36 & 19063 (11.9) & 4.38 \\
1600 & -0.95 & 3.02 & 11.36 & 19063 (11.9) & 4.40 \\
1700 & -1.07 & 3.59 & 11.46 & 39226 (12.0) & 4.72 \\
1800 & -1.10 & 3.74 & 11.51 & 39226 (12.0) & 4.78 \\
1900 & -1.07 & 3.54 & 11.50 & 39226 (12.0) & 4.74 \\
2000 & -1.23 & 4.39 & 11.64 & 39226 (12.0) & 5.02 \\
    \hline
    \end{tabular}
  \end{center}
\label{tab_res_frm}
\end{table*}
\begin{table*}
  \begin{center}
    \begin{tabular}{cccccc} \hline
      $v_{\rm max}({\rm km}$ ${\rm s}^{-1})$ & $n_{\rm c}$ & $\log_{10}A_{\rm c}$ & $\log_{10}v_{\rm c}({\rm km}$ ${\rm s}^{-1})$ & $v_{\rm circ,c}$ ($\log_{10}M_{\rm stl,c}^{\rm tot}$) & max error (\%) \\\hline
100 & -3.78 & 1.66 & 2.09 & 77 (9.6) & 4.61 \\
200 & -2.86 & 1.33 & 2.22 & 170 (10.8) & 7.51 \\
300 & -1.10 & 0.19 & 1.82 & 353 (11.1) & 12.00 \\
400 & -0.72 & -0.09 & 1.66 & 705 (11.3) & 12.80 \\
500 & -0.58 & -0.26 & 1.64 & 1185 (11.4) & 12.37 \\
600 & -0.49 & -0.40 & 1.61 & 2171 (11.5) & 11.69 \\
700 & -0.45 & -0.50 & 1.68 & 3144 (11.6) & 11.23 \\
800 & -0.40 & -0.60 & 1.64 & 4977 (11.7) & 9.95 \\
900 & -0.38 & -0.69 & 1.69 & 6592 (11.7) & 9.36 \\
1000 & -0.36 & -0.76 & 1.73 & 8301 (11.8) & 8.81 \\
1100 & -0.34 & -0.83 & 1.75 & 9570 (11.8) & 8.11 \\
1200 & -0.33 & -0.89 & 1.83 & 9570 (11.8) & 7.83 \\
1300 & -0.32 & -0.95 & 1.87 & 9570 (11.8) & 7.43 \\
1400 & -0.31 & -1.01 & 1.88 & 9570 (11.8) & 6.88 \\
1500 & -0.30 & -1.06 & 1.89 & 9570 (11.8) & 6.47 \\
1600 & -0.30 & -1.10 & 1.94 & 9570 (11.8) & 6.50 \\
1700 & -0.29 & -1.15 & 1.97 & 9570 (11.8) & 6.01 \\
1800 & -0.28 & -1.20 & 1.98 & 9570 (11.8) & 5.61 \\
1900 & -0.27 & -1.24 & 2.02 & 9570 (11.8) & 5.55 \\
2000 & -0.27 & -1.28 & 2.02 & 9570 (11.8) & 5.19 \\
      \hline
    \end{tabular}
  \end{center}
\end{table*}

%--- 8 ---
\begin{table*}[ht]
\caption{
Parameters of M82. 
The distance of M82 is assumed to be 3.63Mpc. 
(1) The stellar mass is estimated from the K-band luminosity and $B-V$ colour \citep{Jarrett03,Dale07,Bell03}. 
(2) We use the K-band effective radius \citep{Jarrett03}. 
(3) To determine the dark matter halo mass distribution, the NFW model is fitted to the observed rotation curve \citep{Greco12} with the stellar mass distribution estimated above. 
}
\begin{center}
 \begin{tabular}{|c|c|c|} 
  \multicolumn{3}{l}{stellar mass $^{\rm (1,2)}$} \\ \hline
  M$_{\rm stl}$    ($M_{\odot}$)    &
  r$_{\rm efct}$   (kpc)            &
  \ \\ \hline
  5.19 $\times$ 10$^8$ &  0.723 & \ \\ \hline
  \multicolumn{3}{l}{halo mass $^{\rm (3)}$}   \\ \hline
  M$_{\rm dmh}$    ($M_{\odot}$)    &
  r$_{\rm dmh}$    (kpc)            &
  r$_{\rm virial}$ (kpc)            \\ \hline
  6.13 $\times$ 10$^{10}$ & 2.12 & 102 \\ \hline
 \end{tabular}
\end{center}
\label{tab_m82_prm}
\end{table*}

%--- 9 ---
\begin{table*}
\caption{
Fitting result of column density in Section \ref{app_ccm} where $\epsilon_{\rm rem}$ represents the residual rate of the total SNe energy after radiative cooling, $\dot{m}_{\infty}$ indicates the final mass flux at infinity, $\eta$ ($\lambda$) shows the mass loading factor (mass loading rate), $v_{\rm term}$ represents the terminal velocity,  and $\dot{e}_{\infty}$ ($\equiv$ (1/2) $v_{\rm term}^2$ $\dot{m}_{\infty}$) indicates the final energy flux at infinity. }
  \begin{tabular}{cccccccc} 
    \hline
    \multicolumn{8}{l}{Transonic Outflow model} \\ 
    (SFR, $\epsilon_{\rm rem}$) & transonic point & $\dot{m}_{\infty}$ & $\lambda$ ($\eta$) & $v_{\rm term}$ & $v_{\rm max}$ & $r$($v_{\rm max}$) & $\log_{10}\dot{e}_{\infty}$ \\
    & (pc) & ($M_{\odot}$ yr$^{-1}$) & & (km s$^{-1}$) & (km s$^{-1}$) & (kpc) & (erg yr$^{-1}$) \\
    \hline
    (2.4, 0.1)  & 782 & 5.96 & 7.10(2.49)  & 66.2 & 146 & 2.90 & 47.4 \\ 
(2.4, 0.3)  & 619 & 10.4 & 12.3(4.31)  & 243  & 249 & 19.6 & 48.8 \\ 
(6.8, 0.1)  & 624 & 10.1 & 4.25(1.49)  & 234  & 241 & 16.3 & 48.7 \\ 
(6.8, 0.3)  & 554 & 15.7 & 6.62(2.32)  & 414  & - & -  & 49.4 \\ 
(23.1, 0.1) & 550 & 16.5 & 2.05(0.716) & 436  & - & -  & 49.5 \\ 
(23.1, 0.3) & 523 & 24.6 & 3.04(1.06)  & 675  & - & -  & 50.0 \\
  \end{tabular}
  \begin{tabular}{cccccccc} 
    \hline
    \multicolumn{8}{l}{Chevalier \& Clegg (1985) model} \\ 
    (SFR, $\epsilon_{\rm rem}$) & transonic point & $\dot{m}$$_{\infty}$ & $\lambda$ ($\eta$) & $v_{\rm term}$ & $v_{\rm max}$ & $r$($v_{\rm max}$) & $\log_{10}\dot{e}$$_{\infty}$ \\
    & (pc) & ($M_{\odot}$ yr$^{-1}$) & & (km s$^{-1}$) & (km s$^{-1}$) & (kpc) & (erg yr$^{-1}$) \\
    \hline
    (2.4, 0.1)  & 200 & 8.66 & 10.3(3.61)  & 228 & - & - & 48.6 \\ 
(2.4, 0.3)  & 200 & 12.5 & 14.9(5.21)  & 328 & - & - & 49.1 \\ 
(6.8, 0.1)  & 200 & 12.3 & 5.16(1.80)  & 322 & - & - & 49.1 \\ 
(6.8, 0.3)  & 200 & 17.7 & 7.44(2.60)  & 464 & - & - & 49.6 \\ 
(23.1, 0.1) & 200 & 18.4 & 2.28(0.796) & 484 & - & - & 49.6 \\ 
(23.1, 0.3) & 200 & 26.2 & 3.24(1.14)  & 703 & - & - & 50.1 \\
    \hline
  \end{tabular}
\label{tab_m82_mfx}
\end{table*}

\begin{longtable}{lccccc}
\caption{
Estimated parameters of local galaxies. 
The stellar mass and redshift values are listed in Table \ref{tab_prm_loc}. 
The estimation methods are shown in Section \ref{sec_mdl_prm}.}
\label{tab_est_loc}

\multicolumn{6}{c}{} \\
      \    & (5)              & (6)                       & (7)                      & (8)                       & (9)                           \\
      name & effective radius & $\log_{10}$ halo mass     & $\log_{10}$ scale radius & $\log_{10}$ virial radius & $\log_{10}$ circular vel. \\ 
      \    & (kpc)            & ($M_{\odot}$)             & (kpc)                    & (kpc)                     & (km s$^{-1}$)                  \\
\hline
\endfirsthead
      \    & (5)              & (6)                       & (7)                      & (8)                       & (9)                           \\
      name & effective radius & $\log_{10}$ halo mass     & $\log_{10}$ scale radius & $\log_{10}$ virial radius & $\log_{10}$ circular vel. \\ 
      \    & (kpc)            & ($M_{\odot}$)             & (kpc)                    & (kpc)                     & (km s$^{-1}$)                  \\
\hline
\endhead
\hline
\endfoot
\endlastfoot
J002101+005248 & 2.53$\pm$0.502 & 11.3$\pm$0.17 & 1.09$\pm$0.082 & 2.2$\pm$0.06 & 1.9$\pm$0.06 \\ 
J005527-002148 & 3.01$\pm$0.680 & 11.5$\pm$0.19 & 1.23$\pm$0.086 & 2.2$\pm$0.06 & 1.9$\pm$0.06 \\ 
J015028+130858 & 4.68$\pm$0.989 & 12.0$\pm$0.42 & 1.44$\pm$0.181 & 2.4$\pm$0.14 & 2.1$\pm$0.14 \\ 
J021348+125951 & 5.03$\pm$1.076 & 12.2$\pm$0.60 & 1.53$\pm$0.255 & 2.4$\pm$0.20 & 2.2$\pm$0.20 \\ 
J080844+394852 & 3.47$\pm$0.760 & 11.6$\pm$0.20 & 1.20$\pm$0.094 & 2.3$\pm$0.07 & 1.9$\pm$0.07 \\ 
J082354+280621 & 1.79$\pm$0.343 & 10.9$\pm$0.20 & 0.94$\pm$0.094 & 2.0$\pm$0.07 & 1.7$\pm$0.07 \\ 
J092159+450912 & 5.94$\pm$1.259 & 12.9$\pm$1.02 & 1.80$\pm$0.437 & 2.7$\pm$0.34 & 2.4$\pm$0.34 \\ 
J092600+442736 & 2.06$\pm$0.409 & 11.2$\pm$0.17 & 1.10$\pm$0.081 & 2.1$\pm$0.06 & 1.8$\pm$0.06 \\ 
J093813+542825 & 2.69$\pm$0.551 & 11.4$\pm$0.17 & 1.11$\pm$0.079 & 2.2$\pm$0.06 & 1.9$\pm$0.06 \\ 
J102548+362258 & 2.32$\pm$0.448 & 11.3$\pm$0.17 & 1.14$\pm$0.079 & 2.1$\pm$0.06 & 1.8$\pm$0.06 \\ 
J111244+550347 & 4.43$\pm$0.946 & 11.9$\pm$0.35 & 1.40$\pm$0.152 & 2.3$\pm$0.12 & 2.1$\pm$0.12 \\ 
J111323+293039 & 2.85$\pm$0.638 & 11.5$\pm$0.18 & 1.21$\pm$0.084 & 2.2$\pm$0.06 & 1.9$\pm$0.06 \\ 
J114422+401221 & 3.64$\pm$0.807 & 11.6$\pm$0.23 & 1.30$\pm$0.106 & 2.3$\pm$0.08 & 2.0$\pm$0.08 \\ 
J141454+054047 & 1.62$\pm$0.317 & 10.9$\pm$0.21 & 0.91$\pm$0.098 & 2.0$\pm$0.07 & 1.7$\pm$0.07 \\ 
J141612+122340 & 3.91$\pm$0.887 & 11.7$\pm$0.26 & 1.33$\pm$0.114 & 2.3$\pm$0.09 & 2.0$\pm$0.09 \\ 
J142856+165339 & 2.82$\pm$0.615 & 11.5$\pm$0.18 & 1.21$\pm$0.084 & 2.2$\pm$0.06 & 1.9$\pm$0.06 \\ 
J142947+064334 & 2.48$\pm$0.509 & 11.4$\pm$0.17 & 1.17$\pm$0.082 & 2.2$\pm$0.06 & 1.9$\pm$0.06 \\ 
J152141+075921 & 2.86$\pm$0.592 & 11.4$\pm$0.17 & 1.13$\pm$0.080 & 2.2$\pm$0.06 & 1.9$\pm$0.06 \\ 
J152521+075720 & 2.76$\pm$0.563 & 11.4$\pm$0.17 & 1.12$\pm$0.080 & 2.2$\pm$0.06 & 1.9$\pm$0.06 \\ 
J161245+081701 & 3.81$\pm$0.857 & 11.7$\pm$0.26 & 1.33$\pm$0.116 & 2.3$\pm$0.09 & 2.0$\pm$0.09 \\ 
J210358-072802 & 6.74$\pm$1.412 & 13.2$\pm$1.20 & 1.96$\pm$0.510 & 2.8$\pm$0.40 & 2.5$\pm$0.40 \\ 
Haro11 & 4.86$\pm$1.023 & 11.9$\pm$0.37 & 1.35$\pm$0.160 & 2.4$\pm$0.12 & 2.0$\pm$0.12 \\ 
VV114 & 6.96$\pm$1.429 & 12.9$\pm$0.99 & 1.79$\pm$0.426 & 2.7$\pm$0.33 & 2.4$\pm$0.33 \\ 
NGC1140 & 2.96$\pm$0.592 & 11.4$\pm$0.17 & 1.13$\pm$0.080 & 2.2$\pm$0.06 & 1.9$\pm$0.06 \\ 
SBS0335-052 & 1.16$\pm$0.231 & 10.4$\pm$0.24 & 0.72$\pm$0.110 & 1.9$\pm$0.08 & 1.5$\pm$0.08 \\ 
Tol0440-381 & 4.20$\pm$0.905 & 11.7$\pm$0.26 & 1.27$\pm$0.116 & 2.3$\pm$0.09 & 2.0$\pm$0.09 \\ 
NGC1705 & 1.87$\pm$0.370 & 10.9$\pm$0.20 & 0.95$\pm$0.093 & 2.1$\pm$0.07 & 1.7$\pm$0.07 \\ 
NGC1741 & 3.56$\pm$0.761 & 11.5$\pm$0.20 & 1.20$\pm$0.094 & 2.3$\pm$0.07 & 1.9$\pm$0.07 \\ 
IZw18 & 0.79$\pm$0.162 & 9.9$\pm$0.25 & 0.51$\pm$0.111 & 1.7$\pm$0.08 & 1.4$\pm$0.08 \\ 
NGC3310 & 3.83$\pm$0.818 & 11.6$\pm$0.21 & 1.23$\pm$0.098 & 2.3$\pm$0.07 & 1.9$\pm$0.07 \\ 
Haro3 & 2.22$\pm$0.433 & 11.1$\pm$0.18 & 1.02$\pm$0.084 & 2.1$\pm$0.06 & 1.8$\pm$0.06 \\ 
NGC3690 & 7.40$\pm$1.528 & 13.2$\pm$1.20 & 1.93$\pm$0.517 & 2.8$\pm$0.40 & 2.5$\pm$0.40 \\ 
NGC4214 & 2.38$\pm$0.455 & 11.2$\pm$0.17 & 1.05$\pm$0.081 & 2.1$\pm$0.06 & 1.8$\pm$0.06 \\ 
IRAS19245-4140 & 2.48$\pm$0.478 & 11.2$\pm$0.17 & 1.07$\pm$0.080 & 2.1$\pm$0.06 & 1.8$\pm$0.06 \\ 
NGC7673 & 4.35$\pm$0.899 & 11.7$\pm$0.25 & 1.28$\pm$0.112 & 2.3$\pm$0.08 & 2.0$\pm$0.08 \\ 
NGC7714 & 4.90$\pm$0.987 & 11.9$\pm$0.36 & 1.36$\pm$0.156 & 2.4$\pm$0.12 & 2.0$\pm$0.12 \\ 
J0826+43 & 4.82$\pm$1.069 & 12.9$\pm$1.16 & 1.80$\pm$0.477 & 2.6$\pm$0.39 & 2.5$\pm$0.39 \\ 
J0905+57 & 4.27$\pm$0.999 & 12.6$\pm$1.02 & 1.71$\pm$0.417 & 2.5$\pm$0.34 & 2.4$\pm$0.34 \\ 
J0944+09 & 4.11$\pm$0.921 & 12.2$\pm$0.62 & 1.54$\pm$0.255 & 2.4$\pm$0.21 & 2.2$\pm$0.21 \\ 
J1104+59 & 4.30$\pm$0.993 & 12.4$\pm$0.81 & 1.60$\pm$0.334 & 2.4$\pm$0.27 & 2.3$\pm$0.27 \\ 
J1506+54 & 4.51$\pm$1.024 & 12.6$\pm$0.98 & 1.69$\pm$0.405 & 2.5$\pm$0.33 & 2.4$\pm$0.33 \\ 
J1506+61 & 3.50$\pm$0.814 & 11.9$\pm$0.30 & 1.37$\pm$0.129 & 2.3$\pm$0.10 & 2.1$\pm$0.10 \\ 
J1558+39 & 4.72$\pm$1.054 & 12.4$\pm$0.74 & 1.59$\pm$0.309 & 2.5$\pm$0.25 & 2.3$\pm$0.25 \\ 
J1613+28 & 6.63$\pm$1.430 & 14.3$\pm$1.76 & 2.40$\pm$0.736 & 3.1$\pm$0.59 & 2.9$\pm$0.59 \\ 
J1713+28 & 4.90$\pm$1.081 & 12.9$\pm$1.16 & 1.81$\pm$0.478 & 2.6$\pm$0.39 & 2.5$\pm$0.39 \\
      \hline
\end{longtable}

%--- 2 ---
\begin{table*}
\caption{
Estimated parameters of high-redshift galaxies. 
Parameters in Table \ref{tab_prm_hiz} are used as mentioned in Section \ref{sec_mdl_prm}. }
  \begin{center}
    \begin{tabular}{lcccccc} \hline
      \    & (5)              & (6)                       & (7)                      & (8)                       & (9)                           \\
      line & effective radius & $\log_{10}$ halo mass     & $\log_{10}$ scale radius & $\log_{10}$ virial radius & $\log_{10}$ circular velocity \\ 
      \    & (kpc)            & ($M_{\odot}$)             & (kpc)                    & (kpc)                     & (km s$^{-1}$)                        \\ \hline
NaID & 4.45$\pm$0.659 & 11.8$\pm$0.20 & 1.30$\pm$0.092 & 2.3$\pm$0.07 & 2.0$\pm$0.07 \\ 
NaID & 4.93$\pm$0.580 & 11.9$\pm$0.23 & 1.37$\pm$0.106 & 2.4$\pm$0.08 & 2.1$\pm$0.08 \\ 
NaID & 5.04$\pm$0.622 & 12.0$\pm$0.24 & 1.39$\pm$0.109 & 2.4$\pm$0.08 & 2.1$\pm$0.08 \\ 
NaID & 5.27$\pm$0.807 & 12.1$\pm$0.38 & 1.47$\pm$0.169 & 2.4$\pm$0.13 & 2.1$\pm$0.13 \\ 
NaID & 5.69$\pm$0.997 & 12.4$\pm$0.53 & 1.58$\pm$0.233 & 2.5$\pm$0.18 & 2.2$\pm$0.18 \\ 
NaID & 6.17$\pm$1.102 & 12.7$\pm$0.73 & 1.72$\pm$0.312 & 2.6$\pm$0.24 & 2.3$\pm$0.24 \\ 
MgI & 2.22$\pm$0.609 & 11.8$\pm$0.28 & 1.35$\pm$0.121 & 2.1$\pm$0.10 & 2.2$\pm$0.09 \\ 
MgI & 3.15$\pm$0.717 & 12.1$\pm$0.51 & 1.49$\pm$0.203 & 2.2$\pm$0.17 & 2.3$\pm$0.17 \\ 
MgI & 3.74$\pm$0.821 & 12.6$\pm$0.98 & 1.65$\pm$0.380 & 2.3$\pm$0.33 & 2.4$\pm$0.33 \\ 
MgII & 2.24$\pm$0.623 & 11.8$\pm$0.25 & 1.36$\pm$0.112 & 2.1$\pm$0.09 & 2.2$\pm$0.09 \\ 
MgII & 3.12$\pm$0.679 & 12.1$\pm$0.49 & 1.49$\pm$0.193 & 2.2$\pm$0.16 & 2.3$\pm$0.16 \\ 
MgII & 3.75$\pm$0.794 & 12.6$\pm$1.03 & 1.66$\pm$0.403 & 2.3$\pm$0.34 & 2.5$\pm$0.34 \\ 
CII & 2.37$\pm$1.200 & 12.2$\pm$1.13 & 1.46$\pm$0.427 & 2.1$\pm$0.38 & 2.4$\pm$0.38 \\ 
CIV & 2.40$\pm$1.176 & 12.2$\pm$1.02 & 1.46$\pm$0.390 & 2.1$\pm$0.34 & 2.4$\pm$0.34 \\ 
SiIIandCII & 1.39$\pm$0.385 & 12.2$\pm$1.19 & 1.17$\pm$0.447 & 1.8$\pm$0.40 & 2.6$\pm$0.40 \\
      \hline
    \end{tabular}
  \end{center}
\label{tab_est_hiz}
\end{table*}

\end{document}